\documentclass[aps,prb,showpacs, twocolumn, amsfonts, amsmath, floatfix, balancelastpage]{revtex4-1}

\usepackage{amsmath}
\usepackage{amssymb}
\usepackage{epsf}
\usepackage{epsfig}
\usepackage{graphicx}
\usepackage{pstricks}
\usepackage[english]{babel}
\usepackage[latin1]{inputenc}
\usepackage{rotating}
\usepackage{color}
\definecolor{darkblue}{rgb}{0.0,0.0,0.4}
\definecolor{darkgreen}{rgb}{0.0,0.4,0.0}
\definecolor{darkred}{rgb}{0.6,0.0,0.0}
\usepackage[bookmarksopen]{hyperref}
\hypersetup{colorlinks,linkcolor=darkgreen,citecolor=darkblue,urlcolor=darkblue}
\usepackage{slashed}
\usepackage{tabularx}
\usepackage{array}
\usepackage{balance}

\usepackage{txfonts}

\newcommand{\bea}{\begin{eqnarray}}
\newcommand{\eea}{\end{eqnarray}}

\newcommand{\eq}[1]{(\ref{eq:#1})}
\newcommand{\Eq}[1]{Eq.~(\ref{eq:#1})}

\newcommand{\Fig}[1]{Fig.~\ref{fig:#1}}
\newcommand{\fig}[1]{\ref{fig:#1}}

\newcommand{\Sect}[1]{Sect.~\ref{sec:#1}}
\newcommand{\sect}[1]{\ref{sec:#1}}

\newcommand{\citeafter}[2]{#1\cite{#2}}

\makeatletter
\newcommand*{\balancecolsandclearpage}{%
  \close@column@grid
  \clearpage
  \twocolumngrid
}
\makeatother

\usepackage{tikz}
\usetikzlibrary{arrows,shapes}
\usetikzlibrary{trees}
\usetikzlibrary{matrix,arrows} 			
\usetikzlibrary{positioning}				
\usetikzlibrary{calc,through}			
\usetikzlibrary{decorations.pathreplacing}  
\usepackage{pgffor}					

\usetikzlibrary{decorations.pathmorphing}	
\usetikzlibrary{decorations.markings}
\tikzset{
    vector/.style={decorate, decoration={snake}, draw},
	provector/.style={decorate, decoration={snake,amplitude=2.5pt}, draw},
	antivector/.style={decorate, decoration={snake,amplitude=-2.5pt}, draw},
    resum/.style={decorate, decoration={snake,amplitude=2pt}, draw=red},
    resumsup/.style={draw=none,
			      postaction={draw=blue,decoration={snake,amplitude=2pt}, decorate}},
    resumsdown/.style={draw=none,
			      postaction={draw=darkgreen,decoration={snake,amplitude=2pt}, decorate}},
    resumtup/.style={draw=none,
    			      postaction={decoration={markings,mark=at position .51 with {\arrow[scale=1,red]{<}}},decorate},
			      postaction={draw=red,decoration={snake,amplitude=1pt}, decorate}},
    resumtdown/.style={draw=none,
    			      postaction={decoration={markings,mark=at position .61 with {\arrow[scale=1,red]{>}}},decorate},
			      postaction={draw=red,decoration={snake,amplitude=1pt}, decorate}},
	provector/.style={decorate, decoration={snake,amplitude=2pt}, draw},
	antivector/.style={decorate, decoration={snake,amplitude=-2pt}, draw},
    fermion/.style={draw=black, postaction={decorate},
        decoration={markings,mark=at position .55 with {\arrow[draw=black]{>}}}},
    fermionup/.style={draw=blue, postaction={decorate},
        decoration={markings,mark=at position .54 with {\arrow[draw=blue]{>}}}},
    fermiondown/.style={draw=darkgreen, postaction={decorate},
        decoration={markings,mark=at position .54 with {\arrow[draw=darkgreen]{>}}}},
    fermionbar/.style={draw=black, postaction={decorate},
        decoration={markings,mark=at position .55 with {\arrow[draw=black]{<}}}},
    fermionbarup/.style={draw=blue, postaction={decorate},
        decoration={markings,mark=at position .54 with {\arrow[draw=blue]{<}}}},
    fermionbardown/.style={draw=darkgreen, postaction={decorate},
        decoration={markings,mark=at position .54 with {\arrow[draw=darkgreen]{<}}}},
    fermionnoarrow/.style={draw=black},
    gluon/.style={decorate, draw=black,
        decoration={coil,amplitude=4pt, segment length=5pt}},
    scalar/.style={dashed,draw=black, postaction={decorate},
        decoration={markings,mark=at position .55 with {\arrow[draw=black]{>}}}},
    scalarbar/.style={dashed,draw=black, postaction={decorate},
        decoration={markings,mark=at position .55 with {\arrow[draw=black]{<}}}},
    scalarnoarrow/.style={dashed,draw=black},
    electron/.style={draw=black, postaction={decorate},
        decoration={markings,mark=at position .55 with {\arrow[draw=black]{>}}}},
	bigvector/.style={decorate, decoration={snake,amplitude=4pt}, draw},
}

\tikzstyle{block} = [draw, rectangle, 
    minimum height=3em, minimum width=6em]
    
\begin{document}

\title{Build-up of the Kondo effect from real-time effective action for the Anderson impurity model}

\author{%
Sebastian Bock,$^{1,3}$ 
Alexander Liluashvili,$^{1,3}$ and 
Thomas Gasenzer$^{1,2,3,}$}
\email{Email for correspondence: t.gasenzer@uni-heidelberg.de}
\affiliation{$^{1}$Institut f\"ur Theoretische Physik,
             Ruprecht-Karls-Universit\"at Heidelberg,
             Philosophenweg~16,
             69120~Heidelberg, Germany}
\affiliation{$^{2}$Kirchhoff-Institut f\"ur Physik,
             Universit\"at Heidelberg,
             Im Neuenheimer Feld 227,
             69120~Heidelberg, Germany}
\affiliation{$^{3}$ExtreMe Matter Institute EMMI,
             GSI Helmholtzzentrum f\"ur Schwerionenforschung GmbH, 
             Planckstra\ss e~1, 
             64291~Darmstadt, Germany} 

\date{24 February 2016}

\begin{abstract}

{
  The nonequilibrium time evolution of a quantum dot is studied by means of dynamic equations for time-dependent Greens functions derived from a two-particle-irreducible (2PI) effective action for the Anderson impurity model.
  Coupling the dot between two leads at different voltages, the dynamics of the current through the dot is investigated.
  We show that the 2PI approach is capable to describe the dynamical build-up of the Kondo effect, which shows up as a sharp resonance in the spectral function, with a width exponentially suppressed in the electron self coupling on the dot.  
  An external voltage applied to the dot is found to deteriorate the Kondo effect at the hybridization scale.
  The dynamic equations are evaluated within different nonperturbative resummation schemes, within the direct, particle-particle, and particle-hole channels, as well as their combination, and the results compared with that from other methods.
  }
\end{abstract}

\pacs{   03.65.Db, 
       05.60.Gg,  
       71.10.-w,  
       73.63.Kv  
}

\maketitle
%
\section{Introduction}\label{Introduction}
Studying electron transport experimentally in nanodevices, including artificially designed quantum dots, nanotubes, and single molecules, has become a standard nanoscale-technology with many applications ranging from advanced materials to medicine\citeafter{.}{Bera2010a.ma3042260,Algar2010a,Carey2015a.chemrev.5b00063,Li2015a.ADFM:ADFM201501250,Kairdolf2013a.annurev-anchem-060908-155136} 
Recently, ultra-cold atomic gases have been proposed as alternative systems to study strongly correlated transport at the quantum level\citeafter{,}{Recati2005a.PhysRevLett.94.040404,Lewenstein2007a,Nishida2013a.PhysRevLett.111.135301,Bauer2013a.PhysRevLett.111.215304,Nishida2016a.PhysRevA.93.011606} and dots considered as a means for quantum computing\citeafter{.}{imamoglu1999,li2003,kroutvar2004,atatuere2006,press2008}
  
The single quantum dot serves as a seemingly simple but inherently versatile testing ground for theoretical methods: 
When coupled to an environment it entails intricate many-body effects as well as nonequilibrium dynamics in a regime where linear response theory does not apply. 
The Kondo effect in a quantum dot coupled to two electrodes leads to an enhanced differential conductance through the dot, at zero bias voltage between the leads and at low temperatures\citeafter{.}{Hewson1993a.KondoHeavyFermions,Wingreen1994,goldhaber-gordon1998a, goldhaber-gordon1998b,cronenwett1998,nygard2000}
The mathematical description of correlation phenomena underlying the Kondo effect and in particular the emergence of the Kondo temperature scale have represented a special challenge. 
In this work, we study the single-impurity Anderson model within a self-consistent functional field theoretic approach and compute the transient dynamics leading to the build-up of the Kondo effect.
Our results are consistent with a minimum time the Kondo effect needs to build up which is inversely proportional to the respective Kondo temperature scale\citeafter{.}{Nordlander1999a.PhysRevLett.83.808}
An external voltage applied to the dot is found to deteriorate the Kondo resonance\cite{Kaminski2000,muehlbacher2011} at the hybridization scale.

The Kondo effect in thermal equilibrium can be described in terms of exact solutions of the single-impurity Anderson model (SIAM) obtained by means of a Bethe ansatz\citeafter{.}{tsvelick1983, andrei1983}
Dynamical properties encoded in the spectral function are well understood in equilibrium\citeafter{,}{bulla2008} but the precise time evolution from a given initial to the stationary state is still a matter of research.
Also the stationary electrical current for large bias voltages between the leads, beyond the linear-response regime is subject to continuing debate. 
Out of equilibrium, the Kondo regime has been investigated within the Kondo model, obtained in the noncrossing approximation at infinite local coupling $U$\citeafter{,}{Wingreen1994,Nordlander1999a.PhysRevLett.83.808} as well as integrability of the Anderson model\citeafter{.}{Konik2002}
Perturbative studies made use of Fermi liquid theory \cite{Oguri2001,shastry2013} and its extensions for computing full counting statistics\citeafter{.}{gogolin2006, schmidt2007}
Further studies were based on the perturbative evaluation of nonequilibrium Green's functions\citeafter{,}{komnik2004} diagrammatic Monte-Carlo methods\citeafter{,}{muehlbacher2011} self-consistent perturbation theory\citeafter{,}{ding2013} renormalized\citeafter{,}{hewson2001, hewson2005, edwards2011, edwards2013} and cluster perturbation theory\citeafter{,}{nuss2012} and an auxiliary master equation approach\citeafter{.}{Dorda2013a.PhysRevB.88.165124,Dorda2015a.PhysRevB.92.125145}

%
\begin{figure}[t]
\centering
\includegraphics[width=0.435 \textwidth]{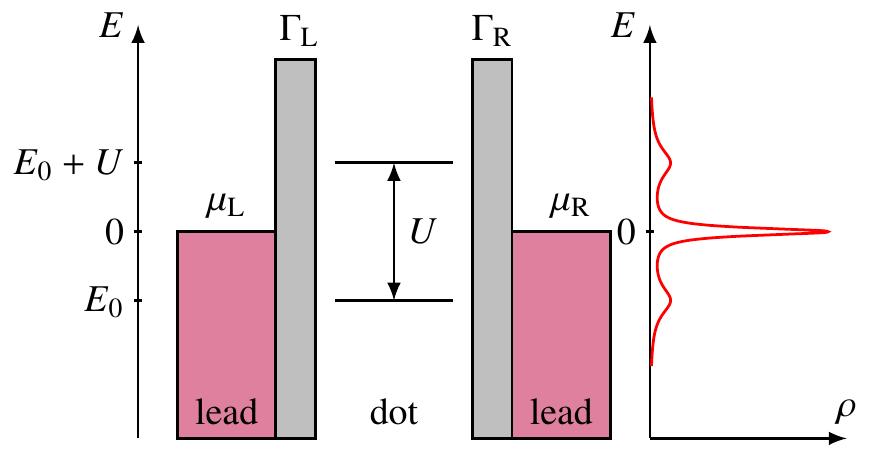}
\caption{(Color online)
Schematic representation of the energy levels in the Anderson model of a quantum dot for the case of a vanishing magnetic field. 
The lower energy level of the dot corresponds to a half-filled dot, the upper level corresponds to a filled one.
These levels are separated by the interaction energy $U$.
The energy zero is set to the  Fermi edges of the leads, $\mu_\mathrm{L}=\mu_\mathrm{R}=0$.
$\Gamma_\mathrm{L,R}$ measure the tunneling rates of electrons between the dot and the leads.
The right panel depicts qualitatively the shape of the spectral function $\rho(E)$ of the dot at low temperatures where a sharp Kondo resonance rises up around the Fermi edge when the dot is tuned to the particle-hole symmetric point, i.e., for $E_{0}=-U/2$. 
This resonance sits between the two side peaks which reflect the tunneling processes into/out of the empty/filled dot.}
\label{fig:dot}
\end{figure}

Several renormalization-group methods have been used to investigate the stationary state of the system: the perturbative real-time renormalization group (RG)\citeafter{,}{Schoeller2000, Andergassen2011, kashuba2013} nonequilibrium extensions of perturbative RG\citeafter{,}{Kaminski2000,Rosch2001,Rosch2003,rosch2003b, rosch2005} and the functional RG approach in its generalization to nonequilibrium situations, see, e.g., Refs.~\onlinecite{Andergassen2006,Gezzi2007,Jakobs2007,Jakobs2010,Karrasch2010,Pletyukhov2010,Karrasch2010b,Andergassen2011, bartosch2009, isidori2010, kopietz2010, schmidt2010, kinza2013, streib2013}.
Further methods include Bethe-Salpeter equations in parquet approximation\citeafter{,}{Janis2007,
Augustinsky2010a.PhysRevB.83.035114} the slave-boson approach\citeafter{,}{smirnov2011a}
and Chebyshev expansion using matrix product states\citeafter{.}{ganahl2014}
The numerical renormalization group (NRG) method has also been successful in describing dynamical correlations of such systems\citeafter{.}{Anders2005,
Roosen2008,Schmitt2010,zitko2009,zitko2011,osolin2013}

The quantum Monte Carlo (QMC) method \cite{egger2000} is numerically exact, but the sign problem permits only short simulation times at small temperatures.
In recent years, the quantum-information inspired time-dependent density matrix renormalization group (tDMRG) approach has been applied to the single-impurity problem\citeafter{,}{al-hassanieh2006, boulat2008, dias2008, heidrich-meisner2009, nuss2013} as well as real-time Monte-Carlo methods \cite{muehlbacher2008, schiro2009, gull2011} and the iterative sum of path integrals (ISPI) approach \cite{weiss2008,
weiss2013} which is numerically exact, but depends on the correlation time of the system being small.
Further methods including the time-dependent Gutzwiller approach\citeafter{,}{lanata2012} excitation states\citeafter{,}{wangpei2012}
and influence functional path integral (INFPI) \cite{segal2011} were employed to quantum impurity models.
Several of these theoretical methods were recently compared in Ref.~\onlinecite{eckel2010} which also contains a concise list of related previous studies.

In this article we use the two-particle-irreducible (2PI) effective action\citeafter{,}{Cornwall1974a} originally termed $\Phi$-functional\citeafter{,}{luttingerward,Baym1962} to derive Kadanoff-Baym dynamic equations describing the transient and stationary transport through a quantum dot. 
This method ensures that, for the case of a closed system, vital symmetries such as the total energy and particle number are conserved during the time evolution, irrespective to the approximation chosen. 
The method has been successfully used to describe thermalization of relativistic  and nonrelativistic systems of bosonic and fermionic fields in different dimensions, see, e.g.~Refs.~\onlinecite{Berges:2001fi,Berges2003a,Gasenzer:2005ze,Arrizabalaga:2005tf,Berges:2007ym,Gasenzer:2008zz,Gasenzer:2010rq,Kronenwett:2010ic,Giraud:2009tn,Berges:2015kfa}.

Here, we use this method to compute the transient filling dynamics of an initially empty dot coupled to zero-temperature leads, the dynamical build-up of the Kondo resonance in the spectral function of the dot electrons, as well as the current-voltage characteristics in the Kondo regime.
We show, in particular, that the width of the Kondo resonance in the stationary state exhibits the expected exponential suppression in the interaction strength on the dot.

Expanding upon our previous work \cite{Sexty:2010ra} we have developed methods for evaluating the resulting equations with higher precision and within different possible approximation schemes.
We use three different truncations, as well as their combination, to approximate the effective action and with it the self-energy of the quantum dot: one resummation scheme includes the contribution of the spin-aligned bubble chains summed to all orders (direct or $s$-channel), as well as two different ones summing ladder chains in the particle-particle and particle-hole (each with opposite spins) channels, also termed $t$- and $u$-channels, respectively.
Summing the three nonperturbative self-energies gives a third, combined (``$stu$'') scheme. 
As shown below, the result of each of these resummations is a frequency-dependent four-point vertex. 
In our approach, we integrate out the leads and thus take them into account in an exact fashion. 
The result of this is a separate contribution to the self-energy of the dot electrons. 

The paper is organized as follows: 
The SIAM and its most important properties are summarized in Section \ref{modelsect}. 
In Section \ref{2pisect}, we lay out the 2PI effective-action approach for Fermions on a quantum dot coupled to two leads. 
In \Sect{NumericalResults} we present our numerical results for the transient and stationary dynamics of the strongly correlated dot.
Here, our  main focus is on comparing results obtained within the different resummation schemes.
\Sect{CurrentVoltage} contains our results for a quantum dot subject to a bias voltage.
We draw our conclusions in Section \ref{consect}.
The appendices comprise technical details and includes additional data allowing for an extensive comparison of the different approximation schemes.

\section{The single-impurity Anderson model }
\label{modelsect}
Our study is based on the Anderson impurity model Hamiltonian describing interacting electrons on a single quantum dot coupled to  free-electron carrying leads,
\begin{equation}
\label{andersonh}
  \begin{split}
    H =&\ H_\textrm{dot} + H_\textrm{leads}+ H_\textrm{tunnel} 
    = \sum_\sigma E_{0\sigma} n_\sigma + U n_{\uparrow} n_{\downarrow} + \\
    &+\ \sum_{k \ell \sigma }  \epsilon_{k\ell}  c^\dagger_{k\ell\sigma}  c_{k\ell\sigma} 
    - \sum_{k\ell\sigma} \left(t_\ell c^\dagger_{k\ell\sigma} d_\sigma + t_\ell^* d^\dagger_\sigma c_{k \ell \sigma} \right)
  \end{split}
\end{equation}
Here,  $\sigma\in\{1,2\} \simeq \{ \uparrow , \downarrow\}$ is the electrons' spin index. 
The index $\ell\in\{+,-\} \simeq \{L,R\} $ labels the leads on the left and right, respectively, and $k$ the momentum of the lead electrons. 
The chemical potentials $ \mu_\ell= \ell eV/2$ of the left and right leads determine the relative bias voltage. 
The single-electron energies on the leads are denoted as $\epsilon_{k\ell}$. 
The occupation-number operator on the dot, $ n_\sigma= d^\dagger_\sigma d_\sigma $, enters the on-dot interaction term $\sim U$ taking into account Coulomb repulsion. 
The one-electron energy on the dot, $E_{0\sigma} = E_0 + \sigma B $, is controlled experimentally through the gate voltage (first term) and the Zeeman shift in a magnetic field $B$ (second term). 
The energy level scheme is shown in \Fig{dot}.

The tunneling strength $t_\ell$ controls the coupling of the dot and the leads which, in this work, we assume to be symmetric, $ \tau =t_L=t_R $.
We assume that the leads are in thermal equilibrium at some temperature $T$ and consider the wide-band limit with a constant density of states $\rho$ around the Fermi surface. 
The dimensionful quantities of the system can be expressed in units of the hybridization as $ \Gamma= 2 \pi |\tau|^2 \rho_L $  which quantifies the dressing of the dot by the leads.
In this article, we give all quantities in units of $\Gamma$, setting the elementary charge, and the Planck and Boltzmann constants to $e=h=k_\mathrm{B}=1$.

We remark that, while universal properties in the Kondo regime are expected to be well described by the above model, real (semi-conductor) quantum dots require a description beyond the two-level approximation\citeafter{.}{Schmid2000a.PhysRevLett.84.5824,Pustilnik2001a.PhysRevLett.87.216601,Garate2011a.PhysRevLett.106.156803,Nah2012a.PhysRevB.85.235311}
Besides the possibility that the spin of the ground state of the dot is $S>1/2$\citeafter{,}{Schmid2000a.PhysRevLett.84.5824,Pustilnik2001a.PhysRevLett.87.216601} the narrow spacing of the dot's single-electron levels as compared to the charging energy causes nonuniversal properties to be  different from those of the Anderson model\citeafter{.}{Garate2011a.PhysRevLett.106.156803,Nah2012a.PhysRevB.85.235311}

In this article, we employ a functional-integral approach to the Anderson quantum field model. 
This requires determining an action functional which for the Hamiltonian (\ref{andersonh}) consists of the terms
\begin{align}
  S_\textrm{dot} =& \int_\mathcal{C} dt \sum_\sigma d^\dagger_\sigma \Big( i \partial_t -E_{0\sigma}\Big) d_\sigma  - U d^\dagger_{\uparrow} d_{\uparrow} d^\dagger_{\downarrow} d_{\downarrow},   \\ 
\label{S:leads}
  S_\textrm{leads} =& \int_\mathcal{C} dt  \sum_{k\ell\sigma} c^\dagger_{k\ell\sigma} \left( i \partial_t - \epsilon_{k\ell} \right) c_{k\ell\sigma}, \\ 
\label{S:tunnel}
 S_\textrm{tunnel} =& \int_\mathcal{C} dt \sum_{k\ell\sigma} \left( t_\ell  c^\dagger_{k\ell\sigma} d_\sigma + 
 t^*_\ell d^\dagger_\sigma c_{k\ell\sigma} \right).
\end{align}
In this paper, we are in particular interested in the Kondo characteristics of the quantum dot coupled to leads at a temperature below the so-called Kondo scale.
This Kondo temperature marks the onset of the Kondo effect which is due to the formation of singlet states of itinerant and localized dot fermions and characterized by a rising resistivity of the dot at low temperatures. 
The exact expression for the Kondo temperature in terms of the Hamiltonian parameters can be obtained by means of a Bethe ansatz\cite{andrei1983} and reproduced by the numerical renormalization group\citeafter{,}{bulla2008}
\begin{equation}
\label{eq:kondotemp}
  T_K = \sqrt{\frac{U \Gamma}{2}} \exp \left(\frac{- \pi U}{8 \Gamma} \right).
\end{equation}
for the particle-hole symmetric system, where $E_0=-U/2$.
 
To reveal the Kondo characteristics we, among other correlations to be introduced below, study the time dependent current through the dot
\bea
I(t) = - {i\pi e } \sum_{k\ell\sigma} \left(
\ell t_\ell \langle c^\dagger_{k\ell\sigma} d_\sigma \rangle 
 - \ell t^*_\ell \langle d^\dagger_\sigma c_{k\ell\sigma} \rangle \right).
\eea 
This can also be written as $I=(I_L-I_R)/2$, where $I_\ell=-e \dot N_\ell(t)$ and $N_\ell(t)= \langle \sum_{k\sigma} c^\dagger_{k\ell\sigma} c_{k\ell\sigma} \rangle $ is the number of electrons on the leads. The stationary current can simply be obtained by waiting for the transient behavior to die out, such that the system is sufficiently close to the final, stationary state.
%

\section{Dynamic equations for the interacting quantum dot coupled to leads}
\label{2pisect}
In this article, we make use of Kadanoff-Baym-type dynamic equations for correlation functions, derived within the two-particle irreducible (2PI) effective-action formalism, also known as the $\Phi$-derivable approach\citeafter{.}{luttingerward,Baym1962,Cornwall1974a} 
In the following, we briefly summarize known basics about this approach and recall its implementation for the Anderson model\citeafter{.}{Sexty:2010ra} 
Since we will be interested in initial-value problems, specifically, in the evolution of two-time correlation functions starting from values given by some initial state, we will work in the Heisenberg picture. 
In the corresponding functional-integral formulation, the time integrations in the action are defined to run along a Schwinger-Keldysh contour $\cal C$, which leads from the initial time $t_0$ to some final time $t$, and then back to $t_0$\citeafter{.}{Keldysh1964a,Chou:1984es} 
%
\subsection{Effective action and equations of motion}
\label{sec:EffActionEOM}
The Kadanoff-Baym equations determine the evolution of the time-ordered two-point function,
\begin{equation}
\label{eq:DefD}
\begin{split}
 D_{\sigma\lambda}\left(t,t'\right) = \ &\Theta_\mathcal{C} \left(t-t'\right) \left\langle d_\sigma\left(t\right) d_\lambda^\dagger \left(t'\right) \right\rangle\\
       -\ &\Theta_\mathcal{C} \left(t'-t\right)  \left\langle d_\lambda^\dagger\left(t'\right) d_\sigma\left(t\right) \right\rangle
\end{split}
\end{equation}
where $\Theta_\mathcal{C} \left(t-t'\right)$ is a $\Theta$-function on the Schwinger-Keldysh contour, and evaluates to 1 (0) if  $t$ is later (earlier) than $t'$ along the contour.
Hence, on the backward contour this amounts to antitime ordering.

Neglecting at first, the leads the 2PI effective action \cite{Cornwall1974a} reads
\begin{equation}
\label{eq:2PIEAwoleads}
  \Gamma_\mathrm{2PI}[ D] = - i\, \textrm{Tr} \left[ \textrm{ln} D^{-1} + D_0^{-1} D \right] + \Gamma_2 \left[D\right] + \textrm{const.}
\end{equation}
where the free inverse propagator is defined as 
\begin{equation} 
 i D_{0,\sigma\lambda}^{-1} \left(t,t'\right) = \Big( i \partial_t - E_{0\sigma}\Big) \delta_\mathcal{C} \left(t-t'\right) \delta_{\sigma\lambda}.
\end{equation}

$\Gamma_2[D]$ is the sum of all closed 2PI diagrams constructed from bare vertices and full propagators. 
Those diagrams which do not fall apart upon cutting two lines are two-particle irreducible\citeafter{.}{Cornwall1974a} 
Taking into account all such diagrams, the effective action in Eq.~(\ref{eq:2PIEAwoleads}) gives exact solutions of the system. 
Here we approximate $\Gamma_2[D]$ by taking into account only certain classes of diagrams in $\Gamma_2[D]$, as laid out in detail in \Sect{resumsec}.

As an example, we show in \Eq{gamma2_mf} the diagram of lowest order in the bare coupling $U$, represented as a black dot. 
\begin{equation}
  \label{eq:gamma2_mf}
  \begin{split}
    \Gamma_2^{(1)} [D] &= \  \begin{tikzpicture}[line width=1.5 pt, scale=1.0, baseline={(0,-0.1)}]
	\draw[fermionup] (0,0) arc (360:0:.5);
	\draw[fermiondown] (0,0) arc (-180:180:.5);
	\draw[fill=black] (0,0) circle (.04cm);
    \end{tikzpicture} \\
    &= - U \int_C dt ~ D_\uparrow(t,t) D_\downarrow (t,t) 
  \end{split}
\end{equation}
Solid lines denote the full propagator or two-point function $D$, with different spin components indicated by different colors.

The Kadanoff-Baym equations of motion result from the Hamilton stationarity conditions for the action
\begin{equation}
  \frac{ \delta \Gamma_\mathrm{2PI} \left[D \right]}{\delta D_{\lambda \sigma }\left(t' ,t \right)} = 0 .
\end{equation}
This equation can be written as the well-known Dyson equation $D^{-1}_{\sigma\lambda}\left(t,t'\right)=D_{0,\sigma\lambda}^{-1}\left(t,t'\right) - \Sigma_{\sigma\lambda}\left(t,t'\right) $, where the self-energy is determined by $\Gamma_{2}$, 
\begin{equation}
  \Sigma_{\sigma\lambda}\left(t,t'\right) = -i \frac{ \delta \Gamma_2 \left[D\right]}{\delta D_{\lambda\sigma} \left(t',t\right) }. 
\end{equation}
We use the decomposition 
\begin{equation}
  D_{\sigma\lambda}\left(t,t'\right) = F_{\sigma\lambda} \left(t,t'\right) - \frac{i}{2}  \rho_{\sigma\lambda}  \left(t,t'\right) \textrm{sign}_\mathcal{C}\left(t-t'\right) ,
  \label{eq:FrhoDecomposition}
\end{equation}
of the time-ordered two-point function \eq{DefD} into commutator and anticommutator contributions, 
\begin{equation}
  \begin{split}
    F_{\sigma\lambda} \left(t,t'\right) &= \frac{1}{2} \left\langle  \left[ d_\sigma(t) , d^\dagger_\lambda (t') \right] \right\rangle , \\
    \rho_{\sigma\lambda} \left(t,t'\right)  &= i \left\langle \left\{ d_\sigma (t) , d^\dagger _\lambda (t') \right\} \right\rangle ,
  \end{split}
\end{equation}
representing the statistical and spectral correlation functions in real-time representation, respectively. 
Here, $\textrm{sign}_\mathcal{C}\left(t-t'\right) = -1 + 2 \Theta_\mathcal{C}\left(t-t'\right) $ is a sign function on the contour. 
For the equal-time arguments the identities
\begin{equation}
   F_{\sigma\sigma}\left(t,t\right)= \frac{1}{2} - n_\sigma\left(t\right) , \qquad
   \rho_{\sigma\lambda} \left(t,t\right) = i \delta_{\sigma\lambda},
   \label{eq:eqtimeFrho}
\end{equation}
hold,
where $n_{\sigma}(t)$ is the mean number of fermions occupying the dot level with spin $\sigma$ at time $t$.
The equation for $\rho(t,t)$ follows from the fermionic equal-time anticommutation relations. 
These definitions imply that the symmetry relations
\begin{equation} 
  F_{\sigma\lambda}\left(t,t'\right) = F_{\lambda\sigma}\left(t',t\right) ^ *  , \qquad 
  \rho_{\lambda\sigma}\left(t,t'\right) = - \rho_{\sigma\lambda}\left(t',t\right) ^ *
  \label{eq:symmetries}
\end{equation}
hold.
Rewriting the Dyson equation for $ D\left(t,t'\right)$ in terms of two equations for $ F\left(t,t'\right)$ and $\rho\left(t,t'\right)$ removes the singularity due to the time ordering on the contour, whereby the time integrals over the contour $\cal \mathcal{C}$ are replaced by simple integrations along the real-time axis. 

We assume that correlations between up and down spins vanish initially. 
Then the number-conserving 2PI equations conserve  $D_{\sigma\lambda} \sim \delta_{\sigma \lambda}$ at all times. 
Introducing the notation $D_{\sigma\sigma} \equiv D_\sigma$ (without summation) the 2PI or Kadanoff-Baym equations of motion result as
\begin{equation} 
\label{eq:eom_diag}
  \begin{split}
    \Big( i\partial_t - M_{\sigma}(t) \Big)\, \rho_\sigma\left(t,t'\right) =& \int_{t'}^t du\,\Sigma_\sigma^\rho \left(t,u\right) \rho_\sigma \left(u,t'\right)\, , \\
    \Big( i\partial_t - M_{\sigma}(t) \Big)\, F_\sigma \left(t,t'\right) =&  \int_0^t du\, \Sigma^\rho_\sigma \left(t,u\right) F_\sigma\left(u,t'\right) \\
    -& \int_0^{t'} du\,\Sigma^F_\sigma \left(t,u\right) \rho_\sigma\left(u,t'\right),
  \end{split}
 \end{equation}
where the self-energy decomposes into local, as well as $F$- and $\rho$-type terms,
\begin{equation}
  \begin{split}
    \Sigma_\sigma\left(t,t'\right) =&-i\, \Sigma_\sigma^{(0)} (t) \delta \left(t-t'\right)  \\
    &+ \Sigma^F_\sigma\left(t,t'\right) - \frac{i}{2} \textrm{sign}_\mathcal{C} \left(t-t'\right) \Sigma^\rho_\sigma \left(t,t'\right). 
  \end{split}
\label{eq:SigmaDecomp}
\end{equation}
The energy term in \eq{eom_diag} local in time is given by
\begin{equation} 
\label{eq:MFMass}
 M_\sigma(t)= E_{0\sigma} + \Sigma_\sigma^{(0)}(t)
\end{equation}
and includes the mean-field shift originating from the double-bubble diagram in \Eq{gamma2_mf}. 
Assuming that $\Sigma$ is derived from the full action $\Gamma_{2}$, the above integro-differential equations are equivalent to the exact Kadanoff-Baym equations and include higher-order correlations through the non-Markovian memory integrals on their right hand side. 

Including only the leading-order `double-bubble' diagram in \Eq{gamma2_mf} yields a perturbative (in the skeleton-graph sense) small-coupling approximation of $\Gamma_2[D]$.
This corresponds to the tadpole approximation of the self-energy, leading to the mean-field approximation of the dynamic equations \eq{eom_diag} (see  \cite{Sexty:2010ra} for more details),
\begin{equation}
\label{eq:meanfieldsigma}
  \Sigma^\textrm{MF}_{\sigma} \left(t,t'\right) = - i U n_{\bar\sigma} \delta \left(t-t'\right).
\end{equation}
The energy shift resulting from this term determines the position of the Kondo resonance. Depending on the mean occupation $n_{\sigma}$ of the dot with an electron of spin $\sigma$ and the strength of the on-site repulsion $U$ it shifts the mean energy for the opposite spin orientation $\bar\sigma=3-\sigma$ away from the bare energy $E_{0\bar\sigma}$; see \Fig{dot}.
%
\subsection{Coupling of the dot to the leads}
\label{leadsect}
The coupling of the strongly interacting spins on the quantum dot with the low-temperature, degenerate Fermi sea of lead electrons plays a crucial role in the emergence of the Kondo effect. 
It allows for the build-up of strong spin-singlet-character correlations of dot and lead fermions if the dot is tuned to the particle-hole symmetric point where the mean energy of the dot electrons is at the level of the Fermi edge of the leads. 
In the following we briefly review the coupling of the single dot system to the leads taking into account the terms (\ref{S:leads}) and (\ref{S:tunnel}) in the action. 
For more details see Ref.~\onlinecite{Sexty:2010ra}. 

The effect of coupling to the leads can be computed exactly, since (\ref{S:leads}) and (\ref{S:tunnel}) include terms only up to quadratic order. 
Integrating these contributions out, the free inverse propagator of the thermal lead electrons appears as part of the free propagator of the dot electrons. 
After decomposition one finds that the effect of a single lead-electron mode of frequency $\epsilon$ can be included into the self-energy of the dot electrons $ \Sigma^F $ and $ \Sigma^\rho$, giving the additional contributions
\begin{align} 
\label{eq:oneleadsigmaf}
    \Sigma^{F(1)}_\textrm{lead} (t,t') &= - |\tau |^2 \left( \frac{1}{2} -  f( \epsilon-\mu)\right) e^{-i \epsilon \left(t-t'\right)} , \\
\label{eq:oneleadsigmarho}
    \Sigma^{\rho(1)}_\textrm{lead}(t,t') &= - i |\tau |^2 e^{-i \epsilon \left(t-t'\right)},
\end{align}
where $f(x) = 1/ \left[1 + \exp\left(\beta x \right)\right] $ is the Fermi function. 
Recall that $\tau$ is the ($L$-$R$-symmetric) tunneling strength.
Integrating over a continuum of lead energy levels one obtains
\begin{equation} 
  \Sigma_\textrm{lead} = \int_{-D}^D d \epsilon\,\rho_L (\epsilon)\Sigma_\textrm{lead}^{(1)} .  
\end{equation}
In the wide-band limit one approximates the leads' density of states $\rho_L(\epsilon)$ as a constant and obtains, with \Eq{oneleadsigmarho},
\begin{equation}
  \Sigma^\rho_\textrm{lead}(t,t') 
  = - 2i | \tau |^2 \rho_L \frac{\sin D(t-t')}{t-t'}  \stackrel{D\rightarrow \infty}
  {=} -  i\,\Gamma \delta(t-t'),
\end{equation} 
where $\Gamma= 2 \pi |\tau|^2 \rho_L $ is called hybridization. 
$\Sigma^F$ can be determined analytically only at zero  temperature where it results as the principal-value contribution
\begin{equation} 
  \Sigma^F_{\textrm{lead}, T=0} (t,t') = i | \tau |^2\, \mathcal{P}  \frac{ e^{-i \mu (t-t')}}{t-t'} .
\end{equation}
At the initial time $t=0$, we assume the quantum dot to be uncorrelated, with mean occupation numbers $n_{\sigma}(0)$ for the two spin components and imagine the coupling to the leads, at equal chemical potentials $\mu=\mu_{L}=\mu_{R}$, to be switched on for $t\ge0$. 
Taking into account only mean-field interactions of the dot electrons, simplified dynamic equations result from Eqs.~\eq{eom_diag} and \eq{meanfieldsigma} in the long-time limit. 
Rewriting these equations in Wigner coordinates $s=t-t'$ and $\mathcal{T}=(t+t')/2$ we obtain the $s$-evolution equation for $\rho(s,\mathcal{T})$ in the limit of large $\mathcal{T}\gg2\pi/\Gamma$ as
\begin{equation} 
 \label{eq:dsrhoMFleads}
  \partial_{s}\rho_{\sigma}(s,\mathcal{T}) 
  = -i\,\Big[M_{\sigma}(\mathcal{T})-i\, \Gamma\mathrm{sgn}(s)\Big]\,\rho_{\sigma}(s,\mathcal{T}).
\end{equation} 
This gives a frequency dependence of the asymptotic spectral function of
\begin{equation}
 \label{eq:rhoMFleads}
  \rho_{\sigma}(\omega) = \frac{2i\Gamma}{(\omega-M_{\sigma})^{2}+\Gamma^{2}}.
\end{equation}
As is required by the conservation of the equal-time anticommutator [cf.~\Eq{eqtimeFrho}],  this Lorentzian distribution is normalized to $\int d\omega\,\rho_{\sigma}(\omega)/(2\pi)=i$.

Analogously, the $\mathcal{T}$-evolution equation for $F_{\sigma}(0,\mathcal{T})=1/2-n_{\sigma}(\mathcal{T})$ yields the dynamic equation for the occupation number in the $\mathcal{T}\gg 2\pi/\Gamma$ limit. In the zero-temperature limit we obtain
\begin{equation} 
  \partial_{\mathcal{T}}n_{\sigma}(\mathcal{T}) = -\Gamma\left(n_{\sigma}(\mathcal{T})-n_{\sigma}^\mathrm{asym}\right),
\end{equation} 
with asymptotic occupation number
\begin{equation}
\label{eq:nsigmaasym}
  n_{\sigma}^\mathrm{asym}  = \frac{1}{2} \left[ 1-\frac{2}{\pi} \arctan\left(\frac{E_{0\sigma} + Un_{\bar\sigma}^\mathrm{asym}} \Gamma \right) \right]\, .
\end{equation}
This set of two coupled transcendental equations determines, for a given set of detunings $E_{0\sigma}$, the two occupations $n_{\sigma}$, i.e. the mean occupation and spin polarization of the dot.

In the following we consider mostly the unpolarized case, with $E_{0\uparrow}=E_{0\downarrow}=E_{0}$. 
In the weak-coupling limit between dot and leads, $\Gamma\to0$, \Eq{nsigmaasym} then evaluates to
\begin{equation}
 \left. n_{\sigma}^\mathrm{asym}\right|_{\Gamma\to0} = \begin{cases}
      1-\theta(E_{0}),& |E_{0}+U/2|>U/2, \\
      -E_{0}/U, &|E_{0}+U/2|<U/2 .
    \end{cases}
\end{equation}
In the case that $U=0$, this is equivalent to the basic result that the dot fills up (runs empty) if its energy is tuned below (above) the Fermi energy of the leads. 
Including the mean-field shift, $U>0$, this behavior remains in place for $E_{0}<-U$ for which the on-site repulsion between the electrons even in a completely filled dot does not lift the total energy above the Fermi edge, as well as for $E_{0}>0$. 
In the interval $E_{0}\in[-U,0]$, however, the shift induces a continuous transition, allowing for any mean occupation between zero and one electron for each spin orientation.

%
\subsection{Nonperturbative self-energy of the dot-electrons}
\label{sec:resumsec}
Computing the build-up of the Kondo effect in real-time evolution requires a nonperturbative determination of the time-dependent self-energy of the dot electrons, which accounts for the strong correlations building up between dot and leads.
In this section we describe the resummation of the classes of diagrams which we found to be necessary to account for Kondo correlations.
The approach used here leads substantially beyond the mean-field approximation introduced above as well as beyond any coupling approximation higher than the leading double-bubble diagram truncation, including, e.g.~to second order in $U$, the `basketball` diagram shown in Eq.~(\ref{eq:gamma2_2nd}). 
We find it necessary to perform resummations of $s$-channel bubble as well as $t$- and $u$-channel ladder chains, often also termed resummations in the direct, particle-particle (pp), and particle-hole (ph) channels. 
In the direct channel, bubbles with alternating spins make up the chains where, in each bubble, the two propagators describe the same spin component. 
This is similar to the next-to-leading-order $1/N$ approximation for $N$-component scalar fields exploited extensively in dynamic 2PI calculations\citeafter{.}{Berges:2001fi,Aarts:2002dj} 
Here, we also require the  $t$- and $u$-channels, for which the chains take the form of a ladder, with the two rails being made up by propagators of opposite spin, with the same spin throughout the rail. 
The two channels differ by the direction of the arrows, i.e., ordering of $d_{\sigma}$ and $d_{\sigma}^{\dagger}$ in the propagator, distinguishing between the $t$-channel (pp) and the $u$-channel (ph) interactions.

%
\subsubsection{Hubbard-Stratonovich transformation}
An elegant way to perform the resummations involves a Hubbard-Stratonovich (HS) transformation. Since the interaction vertex couples ``up''-spins with ``down''-spins, the bubbles in the $s$-channel chains have alternating spins, while the rails in the $t$- and $u$-channel ladders have opposite spins. The action
\begin{equation} 
  S_\mathrm{dot} = \int_\mathcal{C} dt \sum_\sigma d^\dagger_\sigma \Big( i \partial_t -E_{0\sigma} \Big) d_\sigma -  U d^\dagger_{\uparrow} d_{\uparrow} d^\dagger_{\downarrow} d_{\downarrow}  
\end{equation}
is rewritten using auxiliary scalar fields $\chi_1$ and $\chi_2$ by use of the substitution
\begin{equation} 
  - J A^{-1} J \rightarrow \chi^T A \chi + 2 J^T \chi 
\end{equation}
where 
\begin{equation}
  \chi = \Bigg( 
    \begin{matrix} \chi_1 \\ \chi_2 
    \end{matrix} 
    \Bigg) , \qquad 
    A= \frac{1}{2 U}\Bigg( 
    \begin{matrix}
      0 & 1 \\
      1 & 0 \\
    \end{matrix} 
    \Bigg) .
\end{equation}
The $J$ operators for the three channels read
\begin{equation}
  J_{s} = \frac{1}{2} \Bigg( 
    \begin{matrix}  d^\dagger_{\uparrow} d_{\uparrow}  \\ d^\dagger_{\downarrow} d_{\downarrow} \end{matrix} 
    \Bigg),   \qquad
  J_{t} = \frac{1}{2}   \Bigg(
    \begin{matrix} d^\dagger_{\uparrow} d^{\dagger}_{\downarrow}  \\ d_{\downarrow}d_{\uparrow} \end{matrix} 
    \Bigg),  \qquad
  J_{u} = \frac{1}{2}  \Bigg( 
    \begin{matrix} d^\dagger_{\uparrow} d_{\downarrow}  \\ d_{\uparrow}d^{\dagger}_{\downarrow} \end{matrix} 
    \Bigg) .  
\end{equation}
The resulting action is
\begin{equation}
\label{hsaction}
  \begin{split}
    S_{\mathrm{dot},\xi} \left[ d_\sigma, d^\dagger _ \sigma, \chi_i \right] =& \int_\mathcal{C} dt \sum_\sigma d^\dagger_\sigma \Big( i \partial_t -E_{0\sigma}\Big) d_\sigma \\  
    +& \frac{1}{U} \chi_1 \chi_2 + S_{\mathrm{int},\xi}\left[ d_\sigma, d^\dagger _ \sigma, \chi_i \right],
  \end{split}
\end{equation}
with the resummation-scheme-dependent interaction term ($\xi=s,t,u$),
\begin{align}
  S_{\mathrm{int},s} &= d^\dagger_{\uparrow} d_{\uparrow} \chi_1 +  d^\dagger_{\downarrow} d_{\downarrow} \chi_2 , \\
  S_{\mathrm{int},t} &= d^\dagger_{\uparrow} d^\dagger_{\downarrow} \chi_1 +  d_{\downarrow}d_{\uparrow}  \chi_2 , \\
  S_{\mathrm{int},u} &= d^\dagger_{\uparrow} d_{\downarrow} \chi_1 +  d_{\uparrow} d^\dagger_{\downarrow} \chi_2 .    
\end{align}
The free inverse propagators are read off from the quadratic part of the action
\begin{align}
\label{eq:G0inv}
  i G_0^{-1}\left(t,t'\right) &= 2  A  \delta\left(t-t'\right) , \\
  \label{eq:D0inv}
  i  D_{0,\sigma}^{-1} \left(t,t'\right) &=  \Big( i \partial_t - E_{0\sigma} + \bar \chi_\sigma \Big) \delta\left(t-t'\right),
\end{align}
where the free propagator $G_0$  of the scalar fields is a $2 \times 2 $  matrix. Accordingly, we call  $G$  the propagator of the scalars, and $\bar \chi_i=\langle\chi_{i}\rangle$ is the one-point function or expectation value of the auxiliary fields which is nonzero only in the $s$-channel case $\xi=s$.

The corresponding 2PI effective action can be written as:
\begin{align}
    \Gamma_{\xi} \left[G,D,\bar \chi \right] &=\ S_{\text{dot},\xi} \left[ d^\dagger_\sigma = d_\sigma=0,\bar \chi \right] - i\,\textrm{Tr} \left[  \textrm{ln} D^{-1} + D_0^{-1} D \right] 
    \nonumber\\ 
    +& \frac{i}{2} \text{Tr} \left[ \textrm{ln} G^{-1} + G_0^{-1} G \right] + \Gamma_{2,\xi} \left[D,G\right] + \textrm{const.}
\end{align}
where $\Gamma_2[D,G]$ contains all closed 2PI diagrams built from the three-point vertices of the action (\ref{hsaction}) and full scalar and fermion propagators. The lowest-order contributions are shown explicitly in Eqs.~(\ref{eq:gamma2_sch}), (\ref{eq:gamma2_tch}), and (\ref{eq:gamma2_uch}).
%
\subsubsection{Schwinger-Dyson equations}
The stationarity conditions give the Schwinger-Dyson equations. 
The field average of the scalar fields $\chi_{i}$, in contrast to that of the fermionic fields, is nonzero, so we have stationarity conditions of the form $ \delta \Gamma / \delta \bar \chi = 0 $. 
The resulting equations read, for the $s$-channel resummation,
\begin{equation} 
  \bar \chi_1 (t)  = U D_{\downarrow\downarrow} (t,t) ,\qquad 
  \bar \chi_2 (t)  = U D_{\uparrow \uparrow} (t,t) , 
\end{equation}
while, for the $t$- and $u$-channel resummations, the mean values $\bar\chi_{i}$ are given by  correlation functions $D_{\sigma\lambda}$ with $\sigma\not=\lambda$ which we have assumed to vanish from the outset.
The Dyson equations for the propagators read
\begin{align} 
   D_0^{-1} D &= \Sigma * D + \delta , \\ 
   G_0^{-1} G &= \Pi * G + \delta   
\label{eq:constrainteq}
\end{align}
where we again suppressed the time arguments, and $*$ stands for convolution on the contour $\mathcal{C}$,
\begin{equation}
 \left(A*B\right) \left(t,t'\right) = \int_\mathcal{C} dz \, A \left(t,z\right) B\left(z,t'\right).
\end{equation}
$\Sigma$ and $\Pi$ are the self-energy of the fermion and the boson fields, respectively:
\begin{equation} 
\label{resumsigma}
  \Sigma_\sigma \left(t,t'\right)  = -i \frac{\delta \Gamma_2[D,G]}{\delta D_\sigma\left(t',t\right) }  , \quad
  \Pi_{\sigma\lambda} \left(t,t'\right)  =  2i \frac{\delta \Gamma_2 [D,G]}{\delta G_{\lambda\sigma}\left(t',t\right) } .
\end{equation}
%
\subsubsection{$s$-channel resummation}
In the direct or $s$-channel resummation the $\Gamma_2$ part of the action, to lowest order in the auxiliary-field-fermion coupling, reads
\begin{equation}
  \label{eq:gamma2_sch}
  \begin{split}
    \Gamma_{2,s} [D,G]&= \frac{i}{2}\Big[\
    \begin{tikzpicture}[line width=1.5pt, scale=1.0, baseline={(0.0,-0.1)}]
      \draw[fermionup] (0,0) arc (180:0:0.5);
      \draw[fermionbarup] (0,0) arc (180:360:0.5);
      \draw[resumsdown] (0,0) to (1,0);
    \end{tikzpicture}
    \ +\ 
    \begin{tikzpicture}[line width=1.5pt, scale=1.0, baseline={(0.0,-0.1)}]
      \draw[fermiondown] (0,0) arc (180:0:0.5);
      \draw[fermionbardown] (0,0) arc (180:360:0.5);
      \draw[resumsup] (0,0) to (1,0);
    \end{tikzpicture}\ \Big]\\
    &= {i\over 2} \sum_\sigma \int_\mathcal{C} dx dy\, D_\sigma(x,y) D_\sigma(y,x) 
    G_{\sigma\sigma} (x,y),
  \end{split}
\end{equation}
where the wiggly lines represent the bosonic propagator corresponding to an infinite sum of odd numbers of fermionic loops with alternating spins,
\begin{align}
  \label{eq:bosonic_sch}
  \begin{tikzpicture}[line width=1.5pt, scale=1.0, baseline={(0.0,-0.1)}]
    \draw[resumsdown] (0,0) to (1,0);
  \end{tikzpicture}
  &=
    \begin{tikzpicture}[line width=1.5pt, scale=1.0, baseline={(0.0,-0.1)}]
      \draw[fermiondown] (0,0) arc (180:0:0.5);
      \draw[fermionbardown] (0,0) arc (180:360:0.5);
      \draw[fill=black] (0,0) circle (0.04cm);
      \draw[fill=black] (1,0) circle (0.04cm);
    \end{tikzpicture}
    \ + \ 
    \begin{tikzpicture}[line width=1.5pt, scale=1.0, baseline={(0.0,-0.1)}]
      \draw[fermiondown] (0,0) arc (180:0:0.5);
      \draw[fermionbardown] (0,0) arc (180:360:0.5);
      \draw[fermionup] (1,0) arc (180:0:0.5);
      \draw[fermionbarup] (1,0) arc (180:360:0.5);
      \draw[fermiondown] (2,0) arc (180:0:0.5);
      \draw[fermionbardown] (2,0) arc (180:360:0.5);
      \draw[fill=black] (0,0) circle (0.04cm);
      \draw[fill=black] (1,0) circle (0.04cm);
      \draw[fill=black] (2,0) circle (0.04cm);
      \draw[fill=black] (3,0) circle (0.04cm);
    \end{tikzpicture}
    \ + \ \mathcal{O}\left(U^6\right)\, ,
    \nonumber\\[1ex]
  \begin{tikzpicture}[line width=1.5pt, scale=1.0, baseline={(0.0,-0.1)}]
    \draw[resumsup] (0,0) to (1,0);
  \end{tikzpicture}
  &=
    \begin{tikzpicture}[line width=1.5pt, scale=1.0, baseline={(0.0,-0.1)}]
      \draw[fermionup] (0,0) arc (180:0:0.5);
      \draw[fermionbarup] (0,0) arc (180:360:0.5);
      \draw[fill=black] (0,0) circle (0.04cm);
      \draw[fill=black] (1,0) circle (0.04cm);
    \end{tikzpicture}
    \ +\ 
    \begin{tikzpicture}[line width=1.5pt, scale=1.0, baseline={(0.0,-0.1)}]
      \draw[fermionup] (0,0) arc (180:0:0.5);
      \draw[fermionbarup] (0,0) arc (180:360:0.5);
      \draw[fermiondown] (1,0) arc (180:0:0.5);
      \draw[fermionbardown] (1,0) arc (180:360:0.5);
      \draw[fermionup] (2,0) arc (180:0:0.5);
      \draw[fermionbarup] (2,0) arc (180:360:0.5);
      \draw[fill=black] (0,0) circle (0.04cm);
      \draw[fill=black] (1,0) circle (0.04cm);
      \draw[fill=black] (2,0) circle (0.04cm);
      \draw[fill=black] (3,0) circle (0.04cm);
    \end{tikzpicture}
    \ + \ \mathcal{O}\left(U^6\right)\, .
\end{align}
Hence, one obtains
\begin{equation}
  \begin{split}
    \Sigma_\sigma \left(t,t'\right) =&\  D_\sigma\left(t,t'\right) G_{\sigma\sigma}\left(t,t'\right),\\
    \Pi_{\sigma\lambda} \left(t,t'\right) = &\  -D_\sigma\left(t,t'\right) D_\sigma\left(t',t\right)\delta_{\sigma\lambda},
  \label{eq:sChSigmaofPi}  
  \end{split}
\end{equation}
where $\sigma,\lambda\in\{ \uparrow ~=1 , \ \downarrow ~=2 \}$ is used for the fermionic field indices. The equations determining the scalar fields represent constraints which do not contain any time derivatives, because the $ \chi$ fields are auxiliary, nondynamical fields. From the constraint equation \eq{constrainteq} for $G$, one can see that
\begin{equation}
\label{eq:scalarProp}
  \begin{split}
    G  =&\ i U \delta 
      \begin{pmatrix}
        0 & 1 \\
        1 & 0 \\
      \end{pmatrix}  
    - U^2 
      \begin{pmatrix}
        \Pi_{22} & 0 \\
        0  & \Pi_{11} \\
      \end{pmatrix}  \\  
    &-\ i  U^3 
      \begin{pmatrix}
        0  & \Pi_{22} * \Pi_{11} \\
        \Pi_{11} * \Pi_{22} & 0 \\
      \end{pmatrix}  \\ 
    &+\ U^4 
      \begin{pmatrix}
        \Pi_{22} * \Pi_{11} * \Pi_{22} & 0 \\
        0 & \Pi_{11} *\Pi_{22} *\Pi_{11} \\
      \end{pmatrix} 
    + \cdots ,
  \end{split}
\end{equation}
where we have again omitted the $\left(t,t'\right)$ arguments. 
Inserting $G_{11}$ and $G_{22}$ into the self-energy $\Sigma$ of the fermions in Eq.~(\ref{resumsigma}), one finds a sum of bubble chains with alternating spins being generated. For the decomposition of $G$ and the self-energies into statistical and spectral parts see Ref.~\onlinecite{Sexty:2010ra}.
%
\subsubsection{$t$-channel resummation}
In particle-particle or $t$-channel resummation, we obtain
\begin{equation}
  \label{eq:gamma2_tch}
  \begin{split}
  \Gamma_{2,t} [D,G] &= \frac{i}{2}\Big[\
  \begin{tikzpicture}[line width=1.5pt, scale=1.0, baseline={(0.0,-0.1)}]
    \draw[fermionup] (0,0) arc (180:0:0.5);
    \draw[fermiondown] (0,0) arc (180:360:0.5);
    \draw[resumtup] (0,0) to (1,0);
  \end{tikzpicture}
  \ +\ 
  \begin{tikzpicture}[line width=1.5pt, scale=1.0, baseline={(0.0,-0.1)}]
    \draw[fermionbarup] (0,0) arc (180:0:0.5);
    \draw[fermionbardown] (0,0) arc (180:360:0.5);
    \draw[resumtdown] (0,0) to (1,0);
  \end{tikzpicture}
  \,\Big]
  \\
  &= {i\over 2} \sum_\sigma \int_\mathcal{C} dx dy\, D_\sigma(x,y) D_{\bar\sigma}(x,y) G_{\sigma\bar\sigma} (x,y),
  \end{split}
\end{equation}
where $\bar\sigma=3-\sigma$ and we draw explicitly the two identical loop graphs. 
Hence, the bosonic propagator $G_{\sigma\bar\sigma}$ is identical to the sum over the unidirectional fermionic loop chains,
\begin{equation}
  \begin{tikzpicture}[line width=1.5pt, scale=1.0, baseline={(0.0,-0.1)}]
    \draw[resumtup] (0,0) to (1,0);
  \end{tikzpicture}
  =
    \begin{tikzpicture}[line width=1.5pt, scale=1.0, baseline={(0.0,-0.1)}]
      \draw[fermionbardown] (0,0) arc (180:0:0.5);
      \draw[fermionbarup] (0,0) arc (180:360:0.5);
      \draw[fill=black] (0,0) circle (0.04cm);
      \draw[fill=black] (1,0) circle (0.04cm);
    \end{tikzpicture}
    \ +\ 
    \begin{tikzpicture}[line width=1.5pt, scale=1.0, baseline={(0.0,-0.1)}]
      \draw[fermionbardown] (0,0) arc (180:0:0.5);
      \draw[fermionbarup] (0,0) arc (180:360:0.5);
      \draw[fermionbarup] (1,0) arc (180:0:0.5);
      \draw[fermionbardown] (1,0) arc (180:360:0.5);
      \draw[fill=black] (0,0) circle (0.04cm);
      \draw[fill=black] (1,0) circle (0.04cm);
      \draw[fill=black] (2,0) circle (0.04cm);
    \end{tikzpicture}
    \ + \mathcal{O}\left(U^4\right).
  \label{eq:bosonic_tch}
\end{equation}
The self-energies read
\begin{equation}
  \begin{split}
    \Sigma_\sigma (t,t') =&\  D_{\bar\sigma}\left(t',t\right) G_{\sigma\bar\sigma}\left(t',t\right),\\
    \Pi_{\sigma\bar\sigma} \left(t,t'\right) =&\ -D_\sigma\left(t',t\right) D_{\bar\sigma}\left(t',t\right) = \Pi_{\bar\sigma\sigma} \left(t,t'\right),\\
    \Pi_{11}\left(t,t'\right) =&\  \Pi_{22}\left(t,t'\right) = 0 .  
  \end{split}
\label{eq:tChSigmaofPi}  
\end{equation}
In analogy to the $s$-channel case we find
\begin{equation}
\label{eq:tscalarProp}
  \begin{split}
  G_{12} = G_{21} =\ & iU\delta-U^{2}\Pi_{12}-iU^{3}\Pi_{12}*\Pi_{12} \\
                   & +U^{4}\Pi_{12}*\Pi_{12}*\Pi_{12} + \cdots .
  \end{split}
\end{equation}
Note that the mean-field tadpole term is now included as the leading-order term in the resummed scalar propagator $G_{12}$.
\subsubsection{$u$-channel resummation}
In particle-hole or $u$-channel resummation, one obtains
\begin{equation}
  \label{eq:gamma2_uch}
  \begin{split}
    \Gamma_{2,u} [D,G]&=  \frac{i}{2}\Big[\
    \begin{tikzpicture}[line width=1.5pt, scale=1.0, baseline={(0.0,-0.1)}]
      \draw[fermionup] (0,0) arc (180:0:0.5);
      \draw[fermionbardown] (0,0) arc (180:360:0.5);
      \draw[resumtup] (0,0) to (1,0);
    \end{tikzpicture}
    \ +\ 
    \begin{tikzpicture}[line width=1.5pt, scale=1.0, baseline={(0.0,-0.1)}]
      \draw[fermionbarup] (0,0) arc (180:0:0.5);
      \draw[fermiondown] (0,0) arc (180:360:0.5);
      \draw[resumtdown] (0,0) to (1,0);
    \end{tikzpicture}
    \,\Big]
    \\
    &= {i\over 2} \sum_\sigma \int_\mathcal{C} dx dy\, D_\sigma(x,y) D_{\bar\sigma}(y,x) G_{\sigma\bar\sigma} (x,y),
  \end{split}
\end{equation}
with
\begin{equation}
  \label{eq:bosonic_uch}
  \begin{tikzpicture}[line width=1.5pt, scale=1.0, baseline={(0.0,-0.1)}]
    \draw[resumtup] (0,0) to (1,0);
  \end{tikzpicture}
  =
    \begin{tikzpicture}[line width=1.5pt, scale=1.0, baseline={(0.0,-0.1)}]
      \draw[fermiondown] (0,0) arc (180:0:0.5);
      \draw[fermionbarup] (0,0) arc (180:360:0.5);
      \draw[fill=black] (0,0) circle (0.04cm);
      \draw[fill=black] (1,0) circle (0.04cm);
    \end{tikzpicture}
    \ +\ 
    \begin{tikzpicture}[line width=1.5pt, scale=1.0, baseline={(0.0,-0.1)}]
      \draw[fermiondown] (0,0) arc (180:0:0.5);
      \draw[fermionbarup] (0,0) arc (180:360:0.5);
      \draw[fermionbarup] (1,0) arc (180:0:0.5);
      \draw[fermiondown] (1,0) arc (180:360:0.5);
      \draw[fill=black] (0,0) circle (0.04cm);
      \draw[fill=black] (1,0) circle (0.04cm);
      \draw[fill=black] (2,0) circle (0.04cm);
    \end{tikzpicture}
    \ + \mathcal{O}\left(U^4\right).
\end{equation}
The self-energies result as
\begin{equation}
  \begin{split}
     \Sigma_\sigma \left(t,t'\right) = &\  D_{\bar\sigma}\left(t,t'\right) G_{\sigma\bar\sigma}\left(t',t\right),\\
     \Pi_{\sigma\bar\sigma} \left(t,t'\right) = &\  -D_\sigma\left(t,t'\right) D_{\bar\sigma}\left(t',t\right) =  \Pi_{\bar\sigma\sigma} \left(t',t\right),\\
     \Pi_{11}\left(t,t'\right)  =&\  \Pi_{22}\left(t,t'\right) = 0 ,
  \label{eq:uChSigmaofPi}
  \end{split}
\end{equation}
and hence the expansion of $G_{12}$ given in \Eq{tscalarProp} applies also here, however, with
\begin{equation}
\label{eq:uscalarProp}
  G_{12}\left(t,t'\right)\  =\ G_{21}\left(t',t\right) .
\end{equation}
%
%
\subsubsection{$stu$-channel resummation}
Above, we discussed the three different chain resummations allowed by the vertex and distinguished by the spin contractions and direction of the fermionic propagators in the loops of the Feynman diagrams. 
All of them are directly derived from the underlying theory and therefore contribute to the physics but the terms can be of different relevance depending on the physical situation. 
To include all possibilities, it is natural to define a new resummation scheme where all the three different channels are taken into consideration at the same time.
While this is possible as the respective diagrams all contribute to $\Gamma_{2}$ one needs to be careful to avoid double-counting of diagrams. 
We thus define
\begin{equation}
  \begin{split}
    \Gamma_{2,stu}\left[D,G\right] =& \Gamma_{2,s} \left[D,G\right] + \Gamma_{2,t}\left[D,G\right] \\
                                   +& \Gamma_{2,u}\left[D,G\right]-\Gamma_{2,\cap}\left[D,G\right],
  \end{split}
\end{equation}
where $\Gamma_{2,\cap}$ denotes
\begin{equation}
  \begin{split}
    \Gamma_{2,\cap} =& \Gamma_{2,s} \cap \Gamma_{2,t} + \Gamma_{2,s} \cap \Gamma_{2,u} \\
                    +& \Gamma_{2,t} \cap \Gamma_{2,u} - \Gamma_{2,s} \cap \Gamma_{2,t} \cap \Gamma_{2,u}
  \end{split}
\end{equation}
which prevents multiple counting of diagrams as the operator $\cap$ singles out all diagrams that belong to both actions it connects. 
Inserting the leading diagram of the sums in Eqs.~(\ref{eq:bosonic_sch}), (\ref{eq:bosonic_tch}), and (\ref{eq:bosonic_uch}) into Eqs.~(\ref{eq:gamma2_sch}), (\ref{eq:gamma2_tch}), and (\ref{eq:gamma2_uch}), respectively, we find that the second-order diagram 
\begin{equation}
  \Gamma_2^\text{2nd}[D] =
  \begin{tikzpicture}[line width=1.5 pt, scale=1.0, baseline={(0,-0.1)}]
	\draw[fermionup] (0,0) arc (180:0:.5);
	\draw[fermionup] (1,0) arc (0:-180:.5);
	\draw[fermiondown] (0,0) to [out=45, in=180] (0.5,0.25)
                            to [out=0, in=135] (1.0,0.0);
	\draw[fermionbardown] (0,0) to [out=-45, in=180] (0.5,-0.25)
                            to [out=0, in=-135] (1,0);
	\draw[fill=black] (0,0) circle (.04cm);
	\draw[fill=black] (1,0) circle (.04cm);
  \end{tikzpicture}
  \label{eq:gamma2_2nd}
\end{equation}
is included in all channels and hence must be subtracted twice from the sum of all channels. 
Hence, the $stu$-channel nonlocal self-energy which enters the dynamic equations  is given by
\begin{equation}
  \begin{split}
    \Sigma^{\text{stu}}_{\sigma}\left(t,t^{\prime}\right) =&\ \Sigma^{s}_{\sigma} \left(t,t^{\prime} \right) + \Sigma^{t}_{\sigma} \left(t,t^{\prime} \right) \\
                        &+\ \Sigma^{u}_{\sigma} \left(t,t^{\prime}\right) - 2 \Sigma^{\text{2nd}}_{\sigma} \left(t,t^{\prime} \right) \, .
  \end{split}
\end{equation}
Note that the time-local self-energy contribution arising from the double-bubble diagram is treated separately.
%

\subsection{Resummed self-energies in the stationary limit}
\label{sec:EffU}
We close this section with a brief analysis of the self-energy in the stationary long-time limit, as obtained within the different resummation schemes and give closed expressions in terms of effective coupling strengths. 
Details of the derivation are given in Appendix \ref{app:EffU}.
From these expressions for $\Sigma(\omega)$, the long-time spectral function $\rho(\omega)$ can be determined using standard kinetic-theory arguments as summarized in Appendix \sect{KineticTheory}.
From the resummed 2PI part of the action we can derive the self-energy of the dot electrons. In the following we give explicit expressions for the spin-symmetric case which is realized when there is no magnetic field and when the initial Gaussian state is characterized by equal occupation numbers $n_\uparrow (0) = n_\downarrow (0)=n(0)$ for the two spin orientations and thus $D_\uparrow (0,0) = D_\downarrow (0,0)$. For all later times $ t , t' > 0 $ this implies  
\bea 
\label{eq:symmD}
  D_\uparrow (t,t') = D_\downarrow (t,t').
\eea
After initial effects have died out, in the stationary state, all two-point functions will depend only on the differences of the time coordinates, and thus $n_\uparrow (t) = n_\downarrow (t)\equiv n$. In this case we can push the initial time to negative infinity and express all quantities in Fourier space.

For the $s$-channel resummation, taking the definition \eq{sChSigmaofPi} of the fermion self-energy $\Sigma$ in terms of the loop $\Pi$, we find the frequency-dependent components [see \Eq{SigmaDecomp}]
\begin{equation}
\label{eq:effsigmas}
  \begin{split}
    \Sigma^{(0)} &= Un, \\
    \Sigma^F &= F *(U_\textrm{eff}\Pi^F) -  \rho*(U_\textrm{eff} \Pi^\rho )/4 , \\
    \Sigma^\rho &= \rho*(U _\textrm{eff}  \Pi^F) + F*(U _\textrm{eff} \Pi^\rho ), 
  \end{split}
\end{equation}
where 
\begin{equation}
\label{eq:Ueffs}
  U_{\textrm{eff},s}(\omega) = U \frac{1+ |\Pi^R(\omega) |^2}{| 1-[\Pi^R(\omega)]^2 |^2 }
\end{equation}
is a real, $\omega$-dependent effective coupling function expressed in terms of the retarded part of the loop function $\Pi$. Details of its derivation are given in Appendix \ref{app:DecompG}. 
As $U_{\textrm{eff},s}(\omega)$ replaces the bare coupling $U$ in the two-loop self-energy, the resummation leads effectively to the appearance of a frequency-dependent four-point vertex.
This vertex depends on a single frequency only because of its structure of consisting of loop chains connecting fields at two points in time.
For the $t$- and $u$-channel resummations, the same expressions for the components of $\Sigma$ apply as those in \Eq{effsigmas}, with the effective couplings now given by
\begin{equation}
  U_{\mathrm{eff},t}  = U_{\mathrm{eff},u}  = \frac{U}{|1-\Pi^R|^2 }.
\end{equation}
Details of their derivation can be found in Appendix \ref{app:EffU}.

\section{Transient dynamics and stationary state of the strongly correlated dot}
\label{sec:NumericalResults}
In the following we present our numerical results for the transient build-up of the population on the strongly interacting quantum dot coupled to two leads without chemical potential difference, i.e., without a voltage driving a current through the dot.
We consider different initial states, self couplings on the dot, detunings of the free dot level from the band-edge of the leads, as well as temperatures and bandwidths. 
As we show, our results underline the necessity to go beyond the perturbative coupling expansion of the 2PI part $\Gamma_{2}$ of the effective action. 
Our results in particular corroborate the necessity of the particle-hole ($u$-)channel resummation as it has been used within different theoretical approaches before.

We point out that our approach makes it possible to compute the buildup of Kondo correlations, i.e., in particular of the exponential narrowing of the Kondo resonance in accordance with the Bethe-ansatz result \eq{kondotemp}.

%
\subsection{Filling dynamics of the quantum dot}
We first consider the transient evolution of the electron population on the quantum dot after a sudden switch-on of the coupling to the leads, given a particular self coupling $U$ on the dot. 
We thereby study the effect of the different resummation schemes introduced in \Sect{resumsec}. 
We choose an initially empty dot, $n_{\uparrow}(0)=n_{\downarrow}(0)=0$. 
These occupation numbers determine the initial values of $F_{\sigma\sigma}(0,0)$, and we choose all other contributions $F_{\sigma\bar\sigma}(0,0)$ to vanish, implying the absence of correlations between the spin orientations. 
If not otherwise stated, we consider a vanishing magnetic field, $B=0$, the detuning $E_{0}=-U/2$ as for the weakly interacting dot at the particle-hole symmetric point, and a vanishing leads temperature of $T=0$.
We evolve with a time-step size $\Delta t=(300\,\Gamma)^{-1}$ up to $t_\mathrm{max}=6.6\,\Gamma^{-1}$, which amounts to a maximum number of $N=2000$ time steps integrated over in the memory integrals, see Appendix \ref{app:NumImp} for details of the numerical implementation.

%
\begin{figure}[t]
\begin{center}
\includegraphics[width=0.435 \textwidth]{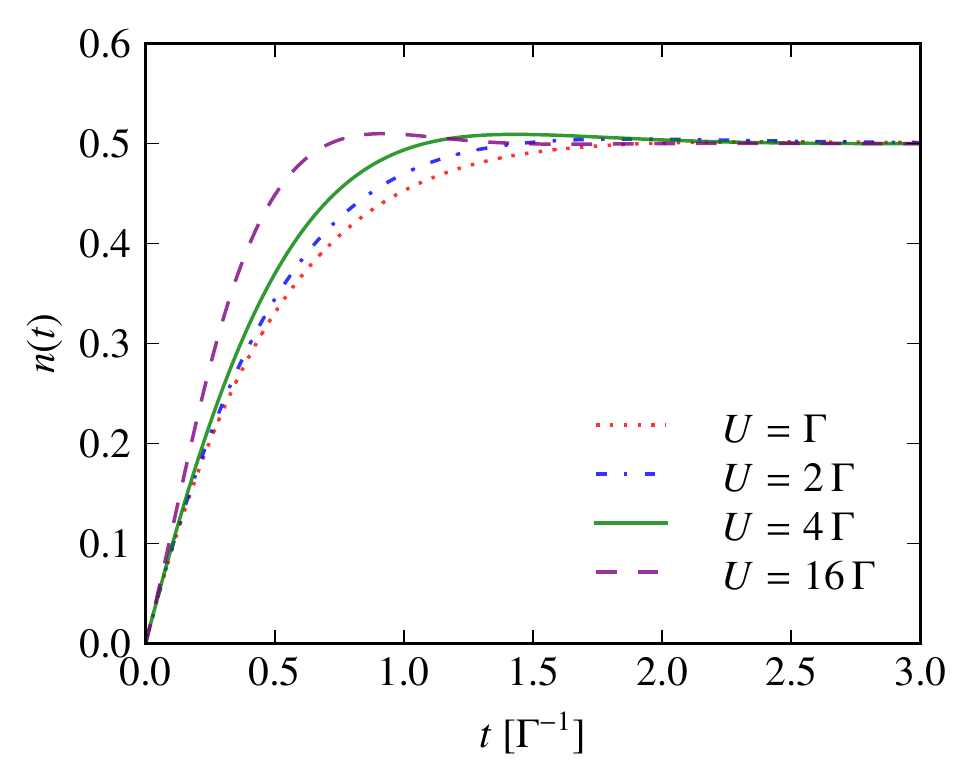}
\includegraphics[width=0.435 \textwidth]{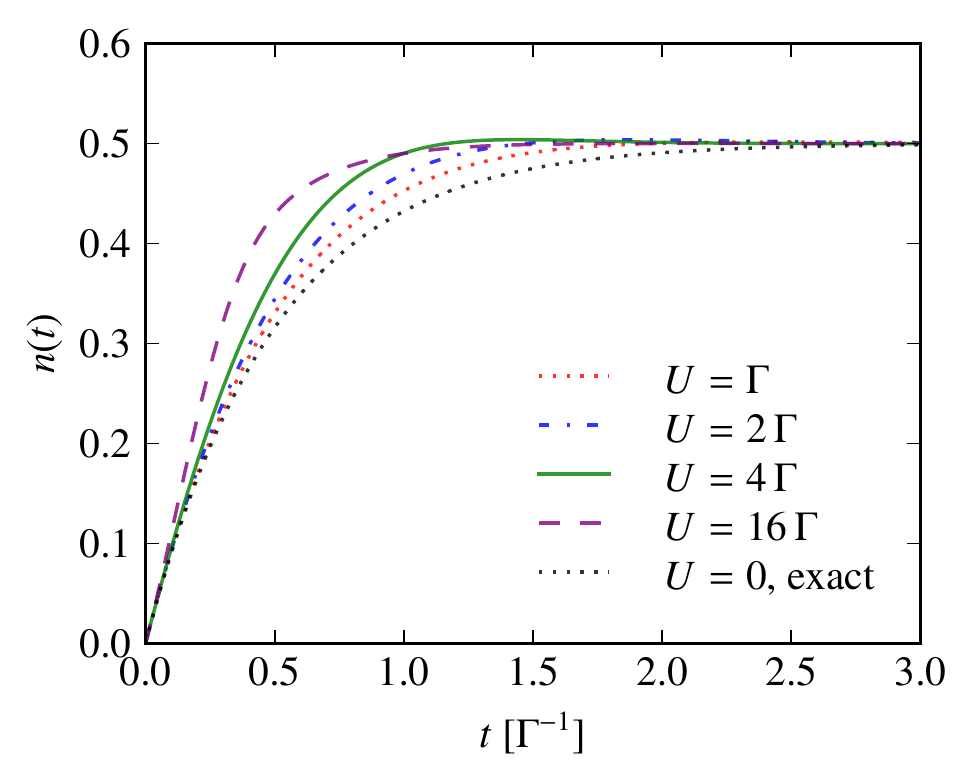}
\caption{(Color online) 
Comparison of the transient populations $n(t)=n_{\uparrow}(t)=n_{\downarrow}(t)$ of the quantum dot at the particle-hole symmetric point starting from zero occupation, after a sudden switch-on of the coupling to the zero-temperature bath represented by the leads, for different interactions between $\uparrow$- and $\downarrow$-spin electrons on the dot. $U$ and $t^{-1}$ are given in units of the hybridization $\Gamma$.
The results are obtained with the dot self-energy in the particle-particle ($t$-)channel (top) and  ($stu$-)channel resummations (bottom). The evolution in the direct ($s$-) and particle-hole ($u$-) channel resummation case was found to not differ from that shown in the bottom panel.}
\label{fig:n_transient}
\end{center}
\end{figure}
%
In Fig.~\ref{fig:n_transient} we compare the time dependence of the population of the dot as obtained  with the full 2PI equations of motion for different resummation schemes applied within the 2PI part $\Gamma_{2}$ of the effective action. 
The different panels show the results obtained with the dot self-energy in (top) particle-particle ($t$-)channel, (bottom)  $stu$-channel resummations.
The evolution in the direct ($s$-) and particle-hole ($u$-) channel resummation case was found to not differ from that shown in the combined $stu$-resummation scheme.
Note that the exact evolution for $U=0$ is given by
\begin{align}
n(t) = \frac12\left(1-e^{-2\,\Gamma t}\right).
\end{align}
We benchmarked our numerics with the exact solution in the noninteracting case and found a numerical error after the first time step of $|n_\mathrm{exact}-n_\mathrm{numerical}|=10^{-9}$, which exponentially decreases during the time evolution to an error of $10^{-15}$, when the occupancy becomes stationary at $n=1/2$. 

In Ref.~\onlinecite{schmidt2008}, a similar analysis was done with perturbation theory and Monte Carlo (MC) methods with results  in good concordance with ours, especially with respect to the MC data. 
The main discrepancy occurs in the time range up to $\Gamma t\simeq0.15$. 
Initially, it seems that the MC data for the occupation number starts with a higher exponent before it continues to increase linearly. 
Note that in contrast to our wide flat-band approximation, in Ref.~\onlinecite{schmidt2008} a more realistic model with a soft cutoff for the leads is taken into account.  

Here we used the particle-hole symmetric setup, which means that the single- and double-occupation levels in the quantum dot are symmetrically situated around the Fermi edge of the leads. 
Therefore, we expect a stationary occupation of $n = 0.5$ which is indeed obtained in all resummation schemes. 
For later times than are shown in \Fig{n_transient}, we find no change in the occupation until our maximum evolution time $t_\mathrm{max} = 6.6\,\Gamma^{-1}$. 
Numerical instabilities observed in the $t$-channel scheme at later times are found to decrease with smaller time stepping.

%
\begin{figure}[t]
\begin{center}
\includegraphics[width=0.435 \textwidth]{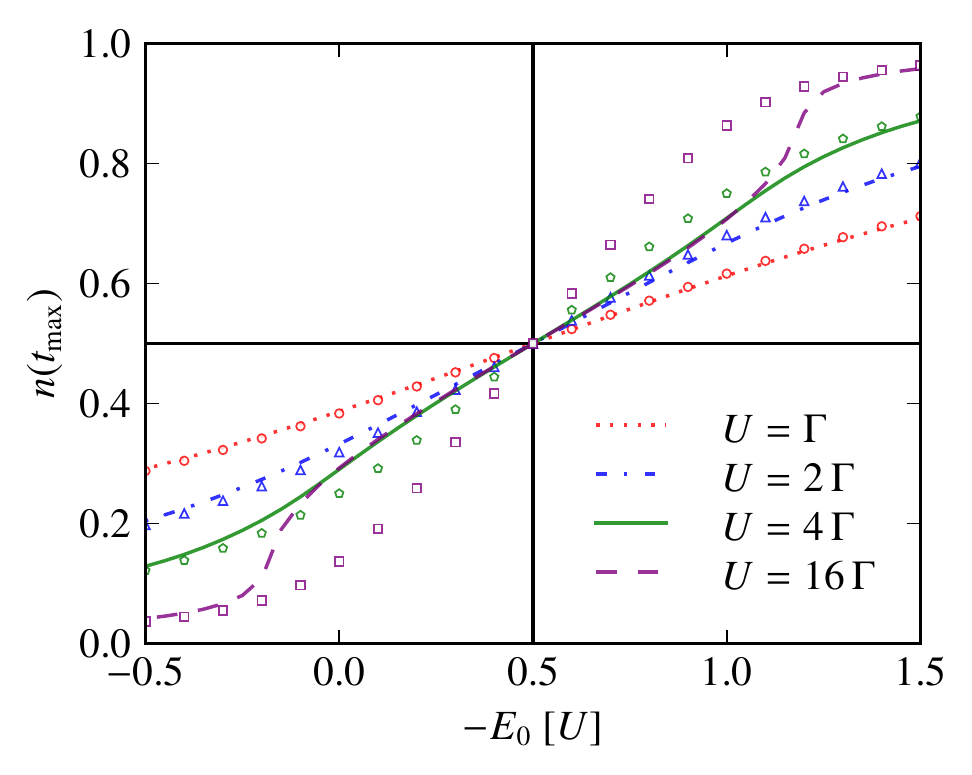}
\includegraphics[width=0.435 \textwidth]{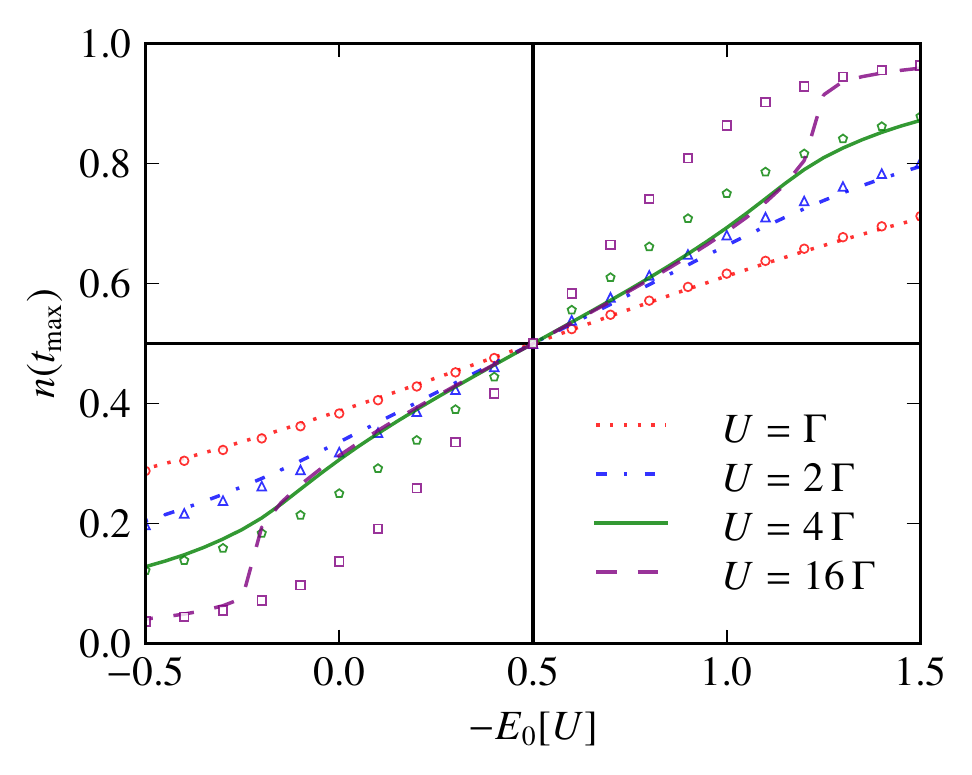}
\caption{(Color online)
The long-time stationary occupation number arising in our real-time evolutions of the Kondo dot coupled to zero-temperature leads, starting  from zero initial occupation, as a function of the gate voltage $E_0$ in units of the interaction strength $U$, for different $U$. 
The data shown as lines are obtained in the particle-particle ($t$-)channel (top panel) and ($stu$-)channel (bottom panel).
Symbols mark the mean-field solution (\ref{eq:mf-nstat}). 
The respective data found in the direct (s-)channel and particle-hole ($u$-) channel approximations do not differ from that in the $stu$-channel. 
The vertical and horizontal black solid lines in each panel mark the particle-hole symmetric setup for which the entire time evolution is shown in Fig.~\ref{fig:n_transient}.}
\label{fig:ntmax_detuning}
\end{center}
\end{figure}
%

%
\subsection{Long-time stationary occupation of the dot}
The stationary occupation number at the end of the finite time evolution is shown, for the different channel approximations and also away from the particle-hole symmetric scenario in Fig.~\ref{fig:ntmax_detuning}. 
The two black solid lines in each panel mark the particle-hole symmetric case discussed before. 
Within all resummation schemes, the occupation number reaches the expected value $n_\uparrow=n_\downarrow=0.5$ for all probed interaction strengths at the particle-hole symmetric point. 

We distinguish three regions: 
First, in the case of a gate voltage much lower than the negative interaction strength, $-E_{0}/U\gg1$, both energy levels lie below the Fermi edges and the quantum dot is fully occupied with two electrons, which means $n = 1$ for the single electron. Second, the dot is empty, $n = 0$, for gate voltages much larger than zero, $-E_{0}/U\ll0$, because both energy levels lie above the chemical potentials of the leads. 
Third, in the regime around the particle-hole symmetric point, $-E_{0}/U\simeq0.5$, the quantum dot is singly occupied, $n = 1/2$. 
The transition between the three regions is determined by the width of the energy levels, which means by the value of the transition strength $\tau$.

Besides the numerical results shown by the lines in each panel, we show the corresponding mean-field solutions for the stationary occupation number, obtained from
\begin{align}
\label{eq:mf-nstat}
n_{\sigma}=\frac12-\frac1\pi \arctan\left(\frac{E_{0}+Un_{\bar\sigma}}{\Gamma}\right)
\end{align}
and denoted by symbols in the respectively same colors. 
In the case of a vanishing magnetic field, \Eq{mf-nstat} simplifies because both occupation numbers are equal. 

For small interaction strengths, the results obtained from the nonperturbative resummation of the 2PI effective action agree very well with the mean-field approximation. 
An explicit deviation, as expected, is visible for larger interaction strengths in the $stu$-channel approximation. 
The particle-particle ($t$-)channel results remain closer to the mean-field values. 
All nonperturbative resummations approach the expected values for large negative values of the gate voltage, where the quantum dot is doubly occupied, as well as for large gate voltages, where the quantum dot is empty. 
For large interaction strengths it is expected that near the particle-hole symmetric point the derivative of the occupancy with respect to the gate voltage $-E_{0}$ is smaller than predicted in the mean-field approximation \eq{mf-nstat}. 
Our data shows a weakening of this derivative for interaction strengths above $U = 4\,\Gamma$, although not as far as found in NRG calculations\cite{Costi2010a.PhysRevB.81.235127} for $U=8\,\Gamma$. 
Note that the consistency checks discussed in \Sect{Friedel} point to finite-size effects for $U \gtrsim 4\,\Gamma$.

%
\subsection{Spectral function: Kondo resonance}
\label{sec:SpecFuncKondo}
\begin{figure}[t]
\begin{center}
\includegraphics[width=0.435 \textwidth]{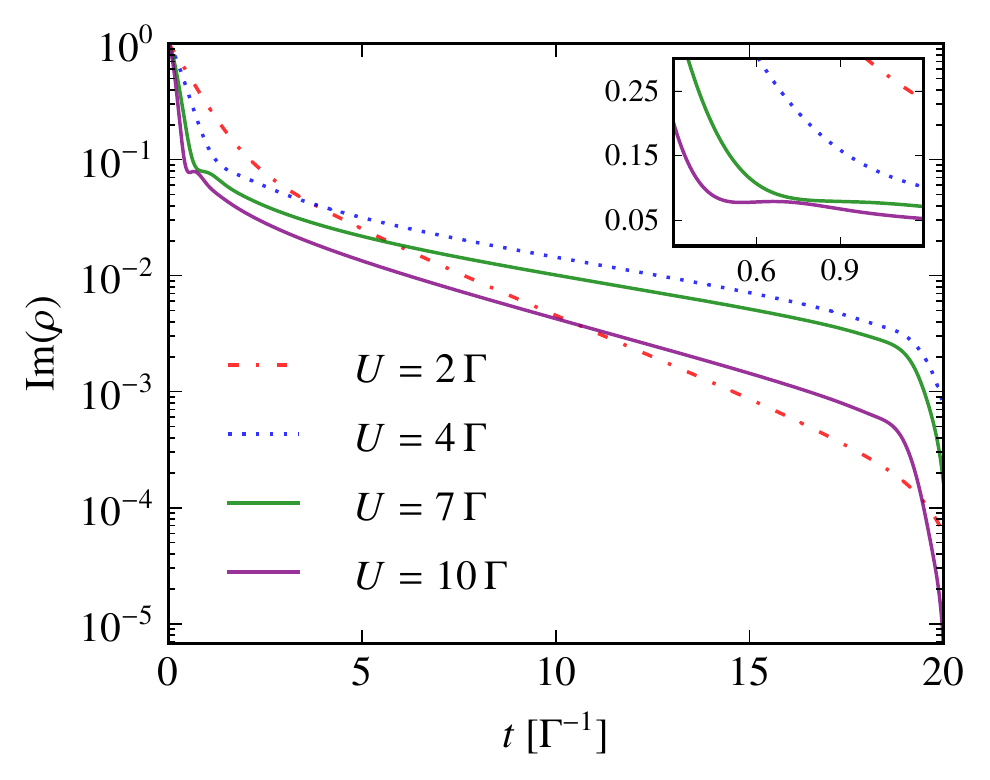}
\includegraphics[width=0.435 \textwidth]{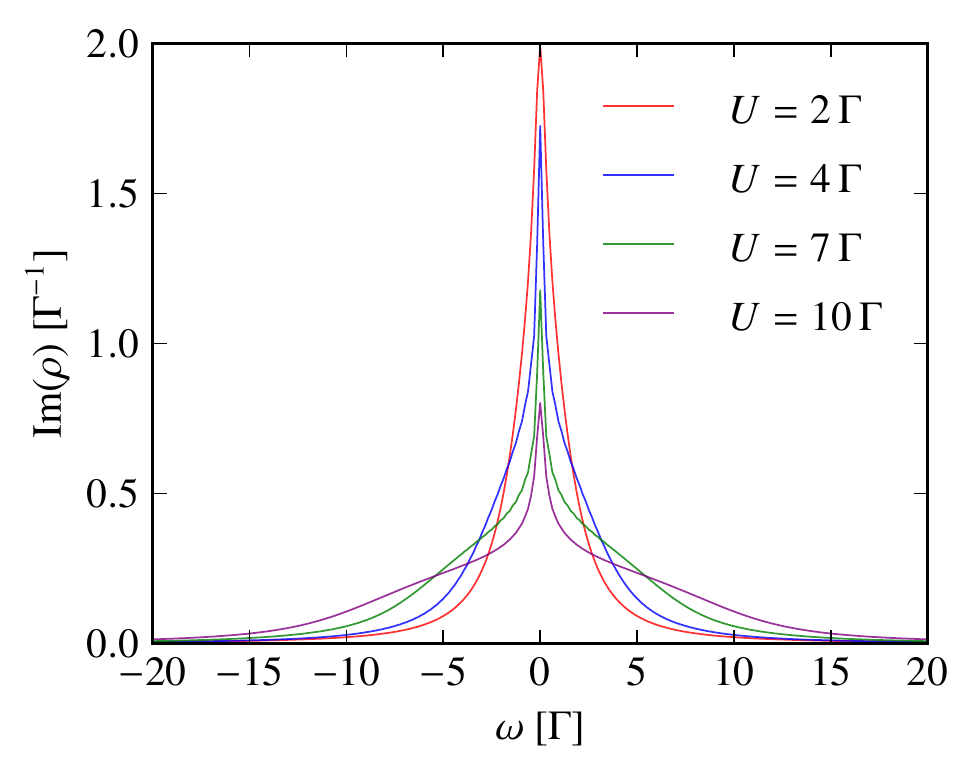}
\caption{(Color online)
(Top) Imaginary part of the spectral function $\rho(t)=\rho(t_\mathrm{max},t_\mathrm{max}-t)$, as obtained in the  $stu$-channel, with $t_\mathrm{max}=20\,\Gamma^{-1}$ and a time-step size of $\Delta t=(800\,\Gamma)^{-1}$, for different interaction strengths $U$ in the particle-hole symmetric setup, $E_{0} = -U/2$. 
The inset shows the transition region from large to small exponential decay. 
(Bottom)
$\mathrm{Im}[\rho(\omega)]$, as obtained by Fourier transform of the functions shown in the top panel. 
The narrowing of the central Kondo resonance is easily seen.
$U$, $t^{-1}$, $\omega$, and $[\mathrm{Im}\,\rho(\omega)]^{-1}$ are given in units of $\Gamma$.
}
\label{fig:rho-omega}
\end{center}
\end{figure}

We begin our discussion of the stationary spectral function with its dependence on the interaction strength $U$ of the on-site Coulomb repulsion. 
We let the quantum dot, which is adjusted to the particle-hole symmetric point, evolve in time until the total time $t_\mathrm{max} = 20\,\Gamma^{-1}$, with a time-step size of $\Delta t = (800\,\Gamma)^{-1}$. 
%
\begin{figure}[t]
\begin{center}
\includegraphics[width=0.435 \textwidth]{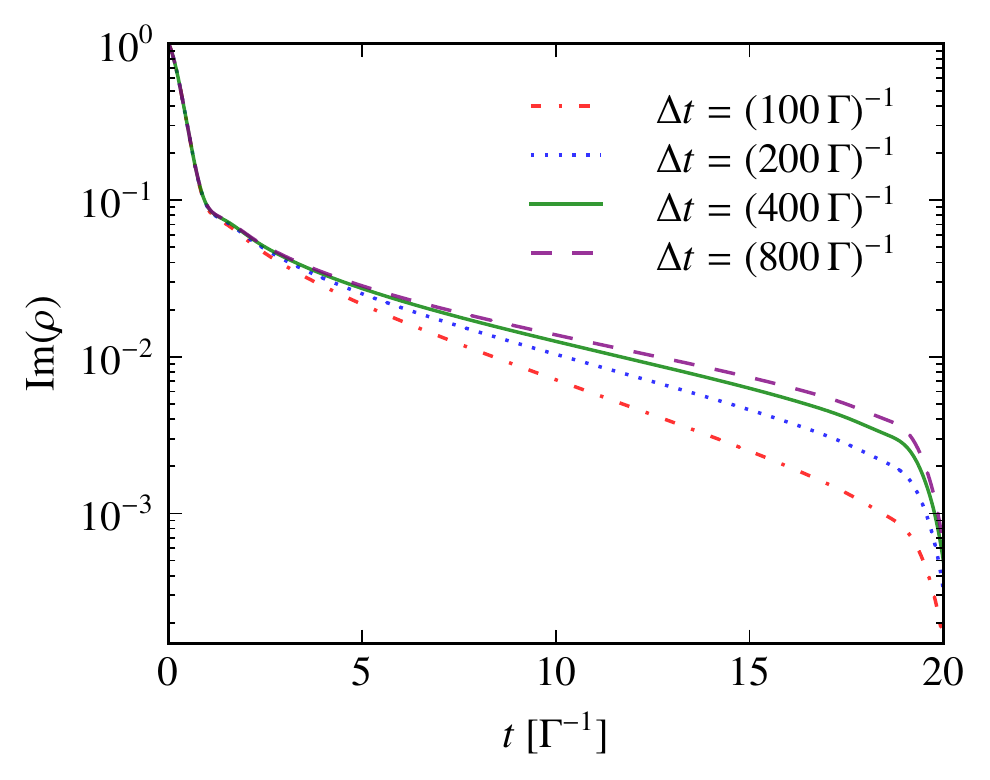}
\includegraphics[width=0.435 \textwidth]{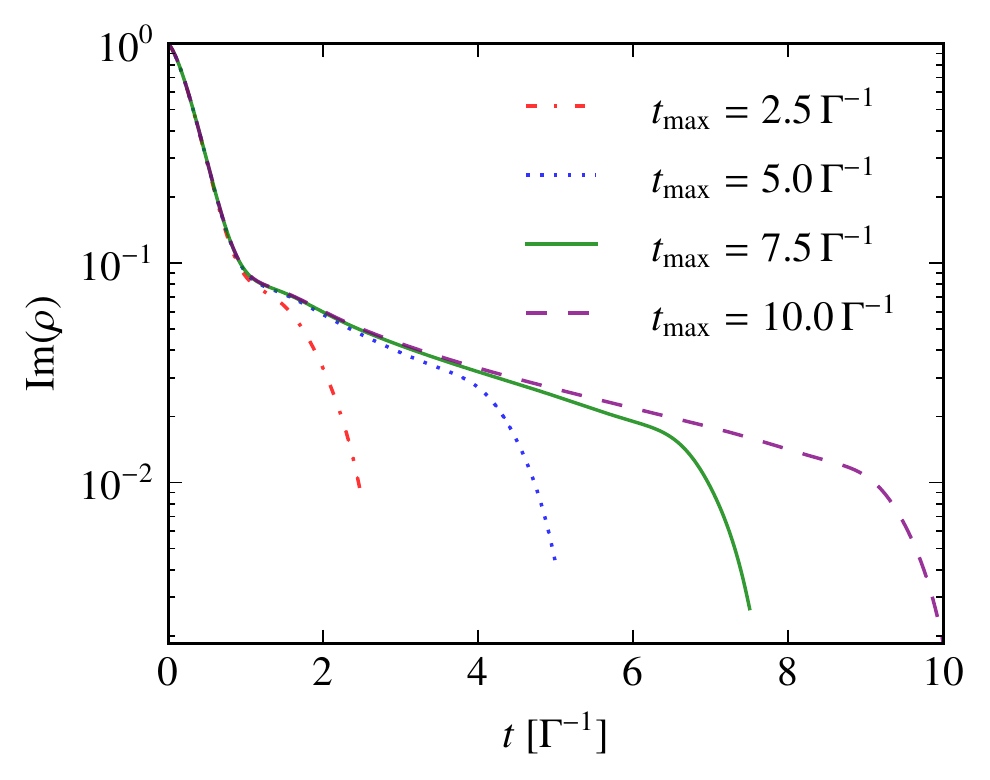}
\caption{(Color online)
(Top) The imaginary part of the spectral function $\rho(t)=\rho(t_\mathrm{max},t_\mathrm{max}-t)$ obtained in the $stu$-channel, for an interaction strength $U=5\,\Gamma$ and a maximum evolution time $t_\mathrm{max}=20\,\Gamma^{-1}$, for different time-step sizes as indicated.
(Bottom)
Same function, as obtained in the $stu$-channel scheme, for $U=5\,\Gamma$, $\Delta t=(800\,\Gamma)^{-1}$, and different maximum times $t_\mathrm{max}$.
The system is chosen in the particle-hole symmetric setup, $E_{0} = -U/2$, at zero temperature, and $t$, $\Delta t$, and $t_\mathrm{max}$ are given in units of $\Gamma^{-1}$, while $\rho(t)$ is dimensionless.
}
\label{fig:rho-t}
\end{center}
\end{figure}
%
In \Fig{rho-t}, the imaginary part of the spectral function $\rho(t)=\rho(t_\mathrm{max},t_\mathrm{max}-t)$, at the end of the time evolution, is shown as obtained in the $stu$-channel approximation, for  $U = 5\,\Gamma$, $t_\mathrm{max}=20\,\Gamma^{-1}$ and different time step sizes $\Delta t$ (top panel) as well as for a fixed $\Delta t=(800\,\Gamma)^{-1}$ and varying maximum evolution times (bottom panel).
In \Fig{rho-omega} (top panel) the imaginary part of the spectral function is shown again as obtained in the $stu$-channel approximation, for small, $U = 2\,\Gamma$, to large interaction strengths, $U = 10\,\Gamma$. 
During the time evolution we calculate the real and imaginary parts of the spectral function in the whole real-time plane. 
In the figure, we show only the imaginary part at the end of the time evolution for positive relative times because the real part is zero and the imaginary part is symmetric according to the symmetry relation in \Eq{symmetries}.
For a comparison of the data shown with that obtained in the other channels, see Appendix \ref{app:NumData}.

%
%
The spectral function for the symmetric single-impurity Anderson model (SIAM) without a bias voltage, at very low temperatures, shows three peaks.
The Hubbard side peaks, corresponding to the single and double-occupation states, have a width of the order of $\Gamma$, while the central, very narrow Kondo peak, located at the Fermi edge has a width of the order of the Kondo temperature $T_\mathrm{K}$. 
Thus, the spectral function in the time domain can be approximately written as
\begin{align}
\rho(t) = Ae^{-\Gamma|t|}\cos(Ut/2)+Be^{-T_\mathrm{K}|t|/2},
\end{align}
where the first term describes the side peaks with width (full width at half maximum) $2\,\Gamma$, located at points $\omega = \pm U/2$, and the second term describes the central Kondo peak.
While the real solution will differ from this functional form its characteristics should be similar.
We see that the first term approaches zero much faster than the second one because the width of the side peaks is much larger than the Kondo resonance, $2\,\Gamma\gg T_\mathrm{K}$. 
While the short-time behavior is dominated by the first term describing the coupling to leads, the late-time characteristics is determined by the Kondo correlations encoded by the second term.

Comparing the above parametrization with the data shown on a log-linear scale in \Fig{rho-omega} (top panel) we find the two exponentials to be clearly visible in $stu$-channel, as well as in the $s$, $u$, and $t$-channels separately; see Appendix \ref{app:NumData}.
The initial drop off depending on the interaction strength follows the behavior predicted by the cosine.
Some indications of oscillations can be seen at the time of transition between the exponentials in the $stu$-channel case, while these are not seen in the channels separately;  see Appendix \ref{app:NumData}.

In the bottom panel of \Fig{rho-omega}, we show the corresponding spectral functions in Fourier space, to the same order as shown in the top panel.
The spectral functions give indications of the expected Hubbard side bands, and the shape of the function does not change much with increasing interaction strength. 
This is different for the other three approximation schemes, see Appendix \ref{app:NumData}.
Besides the Kondo peak at the origin, there are side peaks located at $\omega=\pm U/2$.
These peaks become visible at interaction strengths of $U\gtrsim4$. 
In the stationary limit, the spectral function can be expressed in terms of the spectral part $\Sigma^{\rho}$ of the self-energy; see Appendix \ref{sec:KineticTheory} for details,
\begin{align}\label{eq:spectralfunckin}
    \rho\left(\omega\right)
    &=\frac{\Sigma^{\rho}}{\left(\omega- \textrm{Re} \Sigma\right)^{2}+\left|{\Sigma^{\rho}}/{2}\right|^{2}},
\end{align}
with
\begin{equation}\label{eq:realsigma}
 \textrm{Re} \Sigma\left(\omega\right)=-\mathcal{P}\int_{-\infty}^{\infty}\frac{d\omega^{\prime}}{2\pi}\frac{\Sigma^{\rho}\left(\omega^{\prime}\right)}{\omega-\omega^{\prime}}.
\end{equation}
Exemplary graphical representations of $\Sigma^{\rho}$ and $\textrm{Re} \Sigma$ are given in Fig.~\ref{fig:SelfEnergies} in Appendix \ref{app:NumData}.

%
\subsection{Dynamical build-up of the spectral function}
%
%
\begin{figure}[t]
\begin{center}
\includegraphics[width=0.49 \textwidth]{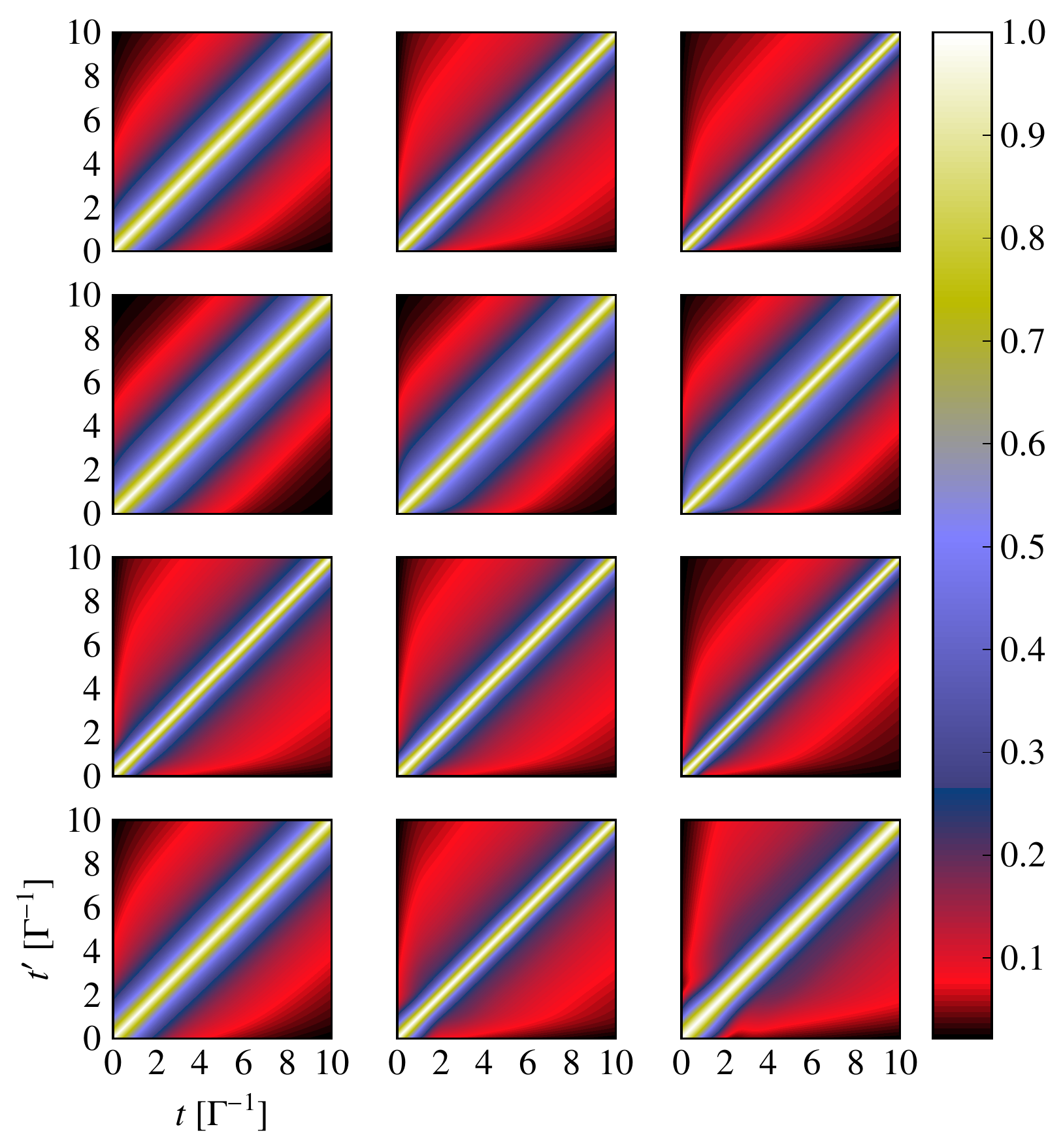}
\caption{(Color online)
The imaginary part of the spectral function in the real time plane, $\mathrm{Im}\rho(t,t')$. The data is obtained within the $s$-, $t$-, $u$-, and $stu$-channel resummation schemes (from top to bottom), for different interaction strength $U/\Gamma = 2$, $4$, $6$ (from left to right), in the particle-hole symmetric setup and for zero-temperature leads.
$\rho$ is normalized to unity on the time diagonal. 
The time-step size was chosen as $\Delta t=(50\,\Gamma)^{-1}$. 
Times are given in units of $\Gamma^{-1}$.}
\label{fig:rho_density}
\end{center}
\end{figure}
%
\Fig{rho_density} shows the imaginary part of the spectral function $\rho(t,t')$ in the real-time plane $(t,t')$ as obtained with the different resummation schemes (top to bottom: $s$, $t$, $u$, $stu$) and for different interaction strengths (left to right: $U/\Gamma = 2$, $4$, $6$), as before in the particle-hole symmetric case, $E_{0}=-U/2$, and for zero temperature. 
As indicated by \Fig{sumrule-deltat}, for early times the results do not depend on the time-step size in the range $\Delta t=(100\,\Gamma)^{-1}\dots(800\,\Gamma)^{-1}$.
Here, we set the step size to $\Delta t=(50\,\Gamma)^{-1}$.
We point out that only in the combined $stu$ case, does the build-up of the Kondo tail with suppressed exponential decay in relative time $|t-t'|$ set in after a delay period of $(t+t')/2 \simeq 2\,\Gamma^{-1}$.
This sharp onset contains signatures of an oscillating spectral function in relative time and thus of the Hubbard side bands which are clearly visible in the $stu$-channel spectral function in \Fig{rho-omega}.

%
\subsection{Friedel sum rule}
\label{sec:Friedel}
The height of the peak of $\rho(\omega)$ at $\omega=0$ is decreasing with increasing interaction strength $U$, implying that the area under the temporal spectral function in time is not conserved.
This is in disagreement with the Friedel sum rule \cite{friedel1952,langer1961,langreth1966} 
which provides an exact relation between the additional states induced below the Fermi energy by a scattering center and the respective scattering phase shift. 
It also holds true for interacting systems.
The generalized Friedel sum rule \cite{nuss2012} connects the quantum dot occupation number with the density of states, i.e., the spectral function in Fourier space at the Fermi edge.
At zero temperature and in the particle-hole symmetric case, the exact relation reads 
\begin{align}
\rho(0) = \frac{2i}{\Gamma}\,\sin^{2}\left(\frac{\pi[n_{\uparrow}+n_{\downarrow}]}{2}\right),
\end{align}
independent of the interaction strength $U$. 
Hence, the sum rule requires a constant height of the peak at the Fermi edge.
Note that the mean-field expression \eq{rhoMFleads} fulfills the sum rule.

%
\begin{figure}[t]
\begin{center}
\includegraphics[width=0.435 \textwidth]{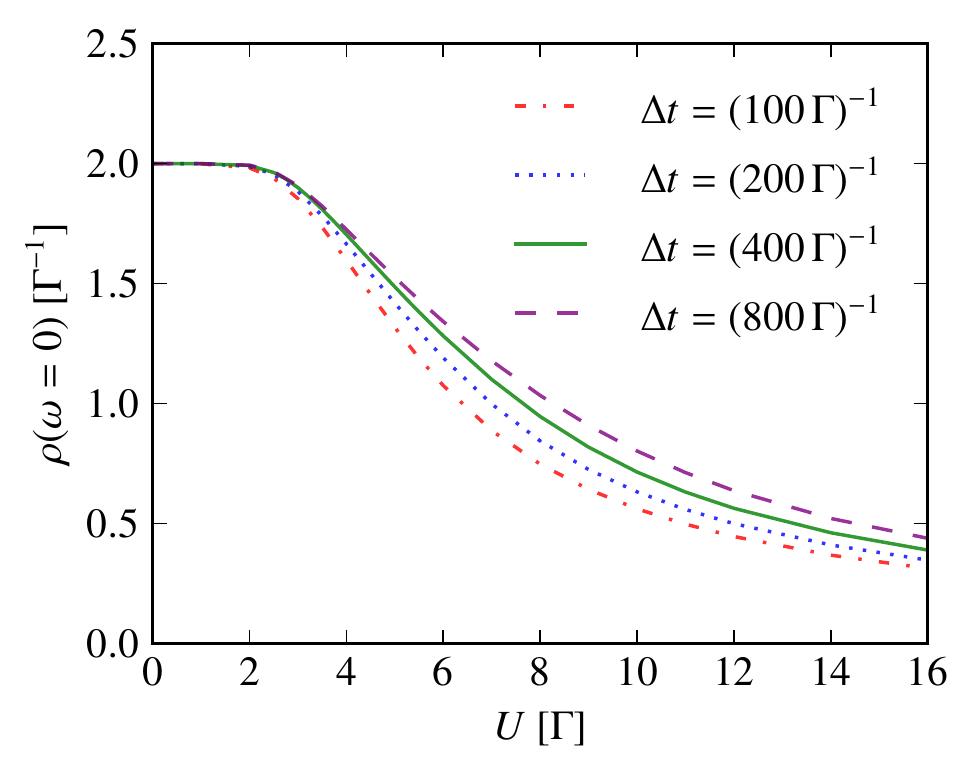}
\includegraphics[width=0.435 \textwidth]{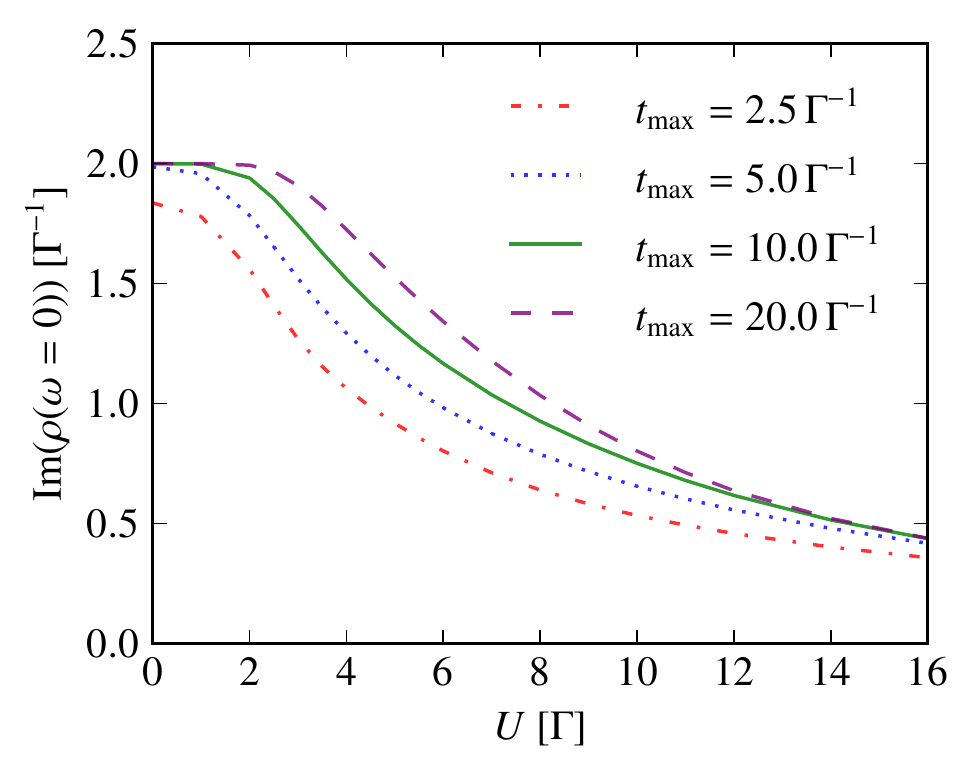}
\caption{(Color online)
Friedel sum rule: The spectral function $\rho(\omega=0)$, as a function of the interaction strength $U$,  obtained in the $stu$-channel resummation scheme, for different time-step sizes and a total evolution time $t_\mathrm{max}=20\,\Gamma^{-1}$  (top) and for different evolution times and a step size of $\Delta t=(800\,\Gamma)^{-1}$ (bottom).
The leads' temperature is $T=0$, and the dot is tuned to the particle-hole symmetric point $E_{0}=-U/2$.
$U$, $t_\mathrm{max}^{-1}$, and $(\Delta t)^{-1}$ are given in units of $\Gamma$.
The figures show that for a given coupling $U$, the Friedel sum rule $\rho(\omega=0)=2$ is the better obeyed the longer the total time at fixed step size or the shorter the step size at fixed total time is.
}
\label{fig:sumrule-deltat}
\end{center}
\end{figure}
%
In \Fig{sumrule-deltat}, we show the results obtained from the nonperturbative resummation of the 2PI effective action in the combined $stu$-channel scheme. 
Results obtained for the $s$- and $u$-channel resummations do not visibly differ from the data shown while in the $t$-channel case the sum rule is obeyed as long as the numerical calculation remains stable, see Appendix  \ref{app:NumData}.
The top panel shows $\rho(\omega=0)$ for different time-step sizes and a total evolution time $t_\mathrm{max}=20\,\Gamma^{-1}$ while in the bottom panel, the evolution time is varied, keeping a constant step size of $\Delta t=(800\,\Gamma)^{-1}$.
The leads' temperature is chosen to vanish, $T=0$, and the dot is tuned to the particle-hole symmetric point $E_{0}=-U/2$.
Our results indicate that the nonperturbative 2PI approach fulfills the sum rule which is expected to apply in the stationary limit, i.e., at long evolution times. 
Deviations from the rule are due to a finite total evolution time as well as time resolution.
We find such deviations irrespective of the correct occupation number at the particle-hole symmetric point.
Increasing the evolution time $t_\mathrm{max}$ by a factor of two extends the range of $U$ for which the sum rule is obeyed by roughly $\Gamma$.
For the spectral functions shown in the previous subsection, the sum rule is obeyed for $U\lesssim2\,\Gamma$.
Increasing the number of evolution time steps by a factor $\alpha$ increases the total computing time by a factor of $\alpha^{3}$ and the required size of memory by a factor of $\alpha^{2}$; see Appendix \ref{app:NumImp} for details of the numerical implementation.

%
\subsection{Kondo temperature}
\label{sec:KondoTemp}
%
%
\begin{figure}[t]
\begin{center}
\includegraphics[width=0.435 \textwidth]{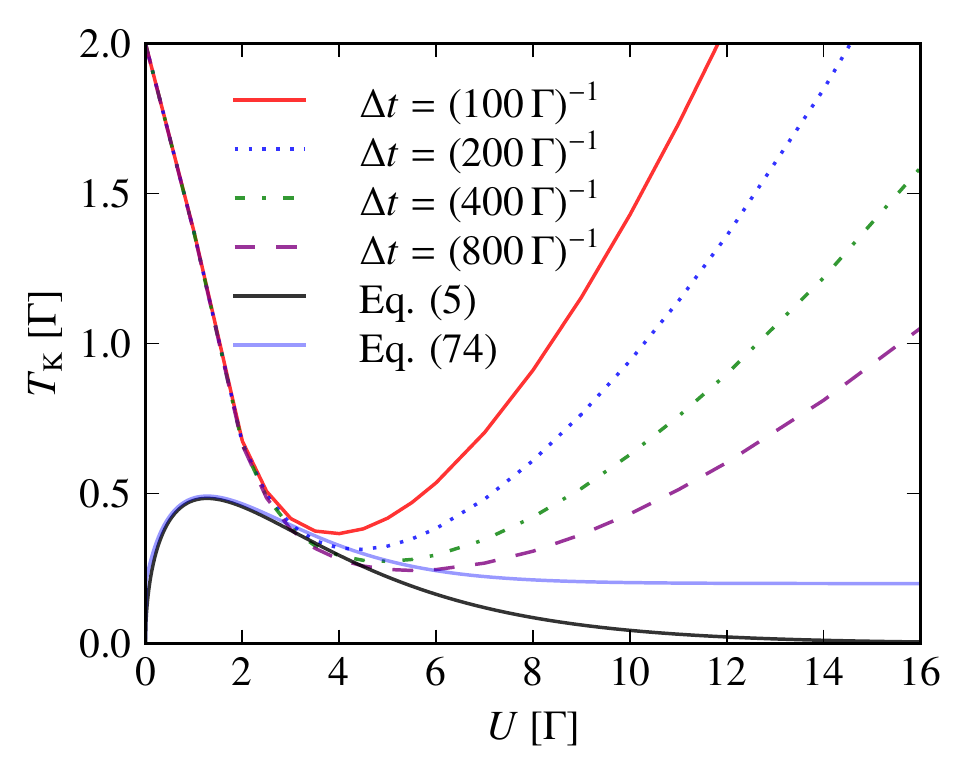}
\caption{(Color online) 
The Kondo temperature, extracted from an exponential fit to the imaginary part of the spectral function (top panel of Fig.~\ref{fig:rho-t}) within $t=(12\dots14)\,\Gamma^{-1}$, as a function of the interaction strength $U$, for different time step sizes. 
$U$, $T_\mathrm{K}$, and $(\Delta t)^{-1}$ are given in units of $\Gamma$.
We compare with the analytical predictions given in Eqs.~(\ref{eq:kondotemp}) (solid black line) and (\ref{eq:kondotempfintime}) (solid pale blue line).
The curves interpolate $13$ data points between $U=0$ to $U=12\,\Gamma$, in unit steps, and intermediate data points at $U/\Gamma=2.5$, $3.5$, $4.5$, $5.5$, $14$, and $16$.
}
\label{fig:TKondo}
\end{center}
\end{figure}
%
We extract the Kondo temperature from an exponential fit to the spectral function in the time domain,
\begin{align}
\rho_\mathrm{fit} = B_{\mathrm{fit}}\exp\{-T^{\mathrm{fit}}_\mathrm{K}|t|/2\}.
\end{align}
We fit this exponential at late evolution times within the range $t=(12\dots14)/\Gamma$ in order to avoid the bending-down near $t=t_\mathrm{max}$.
The Kondo temperature we could extract in this manner is depicted in \Fig{TKondo} as a function of $U$, for the $stu$-channel case and different time steps $\Delta t$ (as usual indicated in units of $\Gamma^{-1}$).
As before, only in the $t$-channel case, are there significantly different predictions, see Appendix  \ref{app:NumData}.
To emphasize the significance of the 2PI results, we compare with the $U$-dependence of the Kondo temperature as given by the standard equilibrium expression  [\Eq{kondotemp} (black line)] as well as by the effective Kondo temperature 
\begin{equation}
\label{eq:kondotempfintime}
  T_{K,\mathrm{eff}}(t) = T_{K}/\tanh[T_{K}t/4],
\end{equation}
predicted in Ref.~\onlinecite{Nordlander1999a.PhysRevLett.83.808}, on the basis of the Kondo Hamiltonian, to apply at a finite evolution time $t$ after a sudden connection of the dot to the leads.
The solid pale blue line shows $T_{K,\mathrm{eff}}(t)$ at the maximum evolution time $t=t_\mathrm{max}=20\,\Gamma^{-1}$.

The above results give evidence that the nonperturbative resummation of the self-energy applied within the 2PI approach to the Anderson-model dynamics quantitatively recovers the dynamic build-up of the Kondo resonance.
Specifically, the expected Kondo temperature, which exponentially decreases with $U$, is recovered within a limited regime of coupling strengths around $U=4\,\Gamma$.
We emphasize that our results are consistent with the prediction (\ref{eq:kondotempfintime}) obtained within the approximations of the Kondo model\cite{Nordlander1999a.PhysRevLett.83.808} that building up the Kondo temperature scale takes a time $t\gg t_K\sim1/T_K$.

At $U\lesssim3\Gamma$, our method of fitting an exponential to the tail of $\rho(t)$ can not distinguish the Kondo and the side-peak resonances such that the Kondo temperature rises to $T_\mathrm{K}=2\,\Gamma$, which corresponds to the noninteracting case and is required by the Friedel sum rule.
For $U\gg4\,\Gamma$, we expect the deviation of our numerical data from the expected exponential suppression be due to the finite time span and resolution, as indicated by the comparison of different choices for $\Delta t$ reachable with the computing resources which were available to us. 
We have checked numerically that, to recover the exponential suppression of $T_\mathrm{K}$ in $U$ [cf.~\Eq{kondotemp}] for $U\simeq4\,\Gamma$, the width of the spectral function must be correct  to order $\mathcal{O}(U^{7})$.

\section{Quantum dot subject to a bias voltage}
\label{sec:CurrentVoltage}
\begin{figure}[t]
\centering
\includegraphics[width=0.35 \textwidth]{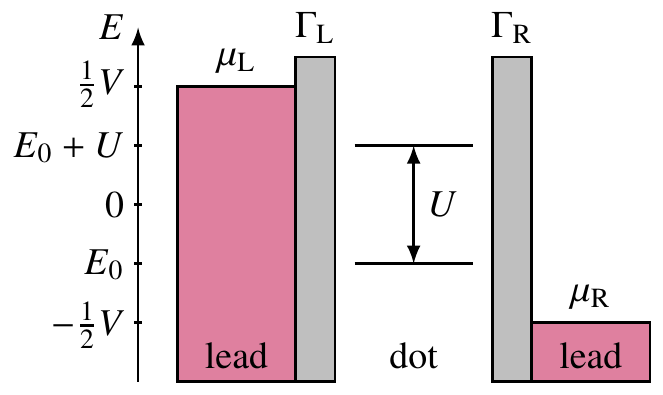}
\hspace*{1cm}
\caption{(Color online)
Schematic representation of the energy levels in the Anderson model of a quantum dot for the case of an applied bias magnetic field, cf.~Fig.~\ref{fig:dot}.
$\mu_\mathrm{L}=-\mu_\mathrm{R}=V/2$ are the energies of the Fermi edges of the leads, separated by the applied voltage $V$.
}
\label{fig:dot_Bfield}
\end{figure}

We now study the nonequilibrium case by applying a bias voltage to the quantum dot as depicted in Fig.~\ref{fig:dot_Bfield}. 
We investigate the transient current through the quantum dot as well as the current-voltage characteristics of the stationary current.
We compare with results obtained with functional renormalization-group (FRG) methods\citeafter{,}{Jakobs2007} as well as the iterative sum of path integrals (ISPI)\citeafter{.}{weiss2008,
weiss2013,becker2012} 
With this we generalize previous studies presented in Ref.~\onlinecite{Sexty:2010ra}, where 2PI results obtained in the direct ($s$)-channel scheme were compared with FRG, ISPI, real-time quantum Monte Carlo  \cite{mak1999,egger2000,muehlbacher2003,werner2009,werner2010} and time-dependent density matrix renormalization group (tDMRG) results\citeafter{.}{heidrich-meisner2009,daley2004,white2004,schmitteckert2004,al-hassanieh2006,boulat2008,dias2008} 
We remark that in Ref.~\onlinecite{heidrich-meisner2009}, FRG and tDMRG results, and in Ref.~\onlinecite{eckel2010}, FRG, tDMRG, ISPI, and rtQMC results were compared in the Kondo as well as the mixed valence regime.

As before, we start with an initially empty quantum dot decoupled from the thermally equilibrated leads.
For this we instantaneously couple the noninteracting metallic leads to the quantum dot by quenching the hybridization parameter $\tau$, interaction strength $U$, and bias voltage $V$. 
The left lead is assumed to have a lower chemical potential than the right one,
$\mu_\text{L} = -\mu_\text{R} = V/2$, implying a bias voltage $V = \mu_\text{L} - \mu_\text{R}$. 
As before we assume the leads to stay in thermal equilibrium over the entire time evolution and adjust the gate voltage such that the quantum dot stays at the particle-hole symmetric point, $E_0=-U/2$.

%
\subsection{Transient electric current}
To determine the transient electric current building up after the dot is coupled to the leads, we evaluate the expression
\begin{align}
  \label{eq:electrical_current}
    I(t) 
    &= {\Gamma}\, \mathrm{Im} \sum_\sigma \int_{-D}^{D} {d}\epsilon \int_0^{t} {d}t' 
    \Big[ f(\epsilon-\mu_\text{L}) - f(\epsilon-\mu_\text{R}) \Big]
    \nonumber\\
    &\qquad\times\,\exp\{-i \epsilon (t-t')\}\, \rho_\sigma(t',t), 
\end{align}
where $D$ denotes the half-width of the band taken into account around the Fermi edge of the leads.
The derivation of \Eq{electrical_current} is given in Appendix \sect{electrical_current}.
In Fig.~\ref{fig:iv_t_stu}, we show the build-up dynamics of the electric current $I(t)$, obtained in the 
$stu$-channel 
for different bias voltages $V$, for an on-site interaction strength $U=4\,\Gamma$ and a time-step size of $\Delta t = (300\,\Gamma)^{-1}$. 
For the other channels we did not find any qualitative differences.
We observe three characteristic regions for all considered resummation schemes.
The initial linear rise of the current is determined by the applied bias voltage,
\begin{equation}
  \left. \frac{{d}I}{{d}t}\right|_{t=0} = 2 \, V \,. 
\end{equation}
Recall further that the initial slope of the rising dot occupation number was found to be independent of the probed Coulomb repulsion strengths up to $U=16\,\Gamma$. 
Since the dot is initially empty, this is also true for the current. 
\begin{figure}[t]
\begin{center}
\includegraphics[width=0.435 \textwidth]{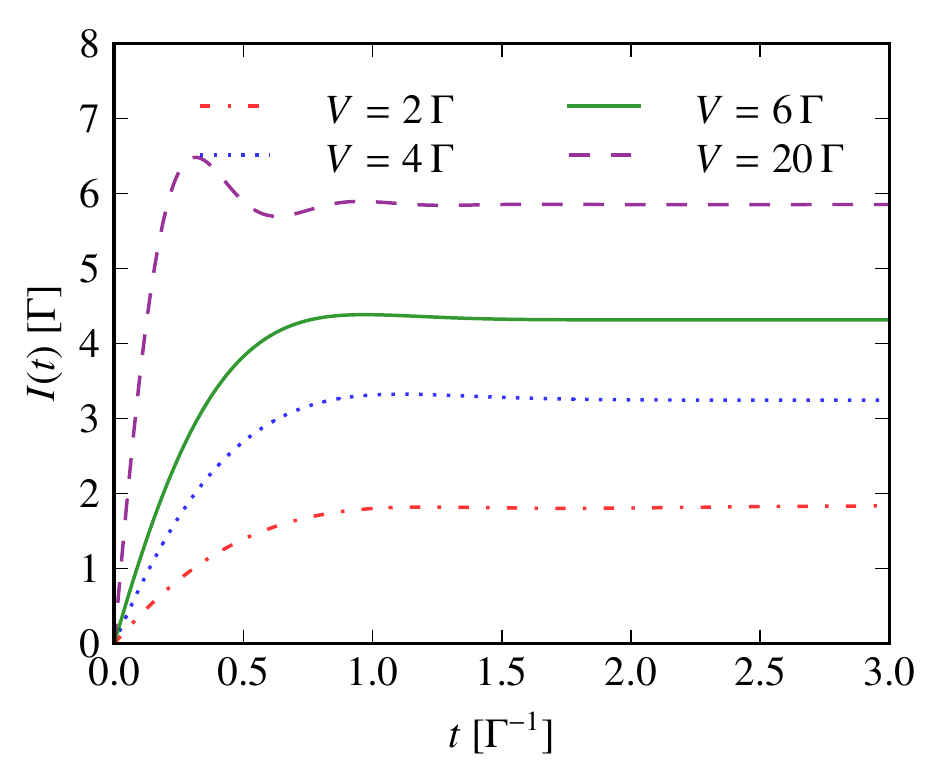}
\caption{(Color online)
The transient current $I(t)$ through a quantum dot coupled to leads at a temperature $T=0.1\,\Gamma$, as obtained within the 
$stu$-channel resummation scheme, for different applied bias voltages $V$.
The interaction strength is $U=4\,\Gamma$ and the time-step size $\Delta t = (300\,\Gamma)^{-1}$. 
The particle-hole symmetric case, $E_0=-U/2$, is assumed.
$t^{-1}$, $I$, and $V$ are given in units of $\Gamma$.
}
\label{fig:iv_t_stu}
\end{center}
\end{figure}

Second, the (over-) damped oscillations following the linear rise are affected by both the interaction strength and the applied bias voltage. 
The time when a steady current is reached is of the order of $t_\text{stat} \sim \Gamma^{-1}$. 

The frequency of the oscillations depends linearly on voltage and interaction strength\citeafter{,}{Andergassen2011,nuss2013}
\begin{equation}
  \label{eq:omega_v_u}
  \omega(V,U) = ({V + U})/{2} \, .
\end{equation}
The time dependence of the electrical current can be calculated analytically at mean-field order, assuming a constant dot occupation. 
In the particle-hole symmetric case, $E_0=-U/2$, and with symmetrically adjusted chemical potentials the stationary occupation number is $n=0.5$. 
Therefore, the effective mass term $M_{\sigma}$ in Eq.~\eq{dsrhoMFleads} for the spectral function at mean-field order vanishes. 
Inserting into \Eq{electrical_current} one obtains the exact expression for the electrical current at zero $U$,
\begin{equation}
  I(t) = 4 \int_0^t \frac{{d}u}{u} \, {\text{e}^{-\Gamma u} \sin\left({V} u\right/2)} \, .
  \label{eq:ItMF}
\end{equation}
This is expected to be a good approximation at early times because of the dot initially being empty, and it provides the voltage dependence of the frequency quoted above.

We extract the frequency, from the time between maxima and minima,
for five values of the bias voltage, $V=(16\dots20)\,\Gamma$, for fixed interaction strength.
A linear fit gives
\begin{align}
    \omega(V,2\,\Gamma) &= 0.51 \, V - 0.10\,\Gamma \, ,
    \nonumber\\
    \omega(V,4\,\Gamma) &= 0.54 \, V - 0.94\,\Gamma \, .
\end{align}
Our results are in fair agreement with Eq.~(\ref{eq:omega_v_u}) concerning the voltage dependence while we cannot confirm the predicted dependence on the interaction strength.

Finally, the system reaches a steady state characterized by a current which depends on both bias voltage and interaction strength. 
We discuss the stationary current in more detail below.

%
\subsection{Stationary electric current and conductance}
%
%
\begin{figure}[t]
\begin{center}
\includegraphics[width=0.435 \textwidth]{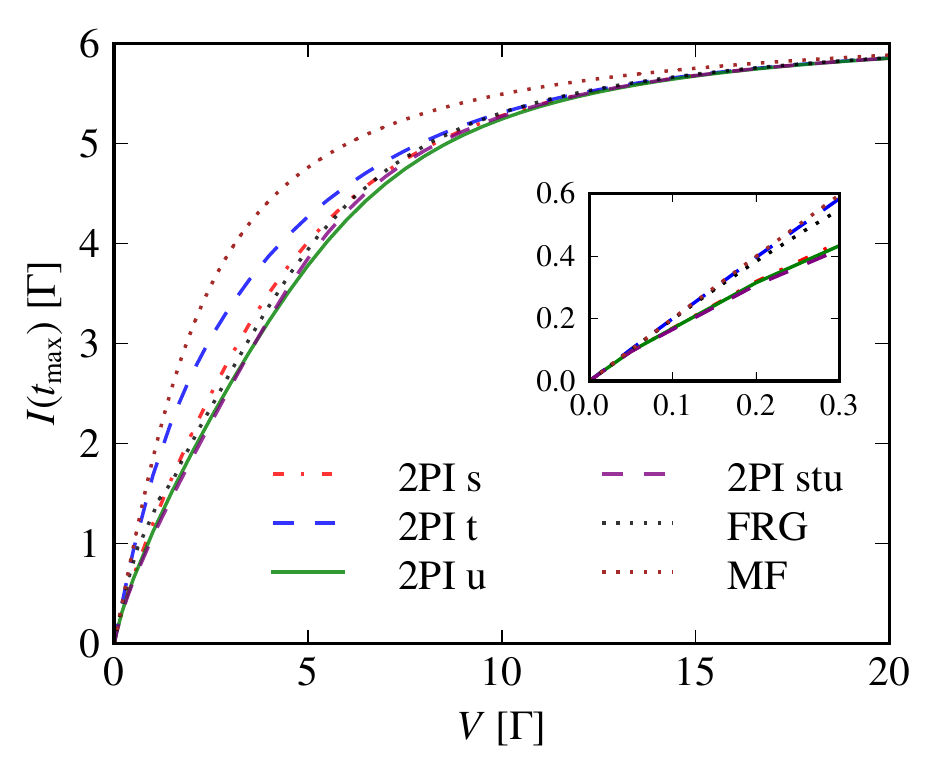}
\includegraphics[width=0.435 \textwidth]{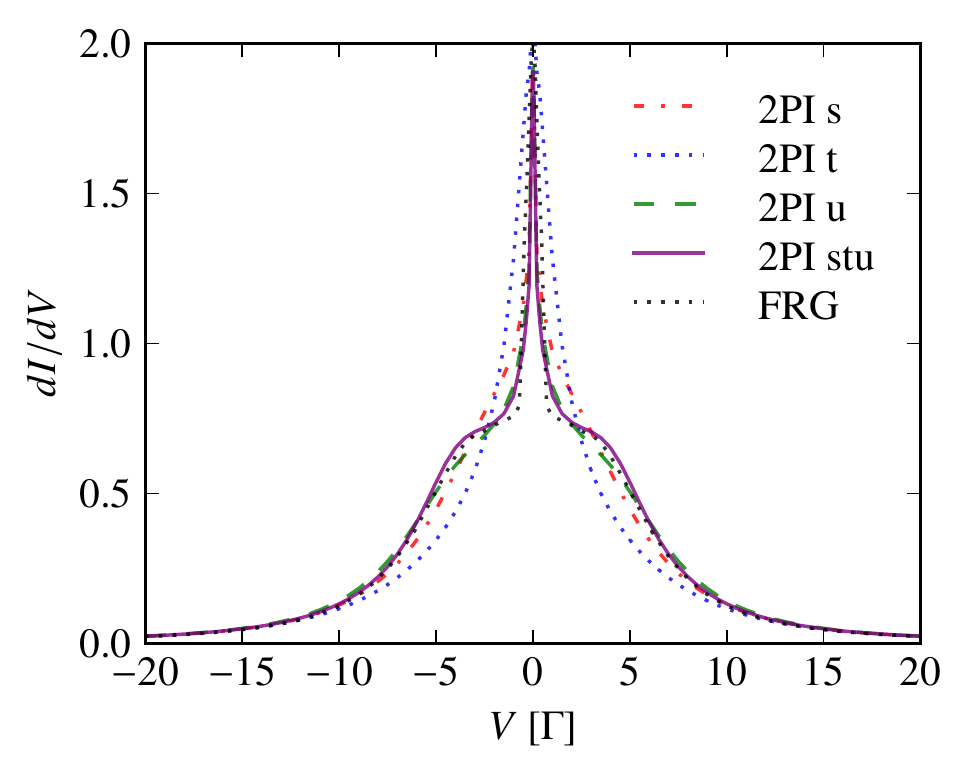}
\caption{
The stationary current $I$ through the dot (top) and the differential conductance (bottom) as functions of the bias voltage $V$, obtained in different resummation channels  in the particle-hole symmetric case, $E_0=-U/2$.
We compare our results those obtained within the functional renormalization-group  (FRG) scheme of Ref.~\onlinecite{eckel2010}.
``MF'' denotes the mean-field result, \Eq{ItMF}, for the current.
The interaction strength is $U = 4\,\Gamma$; the temperature is $T=0.1\,\Gamma$. 
The system was evolved to the total time $t_\text{max}=40\,\Gamma^{-1}$, for bias voltages $V \leq 0.5\,\Gamma$ and $t_\text{max}=6\,\Gamma^{-1}$ otherwise, with a time-step size of $\Delta t = (300\,\Gamma)^{-1}$.
For smaller $U$, the agreement between the different approximation schemes increases, see Appendix \ref{app:NumData}.
}
\label{fig:i_v_u}
\end{center}
\end{figure}
The stationary electric current is given by the Meir-Wingreen formula
\begin{equation}
  \label{eq:stationary_current}
    I = \frac{\Gamma}{2} \mathrm{Im} \sum_\sigma \int^D_{-D} {d}\epsilon\, 
    \Big[ f(\epsilon-\mu_\text{L}) - f(\epsilon-\mu_\text{R}) \Big]\, \rho_\sigma(\epsilon,T) \, ,
\end{equation}
where $D$ denotes the half-width of the band taken into account around the Fermi edge of the leads.
The derivation of \Eq{stationary_current} is given in Appendix \sect{electrical_current}.
In the top panel of Fig.~\ref{fig:i_v_u}, we show our results for the electrical current obtained in the direct ($s$), particle-particle ($t$), particle-hole ($u$), and $stu$-channel cases, for interaction strength  $U=4\,\Gamma$ and compare them with results obtained within the functional renormalization-group (FRG) scheme of Ref.~\onlinecite{eckel2010}.
In the bottom panel, we show the corresponding differential conductance, where we continue the values to negative voltages using the antisymmetry of $I$.

In Appendix \ref{sec:electrical_current}, we recall that the maximum electrical current through the quantum dot is $I_\text{max}=2\pi\,\Gamma$. 
Any truncation of the 2PI effective action conserves the value of the equal-time spectral function $-i\rho(t,t)=1$ 
and thus leads to results obeying this upper limit for the current. 
For large bias voltages, $V > U + 2\,\Gamma$, our numerical results approach the maximum current asymptotically and fit well with the FRG results for arbitrary interaction strengths.
The best agreement is found with the $stu$-channel scheme. 
Note that a significant electrical current can flow only when the bias voltage is of the order of the interaction strength, $V\approx U$, because both states in the quantum dot are then energetically accessible. 
This is an important requirement for charge transport.
However, from the numerical data, one sees that already a small bias voltage induces an electric current. 
This behavior is highly nontrivial and governed by the Kondo effect, 
which gives rise to additional states at the Fermi edges of the leads. 
As a consequence, the conductance reaches its maximum, the conductance quantum defined as $G_0=2{e^2}/{h}$
for zero bias voltage, $G_\text{diff}(0)=G_0$. 
In our data, the peak in the differential conductance at zero bias voltage reaches almost $G_\text{diff}(0)=2$.

Compared with the smaller interaction strengths (see Appendix \ref{app:NumData}), the electrical current shows more varied characteristics. 
Two side peaks are seen in the conductance at $V =\pm U$ because the two energy levels in the quantum dot are separated by the interaction strength $U$. 
When the bias voltage reaches that value, one of the two chemical potentials of the leads is located at the gate voltage and the other at the doubly occupied level, implying an enhanced differential conductance. 

%
\subsection{Voltage-induced deterioration of the Kondo peak}
%
%
\begin{figure}[t]
\begin{center}
\includegraphics[width=0.435 \textwidth]{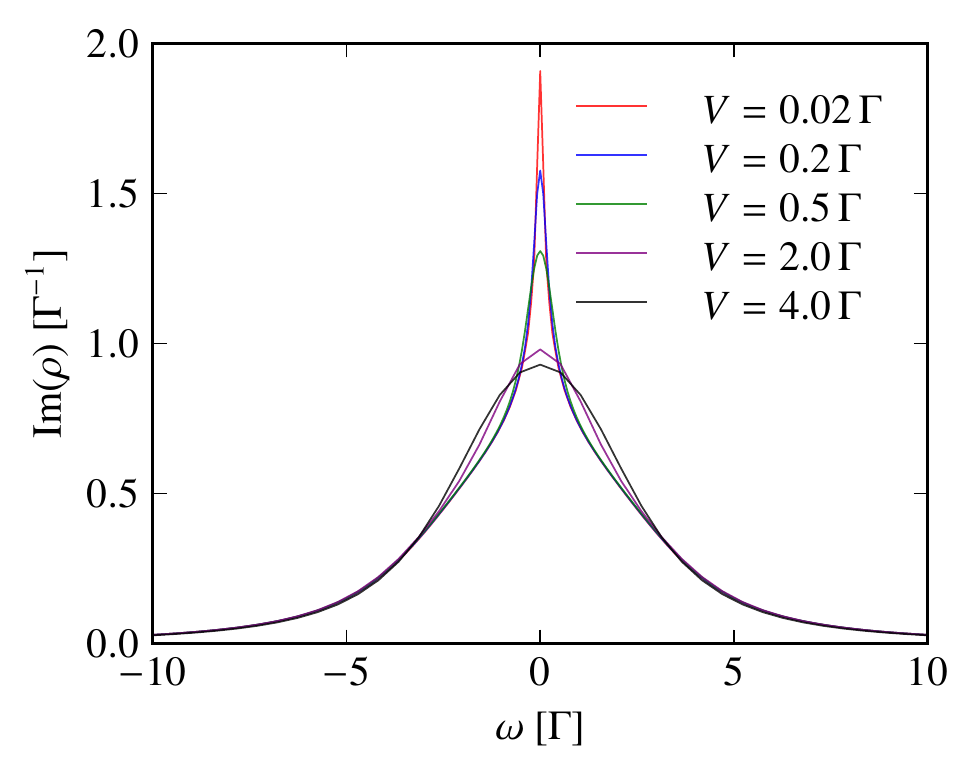}
\caption{(Color online)
Imaginary part of the frequency-dependent spectral function, $\mathrm{Im}[\rho(\omega)]$, as obtained in the  $stu$-channel, for $U=4\,\Gamma$, with $t_\mathrm{max}=40\,\Gamma^{-1}$ ($6.7\,\Gamma^{-1}$ for $V\ge\,\Gamma$) and a time-step size of $\Delta t=(300\,\Gamma)^{-1}$, for different voltages $V$ applied to the dot, in the particle-hole symmetric setup, $E_{0} = -U/2$. 
The narrow Kondo resonance decreases with increasing voltage and is strongly reduced already at $V\gtrsim T_K(U)\simeq0.3\,\Gamma$, while it entirely disappears only for $V\gtrsim\Gamma$, cf.~the results reported in Ref.~\onlinecite{muehlbacher2011}.
No satellite peaks\cite{Kaminski2000} are observed.
}
\label{fig:kondovsV}
\end{center}
\end{figure}
The strong decay of the stationary differential conductance with increasing voltage applied to the dot as shown in Fig.~\ref{fig:i_v_u} indicates that the Kondo resonance is strongly affected by the voltage.
In Fig.~\ref{fig:kondovsV} we show, for the same situation, the imaginary part of the spectral function, $\mathrm{Im}[\rho(\omega)]$, for $U=4\,\Gamma$, for different voltages between $V=0.02\,\Gamma$ and $V=4.0\,\Gamma$, in the stationary limit when the current no longer changes in time.
We find that  the narrow Kondo resonance decreases with increasing voltage and is strongly reduced already at $V\gtrsim T_K(U)\simeq0.3\,\Gamma$ while it entirely disappears only for $V\gtrsim\Gamma$. 
Hence, our results partly corroborate the conclusion drawn on the basis of diagrammatic Monte-Carlo methods applied in Ref.~\onlinecite{muehlbacher2011} that voltages on the order of $V\simeq T_K(U)$ lead to a strong deterioration of the Kondo resonance.
However, as the figure demonstrates, $\Gamma$ is found to play a role even in the presence of the Kondo effect, in addition to the scale $T_K(U)$. 
We furthermore emphasize that we do not find signs of satellite Kondo peaks arising at nonzero voltage\cite{Kaminski2000} which may be due to the relatively small values of $U/\Gamma$ considered here.

%
\subsection{Temperature dependence of the stationary current}
%
%
\begin{figure}[t]
\begin{center}
\includegraphics[width=0.435 \textwidth]{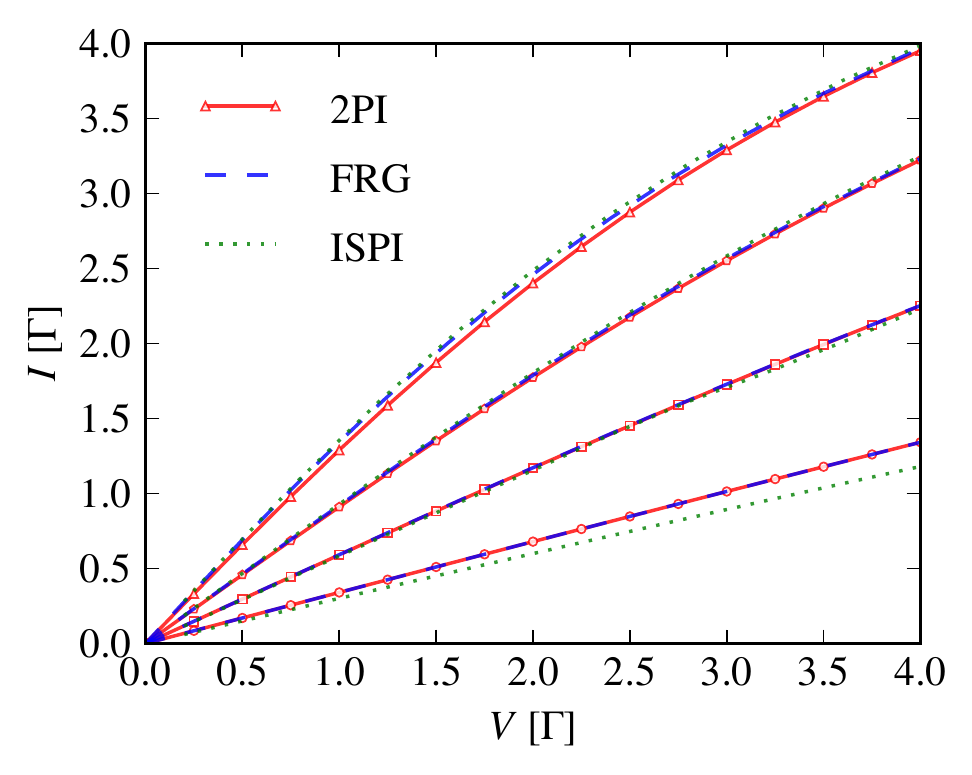}
\caption{(Color online)
Stationary current $I$ through the quantum dot coupled to leads at a temperature of $T/\Gamma=0.4$, $1.0$, $2.0$, $4.0$ (from steep to flat slopes) in the particle-hole symmetric case, $E_0=-U/2$, as functions of the applied voltage $V$, obtained in the 
$stu$-channel scheme, with an interaction strength $U=2\,\Gamma$.
We compare with the FRG and ISPI results of Ref.~\onlinecite{eckel2010}. 
The system was evolved to the total time $t_\text{max}=6\,\Gamma^{-1}$,  with a time-step size of $\Delta t = (300\,\Gamma)^{-1}$.}
\label{fig:i_v_t}
\end{center}
\end{figure}
Keeping the same physical setup as in the previous cases we repeated our computations for different temperatures, $T/\Gamma=0.4$, $1.0$, $2.0$, $4.0$. 
In Fig.~\ref{fig:i_v_t}, we show the results obtained in the 
$stu$-channel scheme for $U=2\,\Gamma$ and compare 
with the FRG and ISPI data from Ref.~\onlinecite{eckel2010}.
With increasing temperature, the electrical current decreases. 
Further data shown in Appendix \ref{app:NumData} indicates that the FRG results are in best agreement with the $s$-channel case.

%
\subsection{Stationary current and conductance in a magnetic field}
\label{sec:CurrentVoltageBfield}
%
\begin{figure}[t]
\begin{center}
\includegraphics[width=0.415 \textwidth]{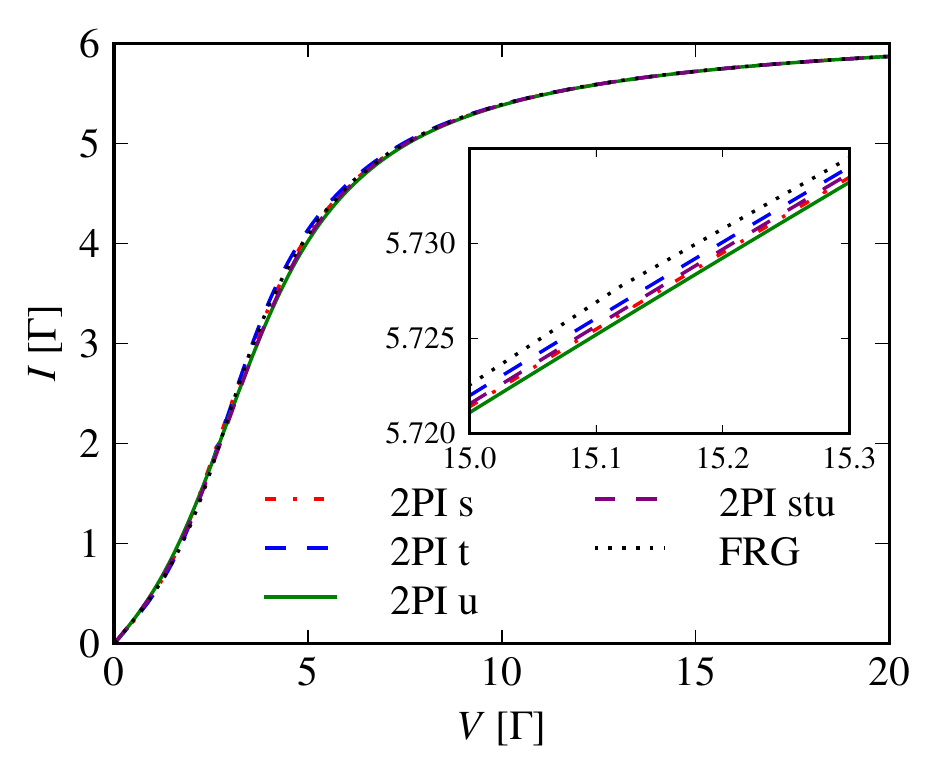}
\includegraphics[width=0.435 \textwidth]{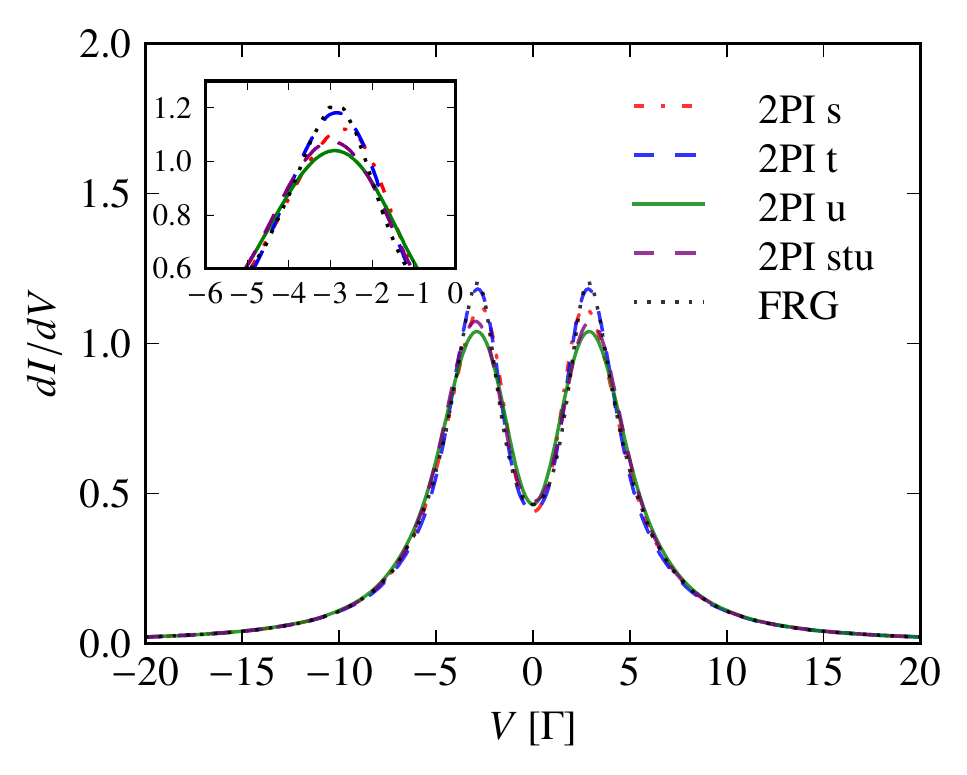}
\caption{Voltage dependence of the stationary current $I(t_\mathrm{max})$ through the dot (top panel) and of the differential conductance (bottom panel), for an interaction strength $U=2\,\Gamma$ and a temperature $T=0.1\,\Gamma$, in the particle-hole symmetric case, $E_0=-U/2$, as obtained with different resummation schemes and in the FRG scheme of Ref.~\onlinecite{eckel2010}.
An external magnetic fields $B = 1.2\,\Gamma$ is applied. 
The system was evolved to the total time $t_\text{max}=6\,\Gamma^{-1}$ with a time-step size of $\Delta t = (300\,\Gamma)^{-1}$.}
\label{fig:i_v_b}
\end{center}
\end{figure}
%
We finally study the influence of a constant external magnetic field applied to the dot on the current-voltage characteristics of the quantum dot; see Fig.~\ref{fig:i_v_b}.
For a small interaction strength $U=2\,\Gamma$ we expect, from our above findings, good agreement with FRG data and thus to be able to isolate the magnetic-field effect on the data.
Under a nonvanishing bias voltage, the Kondo resonance splits into two maxima due to the Zeeman effect. 
The arising peaks are shifted away from the Fermi energies of the two leads. 
This leads to a smaller electrical current for small bias voltages and a smaller differential conductance even for zero bias voltage. 

The results obtained within the different resummation schemes are very close to each other. 
Deviations show up in the close-up view of the split Kondo peak.
The deviation between the 2PI and the FRG results are at the maxima and minima of the differential conductance. 
However, in general, both methods show good agreement with each other.

%
%
\section{Conclusions}
\label{consect}
We have studied the real-time evolution of a quantum dot out of equilibrium, after coupling the initially empty dot to two leads at zero or nonvanishing low temperature.
The dynamics is evaluated in terms of time-dependent Green's functions solving Kadanoff-Baym equations of motion.
These dynamic equations are derived from a 2PI effective action for the Anderson impurity model derived in the framework of a Schwinger-Keldysh functional integral. 
In this way, correlations between the dot and the leads are taken into account in a nonperturbative manner, by determining the frequency-dependent self-energy of the dot electrons by means of three different resummations.
These include summations over bubble- and ladder-chain diagrams, forming the direct ($s$-channel), particle-particle ($t$), and particle-hole ($u$) channels, as well as their combination ($stu$), and are introduced through Hubbard-Stratonovich transformations.
While this approach requires a considerable numerical effort, it is nevertheless possible with state-of-the-art computers.
Considering the effective coupling, one finds that the resummation can be thought of as using a frequency-dependent four-point vertex.
The leads were taken into account by integrating them out exactly within a grand-canonical scheme. 

Our central result is the demonstration that, with our nonperturbative approach, we can describe the dynamical buildup of the Kondo effect.
This shows up, in the case that the dot is tuned to the particle-hole symmetric configuration, in the development of a sharp resonance peak in the spectral function at zero frequency.
At long evolution times, inverse proportional to the Kondo scale $T_K$, which measures the width of the peak, $T_K$ is found to be suppressed exponentially in the dot coupling $U$, in agreement with the perturbative renormalization-group prediction obtained for the simpler Kondo model.
Furthermore, a similar resonance is seen in the differential conductance near zero bias voltage, leading to stationary currents in good agreement with expected values. 
We find little quantitative differences in the results obtained within the $s$, $u$, and $stu$ schemes, while the $t$-channel scheme does not give the Kondo scaling.

%
\begin{acknowledgements}
We would like to thank A.~Komnik, J.~M.~Pawlowski and D.~Sexty for useful discussions.
We are grateful to J.~Eckel for providing the data published in Ref.~\onlinecite{eckel2010}, to A. Komnik for pointing out Ref.~\onlinecite{Nordlander1999a.PhysRevLett.83.808}, and to D.~Sexty for previous collaboration on this and for providing us with the first version of the numerical code which served as a basis for the work presented here.
We we would furthermore like to thank E.~Bittner for his support concerning computing infrastructure and M. Karl for technical help. 
We thank the anonymous referees for their valuable comments and for pointing out to us literature we had been unaware of.
The authors acknowledge the support by the Deutsche Forschungsgemeinschaft (Grant No.~GA 677/7, 8), by the Helmholtz Association (Grant No.~HA216/EMMI), by the Landesgraduiertenf\"orderung Baden-W\"urttemberg, and by the Excellence Programme FRONTIER of Heidelberg University.
\end{acknowledgements}

\begin{appendix}

\section{Numerical implementation}
\label{app:NumImp}
In this appendix we summarize the numerical methods, used to solve the equations of motion \eq{eom_diag}, which we refer to in the form
\begin{equation}
\label{eq:A1}
  \begin{split}
    \partial_{t}\rho\left(t,t^{\prime}\right)
    &=f_{1}\left(t,t^{\prime}\right),
    \\
    \partial_{t}F\left(t,t^{\prime}\right)
    &=f_{2}\left(t,t^{\prime}\right),
  \end{split}
\end{equation}
with $f_{1}$ and $f_{2}$ containing the memory integrals:
\begin{align}
\label{eq:A2}
    f_1\left(t,t^{\prime}\right)
    =&\ -i M \left(t\right) \rho\left(t,t^{\prime}\right) - i \int_{t^{\prime}}^{t} du \, \Sigma^{\rho} \left(t,u\right) \rho \left(u,t^{\prime}\right) , 
    \\
    f_2 \left(t,t^{\prime}\right) 
    =&\ -i M \left(t\right) F\left(t,t^{\prime}\right)   - i \int_{0}^{t}du\, \Sigma^{\rho}\left(t,u\right)F\left(u,t^{\prime}\right)              
    \nonumber\\
       &+\ i \int_{0}^{t^\prime} du \, \Sigma^{F} \left(t,u\right) \rho \left(u,t^{\prime} \right).
\end{align}
\Eq{A1} forms a system of partial integro-differential equations which we solve by means of multistep methods which contain information from the past time evolution. 
In each time step we use a third-order Adams--Bashforth predictor-corrector method which makes use of the values at the three previous points in time. 
For example, the discretization of the spectral function is given by
\begin{align}
\label{eq:Adams-Bashforth}
    \rho\left(t_{k+1},t_j\right) 
    =&\ \rho\left(t_k,t_j\right) + \frac{\Delta t}{24} \, \Big[ 55 f_1 \left(t_k,t_j\right) - 59 f_1 \left(t_{k-1},t_j\right) 
    \nonumber\\
     &+\ 37 f_1 \left(t_{k-2},t_j\right) - 9 f_1 \left(t_{k-3},t_j\right) \Big],
\end{align}
with $t_{k}=k\Delta t$ and $\Delta t$ being the step size. 
After the point $\rho\left(t_{k+1},t_{j}\right)$ has been calculated, we implement an Adams--Moulton step as a corrector, i.e., we use the calculated data point and and determine it again by means of
\begin{align}
\label{eq:Adams-Moulton}
  \rho\left(t_{k+1},t_{j}\right) 
  =& \rho\left(t_{k},t_{j}\right) + \frac{\Delta t}{24}\,  \Big[ 9 f_1 \left(t_{k+1}, t_j\right) + 19 f_1 \left( t_k,t_j \right) 
  \nonumber\\
   &-\ 5 f_1 \left(t_{k-1}, t_j \right) + f_1 \left(t_{k-2}, t_j \right) \Big].
\end{align}
When solving the 2PI equations, we set $\rho(t,t)=i$ explicitly, as prescribed by the anticommutation relations, and set the initial occupation at $n_{0}$. 
The memory integrals are in Volterra form, i.e., for a particular time $t$ they require only field values in the past which have been computed already before. 
Therefore, our  scheme allows an explicit calculation of each time step from known quantities. 
The integrations are performed by means of a seventh-order closed Newton--Cotes formula. 
Our typical choice of the step size is within the range of $\Delta t\in\left[0.0025,0.01\right]$.

The numerical effort of the presented method was as follows.
Choosing a maximum number of time steps $N=t_\mathrm{max}/\Delta t$, the correlation functions $F$, $\rho$, $\Sigma$, etc., each require memory space of $N^{2}$ real or complex numbers with double precision.
Since performing a single memory integral involves $N$ complex multiplications, the total computing time scales as $N^{3}$.
Our computations were performed on machines equipped with AMD\textsuperscript{\textregistered} Opteron\textsuperscript{\textregistered} processors 6282 (16 cores) and 256GB RAM, using a maximum of $N=16000$ time steps, each run requiring between a few days and several weeks.

\section{The auxiliary-field propagator $G$}
\subsection{Decomposition of $G$ into statistical and spectral parts}
\label{app:DecompG}
%
\begin{table*}[t]
\caption{Decomposition of the fermionic and bosonic propagators into statistical and spectral parts for the different resummation schemes.}
\centering
\newcolumntype{C}{>{\setlength\hsize{1\hsize}\centering\arraybackslash}X} %
\begin{tabularx}{\textwidth}{|C|C|C|}
\hline
&&\\[-1.8ex]
 $s$-channel & $t$-channel & $u$-channel\\[0.5ex]
\hline
&&\\[-1.5ex]
\(\begin{array}{rcl}
   \Sigma^{\rho}_{\sigma}&=&F_{\sigma}\bar{G}_{\sigma\sigma}^{\rho}+\rho_{\sigma}\bar{G}_{\sigma\sigma}^{\text{F}}    
   \\
   \Sigma^{\text{F}}_{\sigma}&=&F_{\sigma}\bar{G}_{\sigma\sigma}^{\text{F}}-\rho_{\sigma}\bar{G}_{\sigma\sigma}^{\rho}/4
   \\[1ex]
   \Pi_{\sigma\sigma}^{\rho}&=&-2 \textrm{Re} \left(\rho_{\sigma}F_{\sigma}^{*}\right)
   \\
   \Pi_{\sigma\sigma}^{\text{F}}&=&-\left|F_{\sigma}\right|^{2}+\left|\rho_{\sigma}\right|^{2}/4 
\end{array}\)
&
\(\begin{array}{rcl}
  \Sigma^{\rho}_{\sigma}&=&\rho_{\bar{\sigma}}^{*}\bar{G}_{\sigma\bar{\sigma}}^{*\text{F}}+F_{\bar{\sigma}}^{*}\bar{G}_{\sigma\bar{\sigma}}^{*\rho}
  \\
  \Sigma^{\text{F}}_{\sigma}&=&F_{\bar{\sigma}}^{*}\bar{G}_{\sigma\bar{\sigma}}^{*\text{F}}-\rho_{\bar{\sigma}}^{*}\bar{G}_{\sigma\bar{\sigma}}^{*\rho}/4
   \\[1ex]
   \Pi_{\sigma\bar{\sigma}}^{\rho}&=&-\Big(F_{\sigma}^{*}\rho_{\bar{\sigma}}^{*}+\rho_{\sigma}^{*}F_{\bar{\sigma}}^{*}\Big)
   \\
   \Pi_{\sigma\bar{\sigma}}^{\text{F}}&=&\rho_{\sigma}^{*}\rho_{\bar{\sigma}}^{*}/4-F_{\sigma}^{*}F_{\bar{\sigma}}^{*}
\end{array}\)
&
\(\begin{array}{rcl}
  \Sigma^{\rho}_{\sigma}&=&\rho_{\bar{\sigma}}\bar{G}_{\bar{\sigma}\sigma}^{*\text{F}}+F_{\bar{\sigma}}\bar{G}_{\bar{\sigma}\sigma}^{*\rho}
  \\
  \Sigma^{\text{F}}_{\sigma}&=&F_{\bar{\sigma}}\bar{G}_{\bar{\sigma}\sigma}^{*\text{F}}-\rho_{\bar{\sigma}}\bar{G}_{\bar{\sigma}\sigma}^{*\rho}/4
   \\[1ex]
   \Pi_{\sigma\bar{\sigma}}^{\rho}&=&-\left(\rho_{\sigma}F_{\bar{\sigma}}^{*}+F_{\sigma}\rho_{\bar{\sigma}}^{*}\right)
   \\
   \Pi_{\sigma\bar{\sigma}}^{\text{F}}&=&\rho_{\sigma}\rho_{\bar{\sigma}}^{*}/4-F_{\sigma}F_{\bar{\sigma}}^{*}
\end{array}\)
\\
&&\\[-1.5ex]
\hline
&&\\[-1.5ex]
\(\begin{array}{rcl}
  \Pi_{11}\left(t,t^{\prime}\right)&=&\Pi_{22}\left(t,t^{\prime}\right) 
  \\
  \Pi_{12}\left(t,t^{\prime}\right)&=&\Pi_{21}\left(t,t^{\prime}\right)=0
  \\[1ex]
  \bar{G}_{21}\left(t,t^{\prime}\right)&=&\bar{G}_{12}\left(t,t^{\prime}\right)
  \\
\bar{G}_{11}\left(t,t^{\prime}\right)&=&\bar{G}_{22}\left(t,t^{\prime}\right)
\end{array}\)
&
\(\begin{array}{rcl}
  \Pi_{12}\left(t,t^{\prime}\right)&=&\Pi_{21}\left(t,t^{\prime}\right)
  \\
  \Pi_{11}\left(t,t^{\prime}\right)&=&\Pi_{22}\left(t,t^{\prime}\right)=0
  \\[1ex]
  \bar{G}_{21}\left(t,t^{\prime}\right)&=&\bar{G}_{12}\left(t,t^{\prime}\right)  
  \\
  \bar{G}_{11}\left(t,t^{\prime}\right)&=&\bar{G}_{22}\left(t,t^{\prime}\right)=0
\end{array}\)
&
\(\begin{array}{rcl}
  \Pi_{12}\left(t,t^{\prime}\right)&=&\Pi_{21}\left(t^{\prime},t\right)
  \\
  \Pi_{11}\left(t,t^{\prime}\right)&=&\Pi_{22}\left(t,t^{\prime}\right)=0
  \\[1ex]
  \bar{G}_{21}\left(t,t^{\prime}\right)&=&\bar{G}_{12}\left(t^{\prime},t\right)
  \\
  \bar{G}_{11}\left(t,t^{\prime}\right)&=&\bar{G}_{22}\left(t,t^{\prime}\right)=0
\end{array}\)
\\
&&\\[-1.5ex]
\hline
&&\\[-1.5ex]
\(\begin{array}{rcl}
  \bar{G}_{22}^{\rho}&=&U\left(\Pi_{11}^{\rho}*\bar{G}_{12}^{\rho}-U\,\Pi_{11}^{\rho}\right)
  \\
  \bar{G}_{12}^{\rho}&=&U\left(\Pi_{11}^{\rho}*\bar{G}_{22}^{\rho}\right)
\end{array}\)
&
\(\begin{array}{rcl}
  \bar{G}_{11}^{\rho}&=&\bar{G}_{22}^{\rho}=0
  \\
  \bar{G}_{12}^{\rho}&=&U\left(\Pi_{21}^{\rho}*\bar{G}_{12}^{\rho}-U\,\Pi_{21}^{\rho}\right)
\end{array}\)
&
\(\begin{array}{rcl}
  \bar{G}_{11}^{\rho}&=&\bar{G}_{22}^{\rho}=0
  \\
  \bar{G}_{12}^{\rho}&=&U\left(\Pi_{21}^{\rho}*\bar{G}_{12}^{\rho}-U\,\Pi_{21}^{\rho}\right)
\end{array}\)
\\
&&\\[-1.5ex]
\hline
&&\\[-1.5ex]
\(\begin{array}{rcl}
  \bar{G}_{22}^{\text{F}}&=&U\left(\Pi_{11}^{\rho}*\bar{G}_{12}^{\text{F}}-\Pi_{11}^{\text{F}}*\bar{G}_{12}^{\rho}-U\,\Pi_{11}^{\text{F}}\right)
  \\
  \bar{G}_{12}^{\text{F}}&=&U\left(\Pi_{11}^{\rho}*\bar{G}_{22}^{\text{F}}\right)-U\left(\Pi_{11}^{\text{F}}*\bar{G}_{22}^{\rho}\right)
\end{array}\)
&
\(\begin{array}{rcl}
  \bar{G}_{11}^{\text{F}}&=&\bar{G}_{22}^{\text{F}}=0
  \\
  \bar{G}_{12}^{\text{F}}&=&U\left(\Pi_{21}^{\rho}*\bar{G}_{12}^{\text{F}}-\Pi_{21}^{\text{F}}*\bar{G}_{12}^{\rho}-U\,\Pi_{21}^{\text{F}}\right)
\end{array}\)
&
\(\begin{array}{rcl}
  \bar{G}_{11}^{\text{F}}&=&\bar{G}_{22}^{\text{F}}=0
  \\
  \bar{G}_{12}^{\text{F}}&=&U\left(\Pi_{21}^{\rho}*\bar{G}_{12}^{\text{F}}-\Pi_{21}^{\text{F}}*\bar{G}_{12}^{\rho}-U\,\Pi_{21}^{\text{F}}\right)
\end{array}\)
\\[-1.5ex]
&&\\
\hline
\end{tabularx}
\label{table1}
\end{table*}
%
%
According to Eqs.~\eq{G0inv} and \eq{constrainteq}, the propagator $ G (t,t')  $ of the bosonic auxiliary field can be written as $  G (t,t')= \bar G (t,t') + i U \sigma_1  \delta (t-t') $, where $\sigma_1$ is the first Pauli matrix.
The equation for $\bar G$ reads
\bea 
\label{eq:sk_eom}
 \left( \begin{array}{cc}
   \bar G_{21}  & \bar G_{22} \\
   \bar G_{11}  & \bar G_{12} \\
 \end{array} \right) 
 = i U \Pi  * \bar G - U^2 
 \left( \begin{array}{cc}
   \Pi_{12}  & \Pi_{11} \\
   \Pi_{22}  & \Pi_{21} \\
\end{array} \right).
\eea
As defined in \Eq{FrhoDecomposition} for fermions, the scalar propagator is decomposed in the same way into a statistical and a spectral part:
\begin{equation}
  \bar G_{\sigma\tau}(t,t') = \bar G^F_{\sigma\tau}(t,t') - \frac{i}{2} \text{sgn}_C(t-t') \bar G^\rho_{\sigma\tau}(t,t'),
\end{equation}
The scalar constraint equation \eq{sk_eom} is then decomposed by means of
\begin{equation} 
  \begin{split}
    (i A * B) (t,t') =& \int_0^{t} dz\, A^\rho (t,z) B^F (z,t') \\
                     -& \int_0^{t'} dz\, A^F(t,z) B^\rho(z,t') \\
                     -&  \frac{i}{2} \text{sgn}_C(t-t') \int_{t'}^t dz\, A^\rho(t,z) B^\rho(z,t').
  \end{split}
\end{equation}
Finally, the fermionic and bosonic self-energies are decomposed as follows. 
\bea 
\label{srsigma}
\Sigma^F_{\sigma} 
&=& -1 ( F_\sigma \bar G^{F}_{\sigma\sigma} - {1\over 4 } \rho_\sigma \bar G^{\rho}_{\sigma\sigma} ),
\nonumber\\
\Sigma^\rho_{\sigma}
&=& -1 ( \rho_\sigma \bar G^{F}_{\sigma\sigma} + F_\sigma \bar G^{\rho}_{\sigma\sigma} ),
\nonumber\\
\Pi^F_{\sigma} 
&=&  |F_\sigma|^2 - {1\over 4} | \rho_\sigma| ^2 ,
\nonumber\\
\Pi^\rho_{\sigma} 
&=&  2 \textrm{Re} (F^*_\sigma \rho_\sigma ) .
\eea
The resulting relations are shown in Table \ref{table1}.

\subsection{Resummed effective coupling}
\label{app:EffU}
In the following we provide the details of the calculation of the effective couplings stated in \Sect{EffU}. 
For vanishing magnetic field, where $E_{0\uparrow} = E_{0\downarrow} $, one can use a symmetric initial condition to save numerical resources. 
The respective symmetries apply to the action truncated at second order in $U$ as well as to the resummed actions. 
We discuss the resummations used in this work in the following. 
Starting from an initial condition, where the following equations are satisfied for $t=t'=0$, they hold for any $ t , t' > 0 $.  
\begin{equation}
\label{eq:app:symmD}
  D_\uparrow (t,t') = D_\downarrow (t,t').
\end{equation}
In $s$-channel resummation we have
\begin{eqnarray} 
\label{eq:s-symm_eq}
  \Pi_{11}(t,t') = \Pi_{22} (t,t') , &\quad&
  \Pi_{12}(t,t') = \Pi_{21} (t,t') =0 , \nonumber \\
  \bar G_{21}(t,t')= \bar G_{12}(t,t') , &\quad& 
  \bar G_{11}(t,t') = \bar G_{22}(t,t'),
\end{eqnarray}
in $t$-channel resummation
\begin{eqnarray} 
\label{eq:t-symm_eq}
  \Pi_{12}(t,t') = \Pi_{21} (t,t') , &\quad& 
  \Pi_{11}(t,t') = \Pi_{22} (t,t') =0 , \nonumber\\
  \bar G_{12}(t,t')= \bar G_{21}(t,t') ,   &\quad& 
  \bar G_{11}(t,t') = \bar G_{22}(t,t')=0,
\end{eqnarray}
and in the $u$-channel
\begin{eqnarray}
\label{eq:u-symm_eq}
  \Pi_{12}(t,t') = \Pi_{21} (t',t) ,  &\quad& 
  \Pi_{11}(t,t') = \Pi_{22} (t,t') = 0 ,  \nonumber\\
  \bar G_{12}(t,t') = \bar G_{21}(t',t) ,   &\quad& 
  \bar G_{11}(t,t') = \bar G_{22}(t,t') = 0.
\end{eqnarray}
In the stationary limit, when transient effects have died out, all two-point functions depend only on the difference of the time coordinates.
In this limit, we can push the initial time to negative infinity, and we can use the following decomposition identities for a convolution of two correlators on the Schwinger-Keldysh contour starting at $t_0=- \infty $,
\begin{equation} 
\label{eq:decompid}
  \begin{split}
    i ( X * Y )^F    &= X^R * Y^F - X^F * Y^A , \\
    i ( X * Y )^\rho &= X^R * Y^\rho - X^\rho * Y^A , \\
    i ( X * Y )^R    &= X^R *Y^R , \\
    i ( X * Y )^A    &= -X^A *Y^A ,
  \end{split}
\end{equation}
where the retarded and advanced two-point functions are defined as
\begin{equation}
  \begin{split}
    G^R(t,t') &= \theta (t-t') G^\rho(t,t'),  \\
    G^A(t,t') &= \theta (t'\!-t) G^\rho(t,t') 
  \end{split}
\end{equation}
such that $G^{\rho}=G^{R}+G^{A}$.

\subsubsection{$s$-channel resummation}
Given the symmetries of Eqs.~\eq{s-symm_eq} one can define
\begin{eqnarray}
  A(t,t') &=& U^{-1} \bar G_{11} (t,t'), \quad 
  B(t,t') = U^{-1} \bar G_{21} (t,t'), \nonumber \\ 
  \Pi(t,t') &=& U \Pi_{11} (t,t'). 
\end{eqnarray}
Hence, the constraint equations \eq{constrainteq} in the form of \eq{sk_eom} can be written as (suppressing time arguments)
\begin{equation} 
\label{eq:s-trinvresum} 
  A =  i \Pi * B - \Pi, \qquad 
  B =  i \Pi * A .
\end{equation}
Inserting the decompositions \eq{decompid} into \eq{s-trinvresum}, after eliminating $B$ and Fourier transforming with respect to $t-t'$, we obtain
\begin{equation} 
  \begin{split}
    A^F     &= \Pi^R \,\Pi^R A^F - \Pi^R\, \Pi^F A^A - \Pi^F + \Pi^F \,\Pi^A A^A ,  \\ 
    A^\rho &= \Pi^R \,\Pi^R A^\rho - \Pi^R\, \Pi^\rho A^A - \Pi^\rho + \Pi^\rho\, \Pi^A A^A, 
  \end{split}
\end{equation}
where we have omitted the arguments $\omega$. 
With the help of the relations \eq{decompid}, \Eq{s-trinvresum} defining the advanced function $A^{A}$, multiplied by $\Pi^{R}-\Pi^{A}$, can be rewritten as
\begin{equation}
  \left( 1 + \Pi^R A^A - \Pi^A A^A \right) \left( \Pi^A \Pi^A -1 \right) = \Pi^R \Pi^A -1 .
\end{equation}
Defining $A$ in terms of an effective coupling $U_{\mathrm{eff},\xi}$,
\begin{equation} 
  UA^F = - \Pi^F U_{\mathrm{eff},\xi}, \qquad 
  UA^\rho = - \Pi^\rho U_{\mathrm{eff},\xi}, 
\end{equation}
($\xi=s,t,u$) we find the expression given in \Eq{Ueffs},
\begin{equation}
  \frac{U_{\mathrm{eff},s}}{U} = \frac{1- \Pi^R \Pi^A }{[ (\Pi^R)^2-1 )][ (\Pi^A )^2 -1 ]}  = \frac{1+ |\Pi^R |^2}{| 1-(\Pi^R)^2 |^2 }
\end{equation}
where we used $\Pi^{R*}(\omega)=-\Pi^{A}(\omega)$.
%

\subsubsection{$t$-channel resummation}
Given the symmetries of Eqs.~\eq{t-symm_eq} we define
\begin{equation} 
  A(t,t')  =  U^{-1} \bar G_{12} (t,t'), \qquad
  \Pi (t,t') = U \Pi_{12} (t,t').
\end{equation}
\Eq{sk_eom} determines
\begin{equation}
\label{eq:t-trinvresum} 
  A =  i \Pi * A - \Pi,  
\end{equation}
Using Eqs.~\eq{decompid} and Fourier transforming this gives
\begin{equation}
  \begin{split}
    A^F    &= \Pi^R A^F  - \Pi^F A^A - \Pi^F  , \\ 
    A^\rho &= \Pi^R A^\rho  - \Pi^\rho A^A - \Pi^\rho .
  \end{split}
\end{equation}
With the help of the relations \eq{decompid}, \Eq{t-trinvresum} for $A^{A}$ reads
\begin{equation}
  \left( 1 + \Pi^A  \right)A^A = -\Pi^A .
\end{equation}
Hence, the $t$-channel effective coupling results as
\begin{equation}
  \frac{U_{\mathrm{eff},t}}{U}  = \frac{1}{(1-\Pi^R)(1+\Pi^{A}) } = \frac{1}{|1-\Pi^R|^2 }
\end{equation}
where we used again that $\Pi^{R*}(\omega) = -\Pi^{A}(\omega)$.
%
\subsubsection{$u$-channel resummation}
Given the symmetries of Eqs.~\eq{u-symm_eq} we define
\begin{equation} 
  A(t,t') =  U^{-1} \bar G_{12} (t,t'), \qquad 
  \Pi (t,t') = U \Pi_{21} (t,t').
\end{equation}
With this the derivation of the effective coupling is analogous to the $t$-channel case, giving the same formal expression
\begin{equation}
  \frac{U_{\mathrm{eff},u}}{U} = \frac{1}{|1-\Pi^R|^2 }.
\end{equation}
%

\section{Spectral function in the stationary limit}
\label{sec:KineticTheory}
In this appendix we derive an expression for the spectral function $\rho(\omega)$ in the long-time limit when transient effects from the initial state have died out. 
This expression is valid as long as changes in the system parameters occur very slowly.
The spectral representation of the stationary propagator which, after initial-state effects have damped out, no longer depends on the central time $\mathcal{T}$ reads
\begin{equation}\label{eq:fullprop}
  \begin{split}
    \lim_{\epsilon\rightarrow0}D\left(\omega\pm i\epsilon\right)&=\lim_{\epsilon\rightarrow 0}\int_{-\infty}^{\infty}\frac{d\omega^{\prime}}{2\pi}\frac{-i\rho\left(\omega^{\prime}\right)}{\omega\pm i\epsilon-\omega^{\prime}}\\
    &=-\mathcal{P}\int_{-\infty}^{\infty}\frac{d\omega^{\prime}}{2\pi}\frac{i\rho\left(\omega^{\prime}\right)}{\omega-\omega^{\prime}}\mp \frac{1}{2}\rho\left(\omega\right),
  \end{split}
\end{equation}
where $\mathcal{P}$ denotes the Cauchy principal value.
Note that due to the antisymmetry in the time domain the spectral function is purely imaginary in Fourier space.
An equivalent relation holds for the self-energy and, hence,
\begin{equation}\label{eq:relation1}
  \begin{split}
    \lim_{\epsilon\rightarrow 0}D\left(\omega\pm i\epsilon\right)&=\textrm{Re} D\left(\omega\right)\mp\frac{1}{2}\rho\left(\omega\right),\\
    \lim_{\epsilon\rightarrow 0}\Sigma\left(\omega\pm i\epsilon\right)&=\textrm{Re} \Sigma\left(\omega\right)\mp\frac{1}{2}\Sigma^{\rho}\left(\omega\right).\\
  \end{split}
\end{equation}
Using the Dyson equation 
\begin{equation}\label{eq:Dysonkin}
  D^{-1}\left(\omega\pm i\epsilon\right)=D_{0}^{-1}\left(\omega\pm i \epsilon\right)-\Sigma\left(\omega\pm i\epsilon\right)=\omega\pm i\epsilon-\Sigma\left(\omega\pm i\epsilon\right),
\end{equation}
the real part of the full propagator results as 
\begin{equation}\label{eq:realpartfullprop}
  \begin{split}
    \textrm{Re} \Big(D\left(\omega\pm i\epsilon\right)\Big)
    &= \textrm{Re} \left(\frac{1}{\omega \pm i\epsilon- \textrm{Re} \Sigma\left(\omega \right)\pm {\Sigma^{\rho}\left(\omega\right)}/{2}}\right)\\
    &=\frac{\omega- \textrm{Re} \Sigma}{\left(\omega- \textrm{Re} \Sigma\right)^{2}+\left|\Sigma^{\rho}/2\right|^{2}}.
  \end{split}
\end{equation}
Combining Eqs.~\eq{relation1}, \eq{Dysonkin}, and \eq{realpartfullprop} we obtain the kinetic expression for the spectral function, Eqs.~\eq{spectralfunckin} and \eq{realsigma},
\begin{equation}\label{eq:app:spectralfunckin}
  \begin{split}
    \rho\left(\omega\right)&=\pm 2\,\Big( \textrm{Re} D\left(\omega\right)-D\left(\omega\pm i \epsilon\right)\Big)\\
    &=\frac{\Sigma^{\rho}}{\left(\omega- \textrm{Re} \Sigma\right)^{2}+\left|{\Sigma^{\rho}}/{2}\right|^{2}},
  \end{split}
\end{equation}
with
\begin{equation}\label{eq:app:realsigma}
 \textrm{Re} \Sigma\left(\omega\right)=-\mathcal{P}\int_{-\infty}^{\infty}\frac{d\omega^{\prime}}{2\pi}\frac{\Sigma^{\rho}\left(\omega^{\prime}\right)}{\omega-\omega^{\prime}}.
\end{equation}
Exemplary graphical representations of $\Sigma^{\rho}$ and $\textrm{Re} \Sigma$ are given in Fig.~\ref{fig:SelfEnergies} in Appendix \ref{app:NumData}.

\section{Electrical current and conductance}
\label{sec:electrical_current}
In this appendix we describe the derivation of the expressions \eq{electrical_current}  and \eq{stationary_current} for the transient and stationary currents, respectively. 
The contribution of one lead to the current through the quantum dot is (keeping track of constants $e$, $h$)
\begin{equation}
  \label{eq:def_current}
  I_\ell(t) = - e \dot{N}_\ell (t) \, , 
  \qquad 
  N_\ell (t) = \sum_{k\sigma} \big\langle c^\dagger_{k\ell\sigma}(t) c_{k\ell\sigma}(t) \big\rangle,
\end{equation}
and the total current is given by the difference 
\begin{equation}
  I = (I_\text{L} - I_\text{R})/{2} \, .
\end{equation}
Only the tunneling part of the Hamiltonian has a nonvanishing commutator with the occupation number $N_{\ell}$, 
\begin{align}
    I(t)  &= -\frac{ie}{2\hbar} \big[ H_\textrm{tunnel}, N_\text{L} - N_\text{R}  \big] 
    \nonumber\\
    &= -\frac{i e}{2\hbar} \sum_{\ell k\sigma} \ell \left(t_\ell  \big\langle c^\dagger_{\ell k\sigma} d_\sigma \big\rangle - t^\ast_\ell  \big\langle d^\dagger_\sigma c_{\ell k\sigma} \big\rangle \right)(t).
\end{align}
To determine this expression within the functional-integral formalism it is convenient to write it as the derivative of the generating functional with respect to a source,
\begin{equation}
  I(t) = -i \frac{\delta}{\delta \eta(t)} \ln Z[\eta] \Big{\vert}_{\eta\equiv0} 
  = \Big\langle \frac{\delta S[\eta]}{\delta \eta(t)}\Big|_{\eta\equiv0} \Big\rangle \, .
  \label{eq:IfromS}
\end{equation}
The required source term to be added to the action reads
\begin{equation}
  S[\eta] = -\frac{ie}{2\hbar}\int_{\mathcal{C}} dt' \sum_{\ell k\sigma} \ell \left( t_\ell  c^\dagger_{\ell k\sigma} d_\sigma - t^\ast_\ell  d^\dagger_\sigma c_{\ell k\sigma} \right)\eta \, .
\end{equation}
One now proceeds in the same way as when integrating out the quadratic leads contribution.
Shifting the tunneling parameter to
\begin{equation}
  \tilde{\tau} = \tau\, \Big[ 1 + \frac{i e \eta}{2\hbar} \ell \delta(t-t') \Big] \, ,
\end{equation}
and completing the squares one obtains the quadratic leads part
\begin{align}
    S_\text{leads} 
    &= -|\tau|^2 \sum_{\ell k\sigma} \int_\mathcal{C} {d}t \, \int_\mathcal{C} 
    {d}t' \, d_\sigma^\dagger(t)\, \Big( A_{\ell k\sigma}(t,t')  
    \nonumber\\
    &+\ \frac{i e}{2\hbar} \ell A_{\ell k\sigma}(t,t') \big[ \eta(t') -\eta(t) \big] \Big)\, d_\sigma(t') \, ,
\end{align}
with $A_{\ell k\sigma}$ denoting the free propagator of the lead electrons defined by $(i\partial_t - \epsilon_{k\ell}(t)) A_{\ell k\sigma}(t,t') = \delta_\mathcal{C}(t,t')$, and where we have dropped the term proportional to $\eta^2$, which does not contribute to the expectation value.
Inserting this into \Eq{IfromS}, we obtain the expression
\begin{align}
  \label{eq:eletrical_current}
          I (t) 
          = -\frac{i e}{2\hbar} |\tau|^2 \sum_{\ell  k\sigma} \ell\, \Big[ 
          &\int_\mathcal{C} {d}t' D_\sigma(t',t) A_{\ell k\sigma}(t,t') 
          \nonumber\\
          + &\int_\mathcal{C} {d}t' D_\sigma(t,t') A_{\ell k\sigma}(t',t) \Big] \, .
\end{align}
Decomposing the product of the dot-electron and lead-electron propagators into statistical and spectral components we find that the former vanishes, leaving 
\begin{align}
    \Big[D_\sigma(t',t) A_{\ell k\sigma}(t,t')\Big]^\rho 
    =&\ \rho_\sigma(t',t) A_{\ell k\sigma}^\text{F}(t,t') 
    \nonumber\\
    &-\ F_\sigma(t',t) A_{\ell k\sigma}^\rho(t,t') \, .
\end{align}
Inserting this into \Eq{eletrical_current} and using symmetry relations to order the time arguments we obtain
\begin{align}
  \label{eq:electrical_current2}
    I(t) = -\frac{e}{2\hbar}& |\tau|^2 \sum_{\ell k\sigma} \int^{t}_0 {d}t' \, \ell \, 
    \nonumber\\
    \times\ \Big[ 
     &A^\text{F}_{\ell k\sigma}(t,t') \rho_\sigma(t',t) 
     - A^\rho_{\ell k\sigma}(t,t') F_\sigma(t',t) 
     \nonumber\\
     +\  &A^{\text{F} \ast}_{\ell k\sigma}(t,t') \rho^\ast_\sigma(t',t) 
     - A^{\rho \ast}_{\ell k\sigma}(t,t') F^\ast_\sigma(t',t) \Big] \, .
\end{align}
The terms containing the statistical propagator $F$ cancel because the spectral part of $A$ does not contain any information about the thermodynamic properties of the leads and thus $A^\rho_\text{L} = A^\rho_\text{R}$. 
Using $|\tau|^2 = \Gamma/(2\pi)$  we obtain
\begin{align}
  \label{eq:electrical_current_A}
    I(t) = -\frac{e \Gamma}{h} \mathrm{Re} \sum_\sigma \int^{t}_0 {d}t' \, 
    \Big[ A^\text{F}_\text{L}(t,t') - A^\text{F}_\text{R}(t,t') \Big]\, \rho_\sigma(t',t) \, ,
\end{align}
and inserting $A_{\ell k\sigma}^\text{F}(t,t') = -i [{1}/{2} - f(\epsilon_{k\ell} - \mu_\ell) ] \exp\{-i \epsilon_{k\ell}(t-t')\}$ for the leads' statistical function leads to \Eq{electrical_current}.

The imaginary part of the  integrand of \Eq{electrical_current} is symmetric under $t'\to-t'$.
Hence,  extending the time integral to negative times and multiplying by $1/2$, in the infinite-time limit gives the stationary current
\begin{align}
  \label{eq:stationary_current_A}
    I(t\to\infty) =
    &\ \frac{e \Gamma}{2 h} \mathrm{Im} \sum_\sigma \int^D_{-D} {d}\epsilon \int^\infty_{-\infty} {d}s\,   
    \rho_\sigma(s,T)
    \nonumber\\
    &\times\ \Big[ f(\epsilon-\mu_\text{L}) - f(\epsilon-\mu_\text{R}) \Big]\,\text{e}^{-i \epsilon s} ,
\end{align}
which is equivalent to \Eq{stationary_current}.
This is known as the Meir-Wingreen formula \cite{meir1992} which is a Landauer formula for the current through an interacting electron region,
combining the transmission amplitude $T$ of a one-level noninteracting scatterer in a conductor with the conductance $G$, which, for low bias voltages, $G=2e^2/h |T|^2$\citeafter{.}{landauer1957}
Given the electrical current, we can introduce the linear and differential electrical conductances
\begin{equation}
  G_\text{lin}(t) = \frac{I(t)}{V} \, , \qquad G_\text{dif}(t) = \frac{{d}I(t)}{{d}V} \, .
\end{equation}

For conciseness we finally give the mean-field stationary current-voltage characteristics at zero temperature which become exact in the $U=0$ limit.
Replacing the statistical self-energy component of the leads with $\Sigma^\text{F}_{\ell,\text{lead}}(\omega) = -\Gamma/2 \, \mathrm{sgn}(\omega-\mu_\ell)$, we obtain \cite{komnik2004}
\begin{align}
  \label{eq:occupation_qdot_voltage}
  n_\sigma = \frac{1}{2} 
  &- \frac{1}{2\pi} \sum_{\ell=\pm1}\arctan\left( \frac{E_0 + U n_{\bar{\sigma}} - {\ell}\,V/2}{\Gamma} \right)\, . 
\end{align}
As we assume the particle-hole symmetric case, $E_0 = -U/2$, the chemical potentials are adjusted symmetrically around the energy zero, $\mu_\text{L} = -\mu_\text{R} = \mu/2$.
With no magnetic field present, Eq.~(\ref{eq:occupation_qdot_voltage}) is solved by $n_{\sigma} = n_{\bar{\sigma}} = 0.5$. 
In this case, the effective mass \eq{MFMass} is zero,
and the stationary current reads
\begin{equation}
  {I(t\to\infty)} = \frac{e \Gamma}{2 h} \sum_\sigma \int^D_{-D} {d}\epsilon \,    
  \frac{2 \Gamma\Big[ f(\epsilon-\mu_\text{L}) - f(\epsilon-\mu_\text{R}) \Big]}{(\epsilon - M_\sigma)^2 + \Gamma^2} \, ,
\end{equation}
and at zero temperature and nonzero bias voltage we get
\begin{equation}
  \label{eq:mf_stationary_current}
    I = \frac{e \Gamma}{h} \int^{\mu/2}_{-\mu/2} {d}\epsilon \, \frac{2 \Gamma}{\epsilon^2 + \Gamma^2} = \frac{4 e \Gamma}{h} \Bigg[\arctan\left( \frac{\mu}{2\,\Gamma} \right) \Bigg] \, .
\end{equation}

The maximum current is obtained in the limit of an infinite chemical potential gradient,
\begin{equation}
  \label{eq:max_current}
  I_\mathrm{max} =  {2\pi\,\Gamma e}/{h} \, ,
\end{equation}
and from \Eq{mf_stationary_current}, we obtain the differential conductance
\begin{equation}
{\mu^2+(2\,\Gamma)^2} \, ,
\end{equation}
in terms of the conductance quantum
$G_0 = 2 e^2/h$\citeafter{.}{kouwenhoven1988}

\section{Numerical results for different resummation schemes}
\label{app:NumData}
\begin{figure*}[h!]
\begin{center}
\includegraphics[width=0.3 \textwidth]{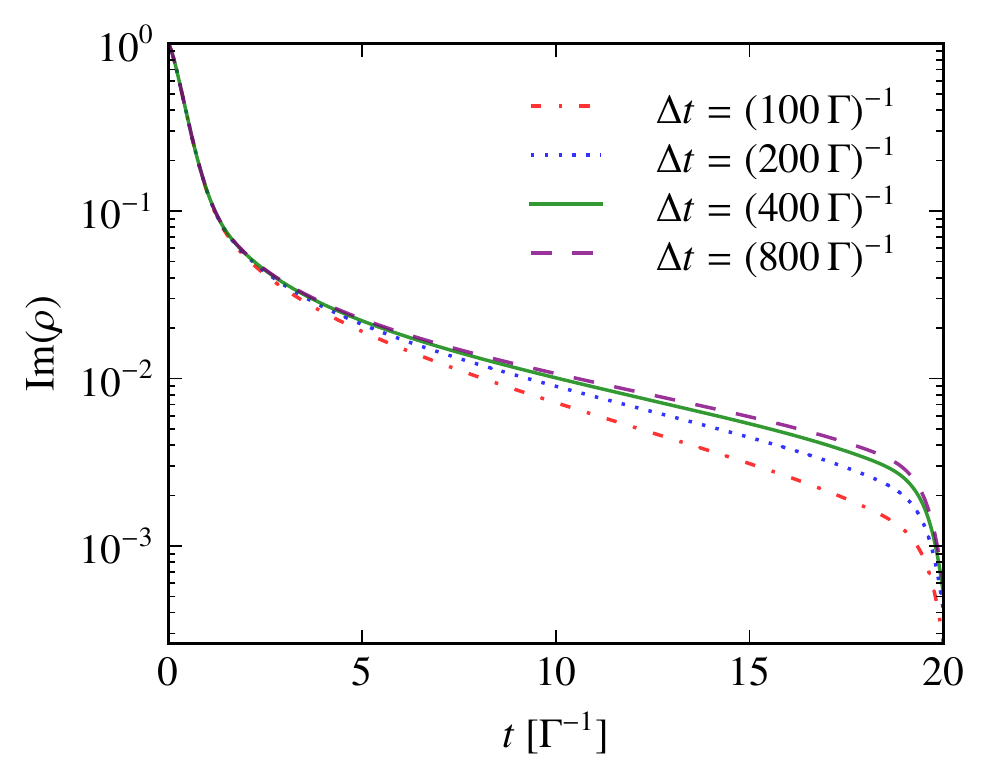}
\includegraphics[width=0.3 \textwidth]{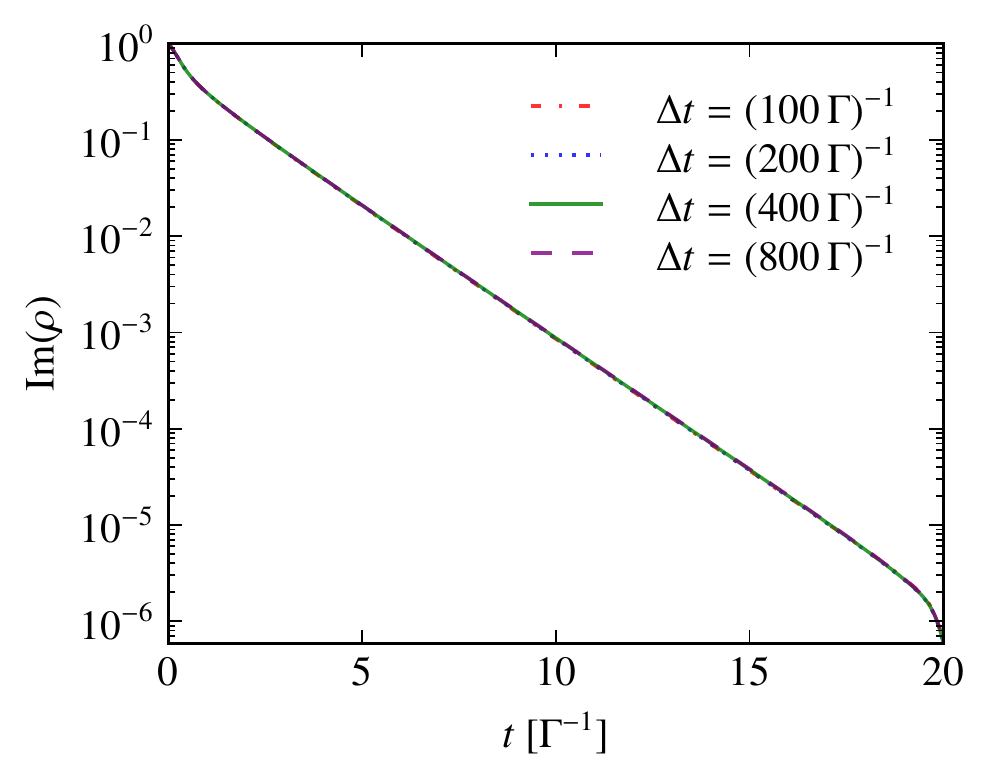}
\includegraphics[width=0.3 \textwidth]{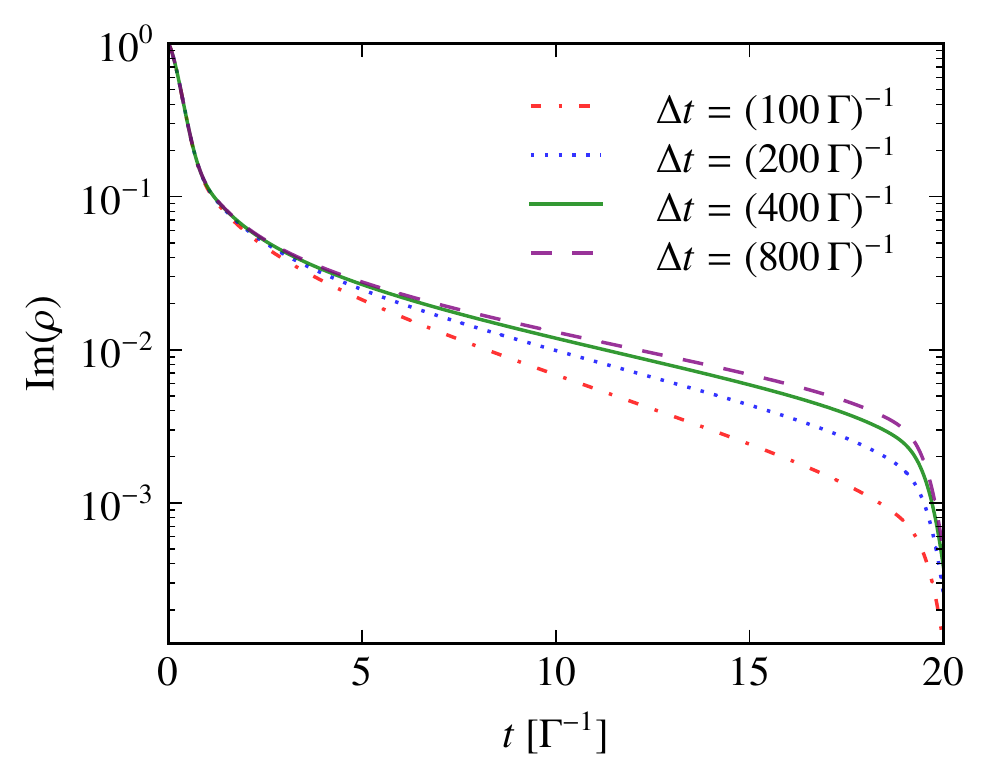}
\\[-1.1ex]
\includegraphics[width=0.3 \textwidth]{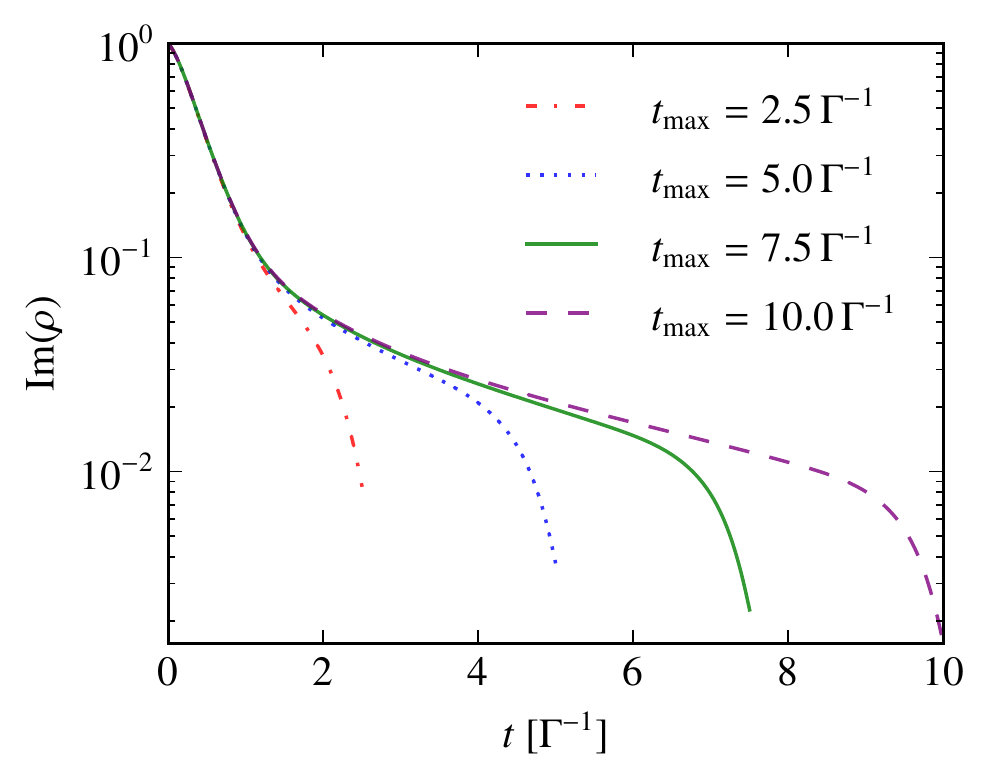}
\includegraphics[width=0.3 \textwidth]{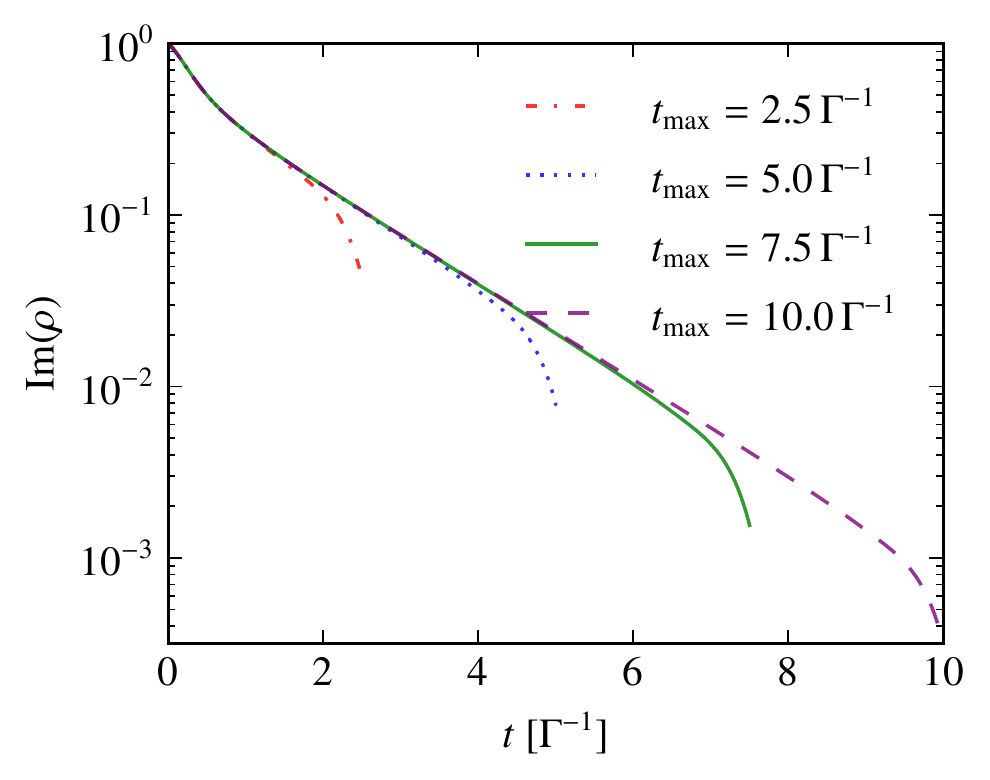}
\includegraphics[width=0.3 \textwidth]{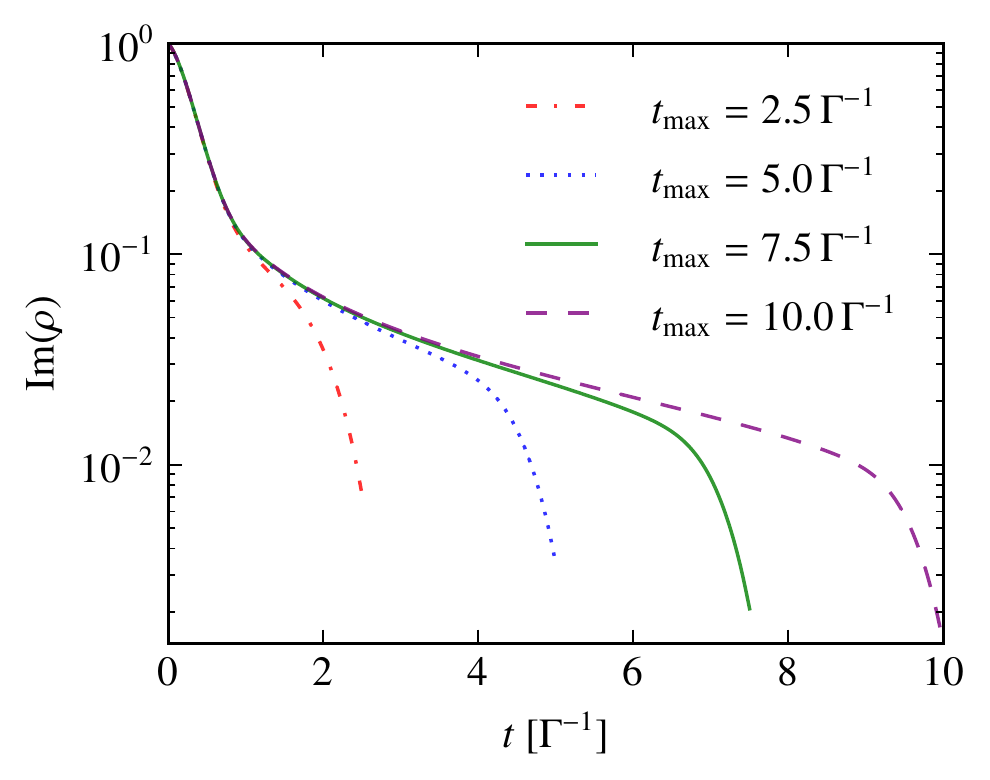}
\\[-1.1ex]
\includegraphics[width=0.3 \textwidth]{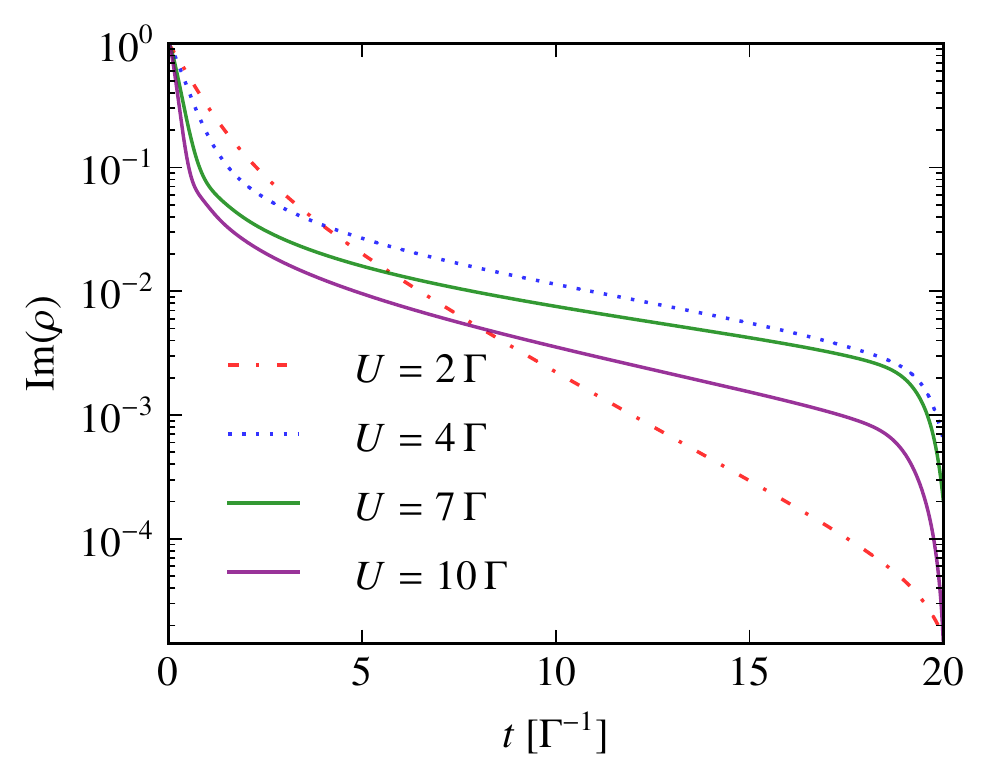}
\includegraphics[width=0.3 \textwidth]{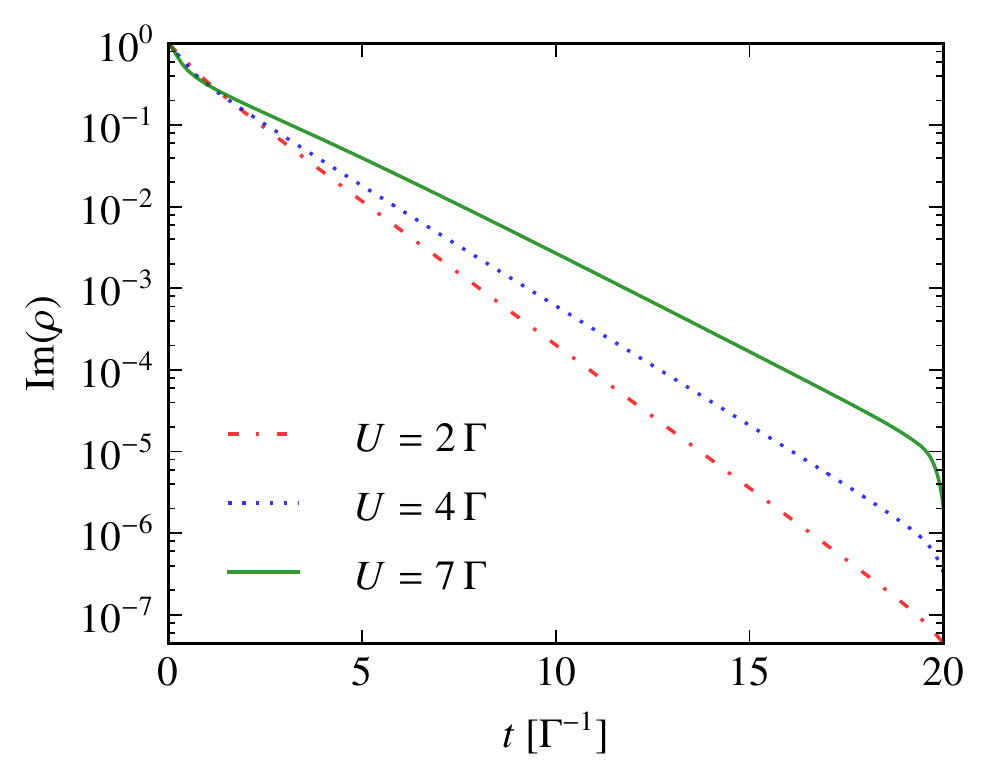}
\includegraphics[width=0.3 \textwidth]{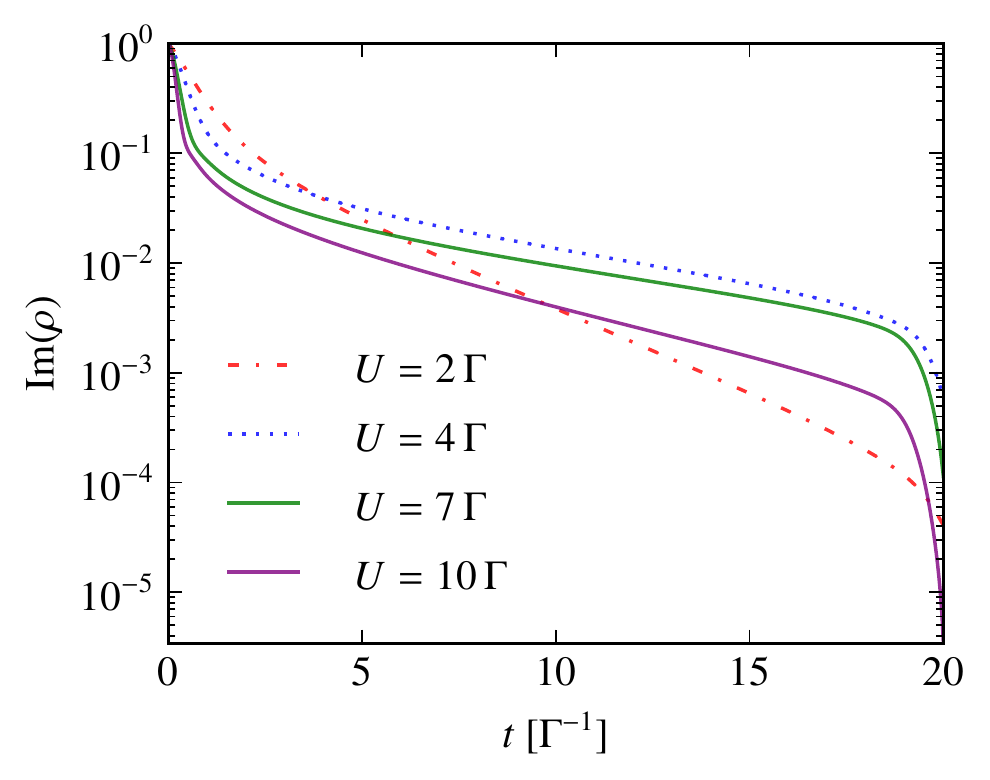}
\\[-1.1ex]
\includegraphics[width=0.3 \textwidth]{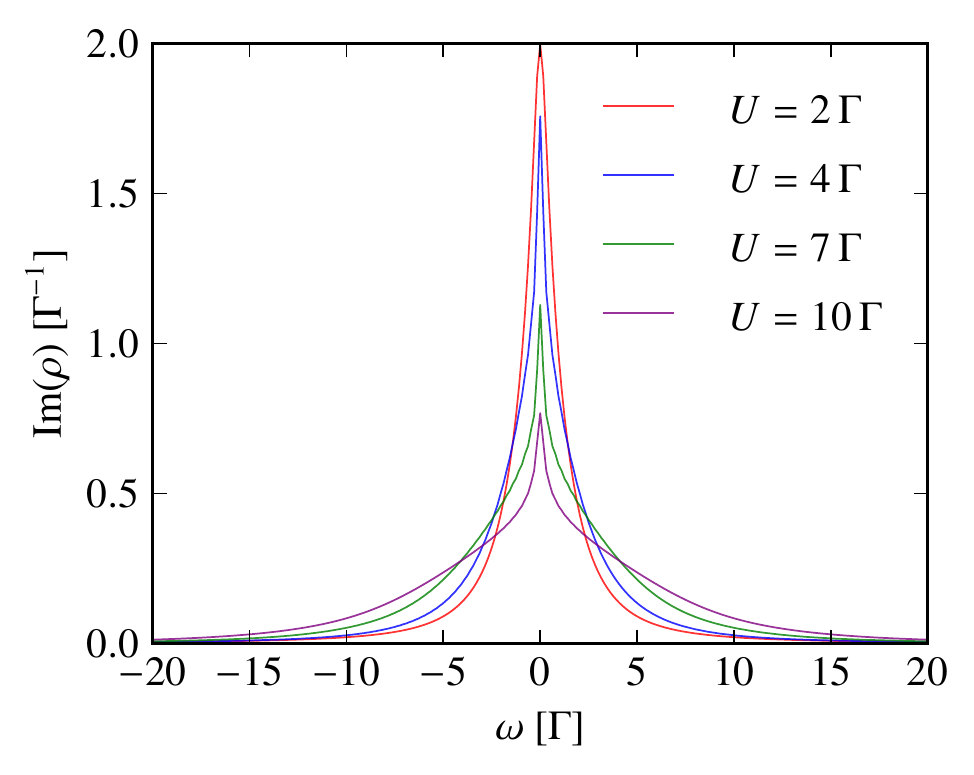}
\includegraphics[width=0.3 \textwidth]{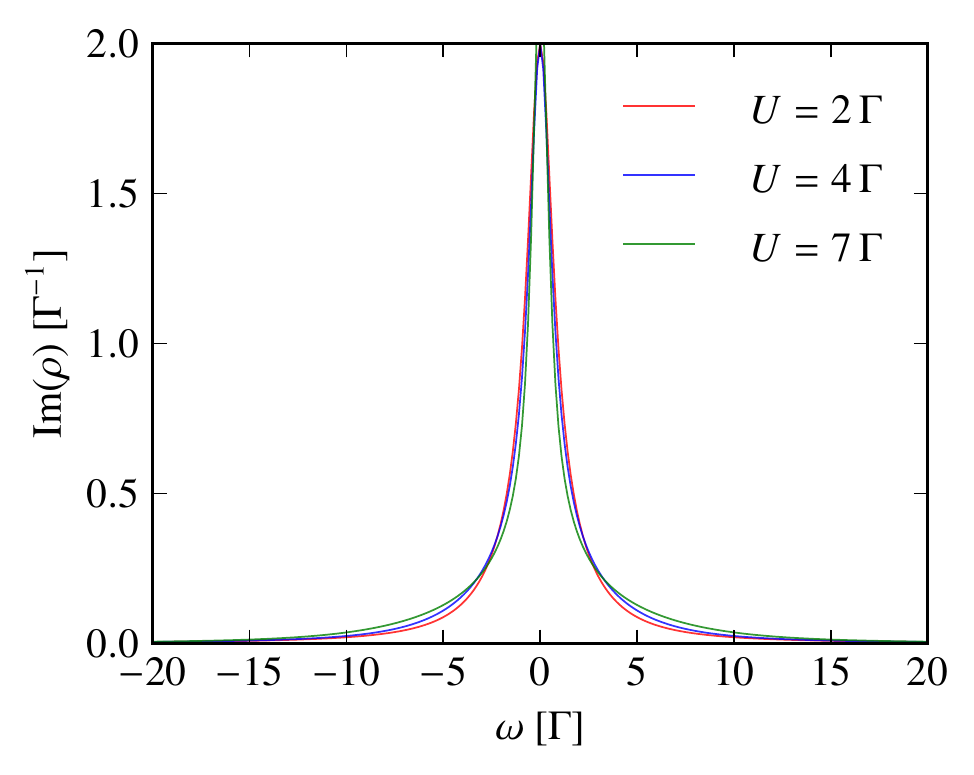}
\includegraphics[width=0.3 \textwidth]{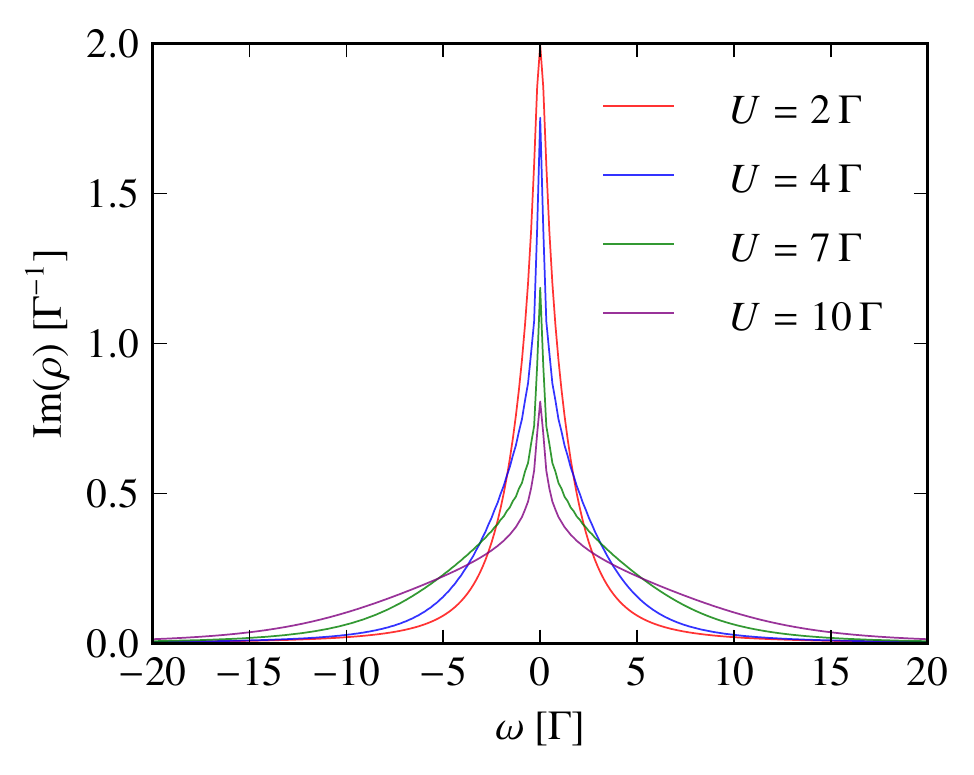}
\\[-1.1ex]
\includegraphics[width=0.3 \textwidth]{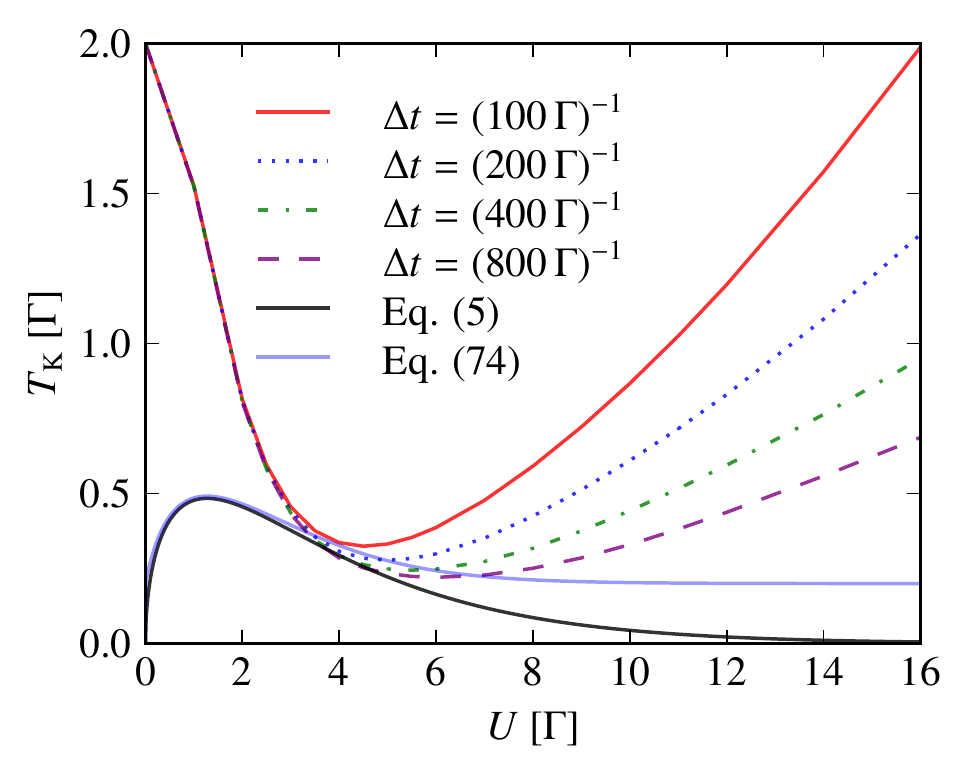}
\includegraphics[width=0.3 \textwidth]{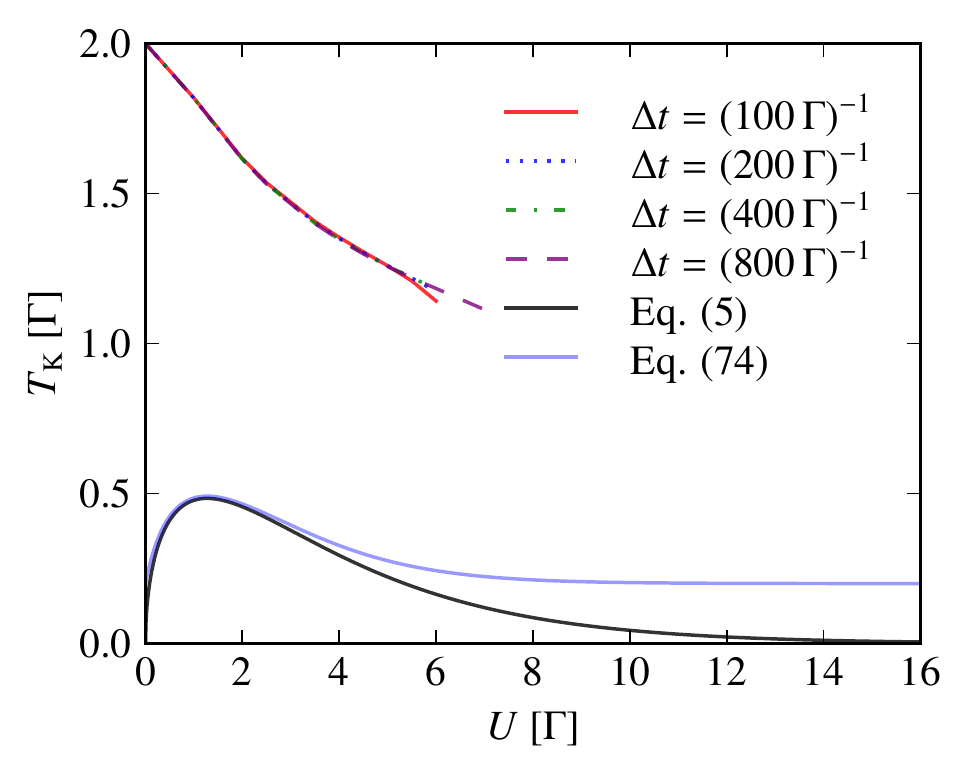}
\includegraphics[width=0.3 \textwidth]{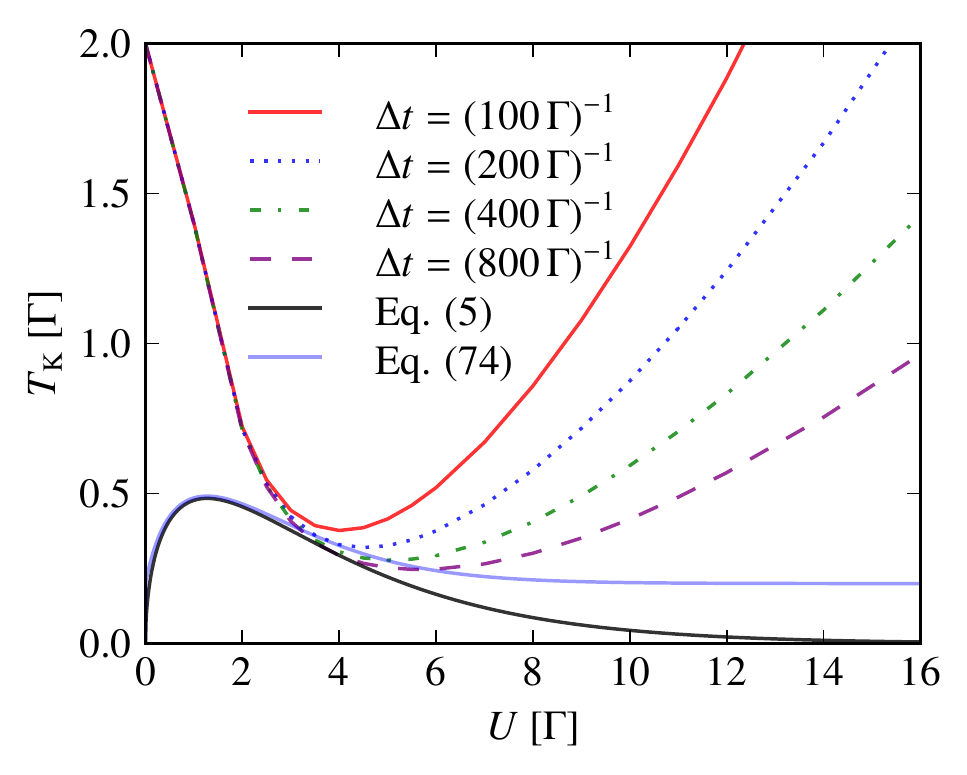}
\caption{(Color online)
The imaginary part of the spectral function, for $E_{0} = -U/2$.
The different  columns show results obtained in the $s$- (left), $t$- (middle), and $u$-channels (right), respectively.
First row: $\mathrm{Im}\rho(t_\mathrm{max},t)$, for $U=5\,\Gamma$, maximum evolution time $t_\mathrm{max}=20\,\Gamma^{-1}$, and different time-step sizes.
Second row: Same function, for $U=5\,\Gamma$, $\Delta t=(800\,\Gamma)^{-1}$, and different $t_\mathrm{max}$.
Third row: Same function, for $t_\mathrm{max}=20\,\Gamma^{-1}$, $\Delta t=(800\,\Gamma)^{-1}$, and different interactions $U$.
Fourth row: $\mathrm{Im}[\rho(\omega)]$, i.e., the Fourier transform of the functions in the third row. 
Narrowing of the central Kondo resonance is seen in the $s$- and $u$-channels.
Fifth row: 
The Kondo temperature $T_\mathrm{K}$, in units of $\Gamma$, as derived from the long-time exponential decay of the functions in the first row.
$U$, $\omega$, and $T_\mathrm{K}$ are given in units of $\Gamma$, $\rho$, $t$, $\Delta t$, and $t_\mathrm{max}$ in units of $\Gamma^{-1}$.
}
\label{fig:rho-t_s-t-u}
\end{center}
\end{figure*}

\begin{figure*}[h!]
\begin{center}
\includegraphics[width=0.305 \textwidth]{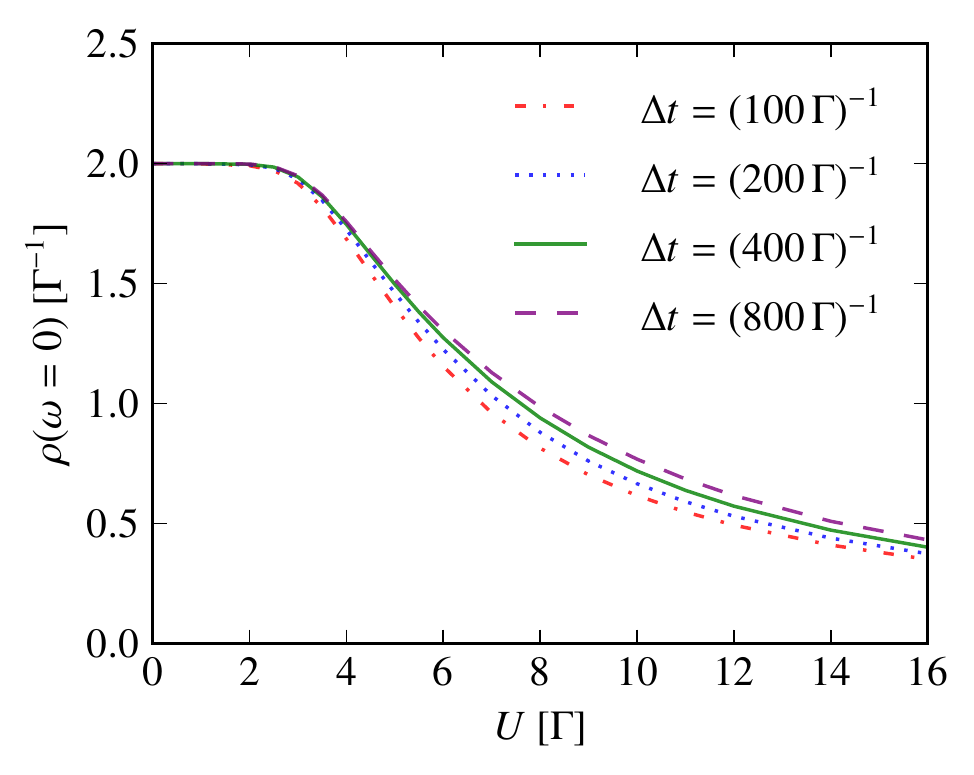}
\includegraphics[width=0.305 \textwidth]{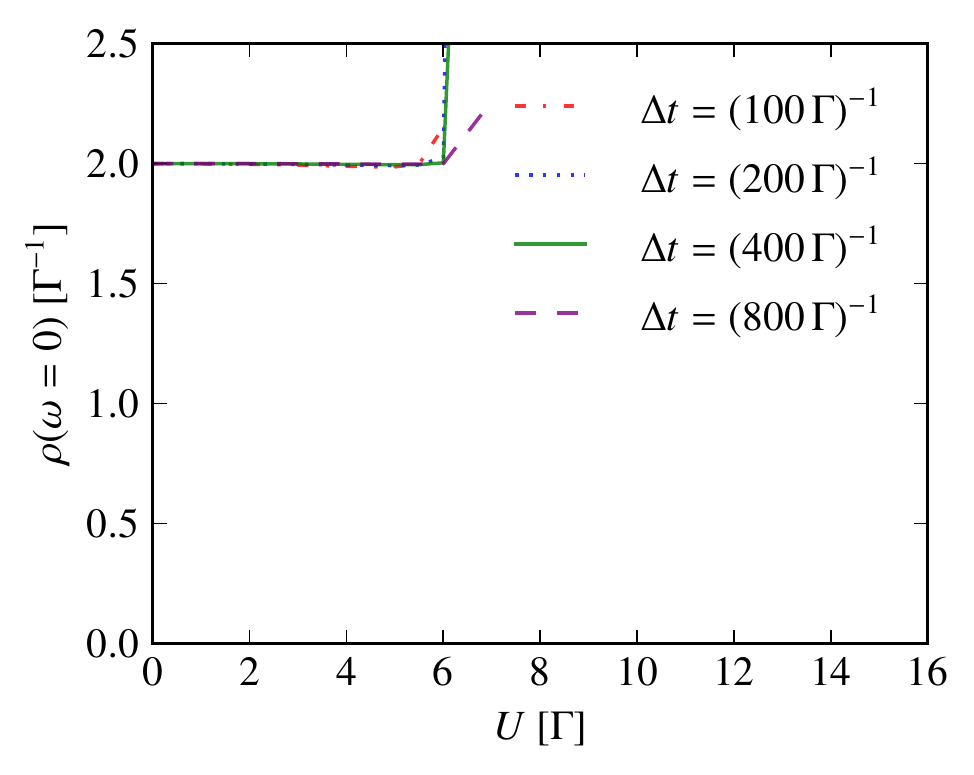}
\includegraphics[width=0.305 \textwidth]{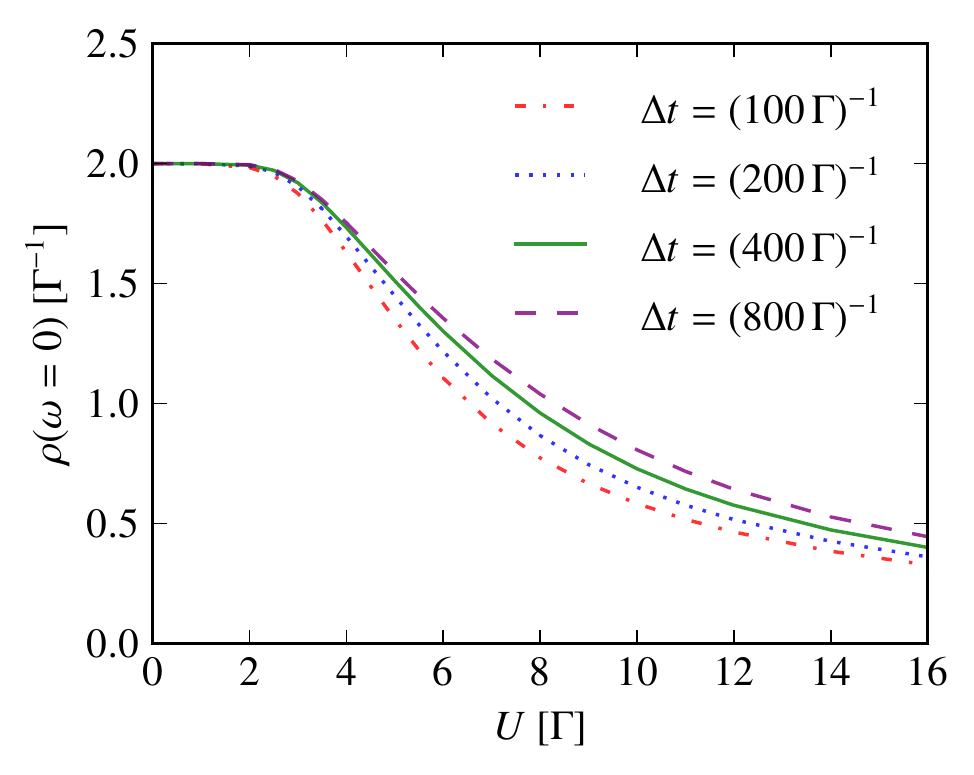}
\\[-0.5ex]
\includegraphics[width=0.305 \textwidth]{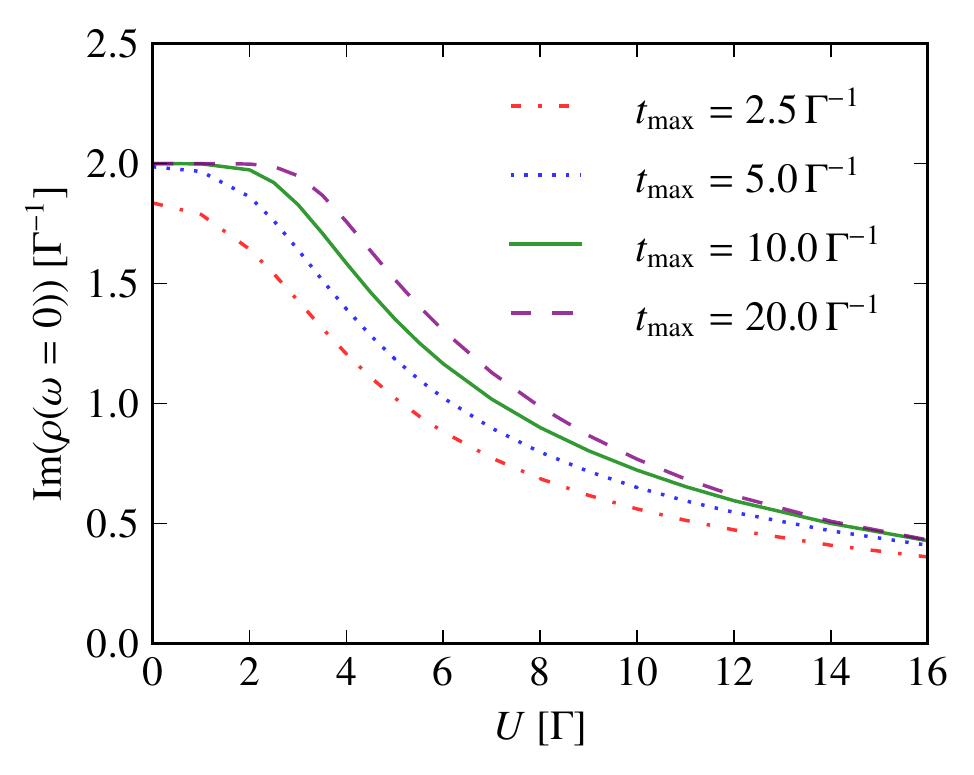}
\includegraphics[width=0.305 \textwidth]{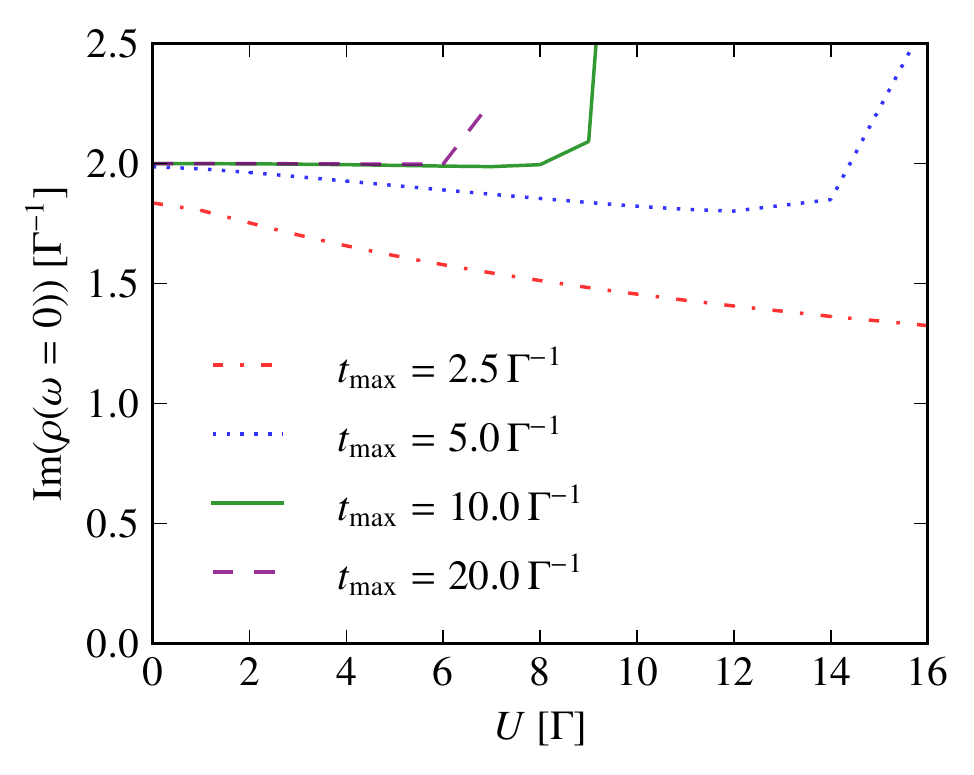}
\includegraphics[width=0.305 \textwidth]{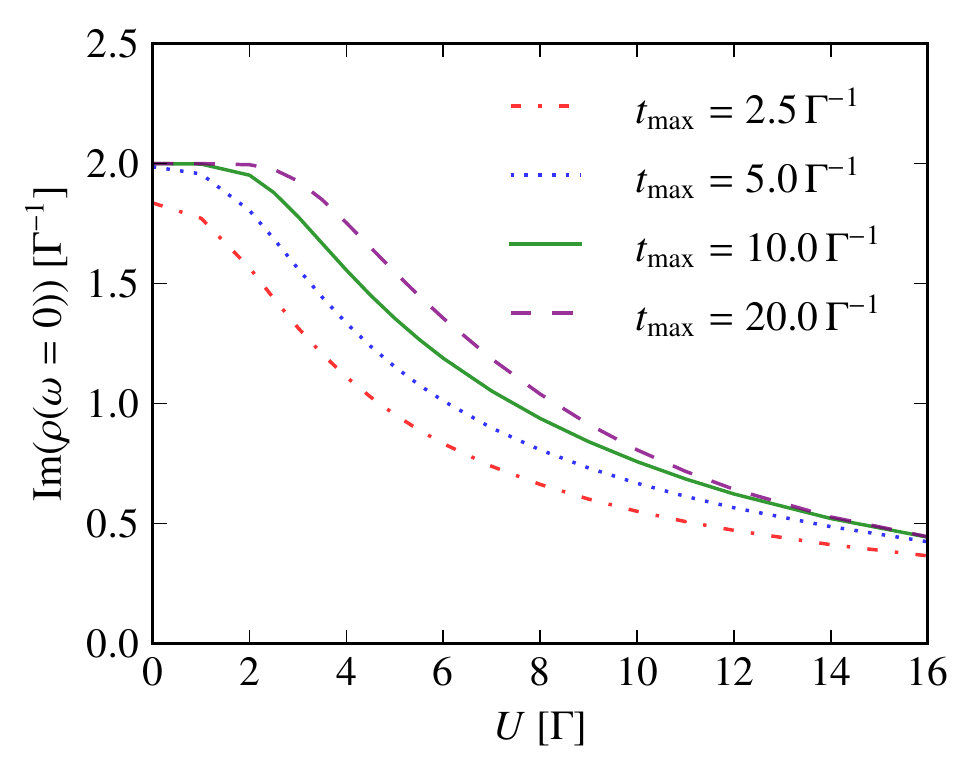}
\caption{(Color online)
Quality of the fulfillment of the Friedel sum rule for the spectral function $\rho(\omega=0)$, as a function of the interaction strength $U$,  obtained in the $s$-channel (left column), the $t$-channel (middle), and the $u$-channel (right) resummation schemes, respectively.
Top row: $\rho(\omega=0)$ for different time-step sizes and a total evolution time $t_\mathrm{max}=20\,\Gamma^{-1}$.
Bottom row: same function, for different evolution times and a step size of $\Delta t=(800\,\Gamma)^{-1}$.
The leads' temperature is $T=0$, and the dot is tuned to the particle-hole symmetric point $E_{0}=-U/2$.
$U$, $t_\mathrm{max}^{-1}$, and $(\Delta t)^{-1}$ are given in units of $\Gamma$.
}
\label{fig:sumrule-deltat_s-t-u}
\end{center}
\end{figure*}

\begin{figure*}[h!]
\begin{center}
\includegraphics[width=0.305\textwidth]{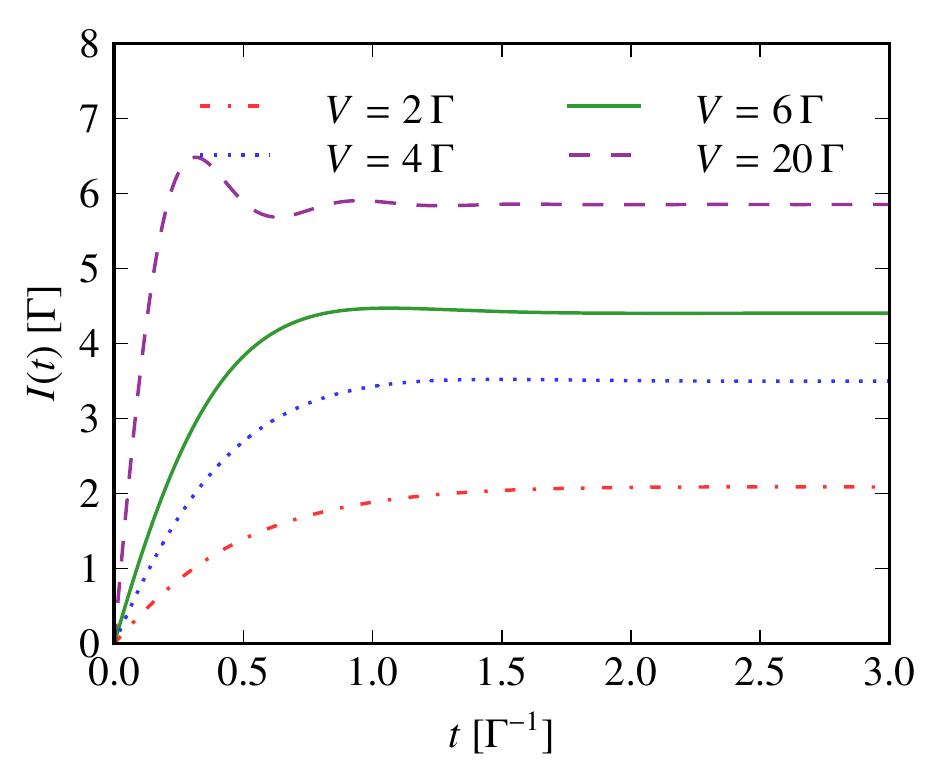}
\includegraphics[width=0.305\textwidth]{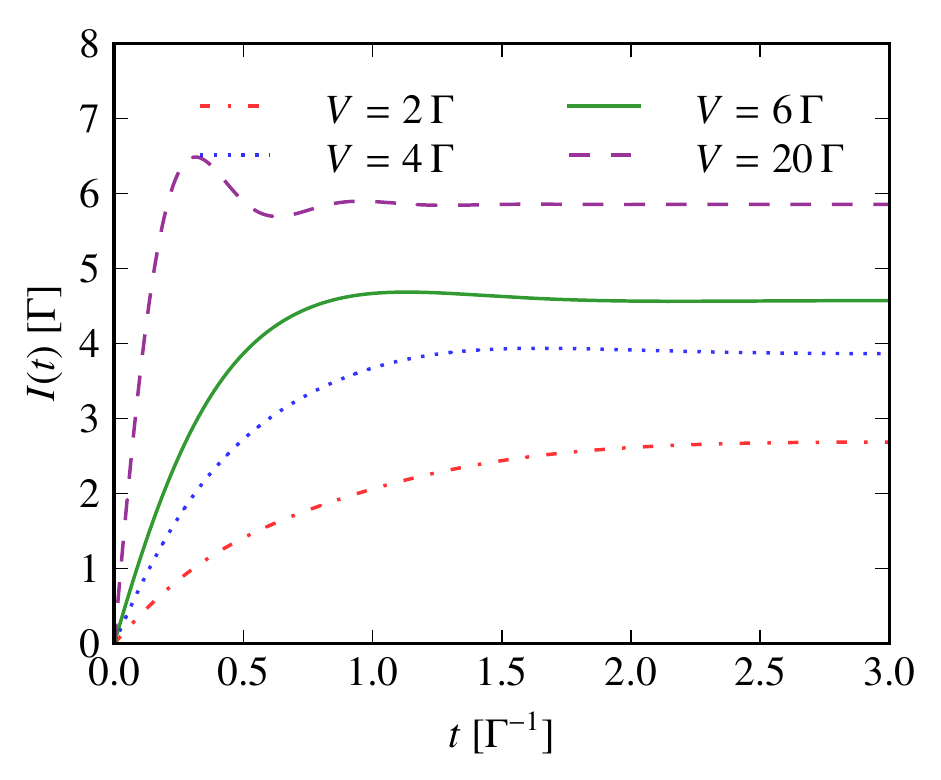}
\includegraphics[width=0.305\textwidth]{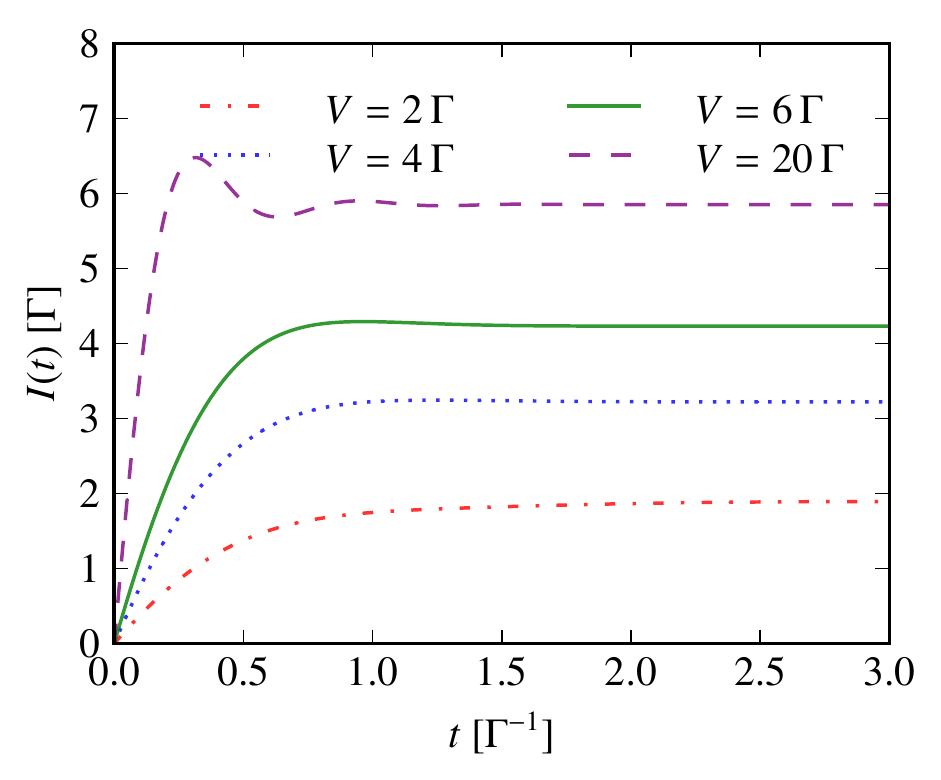}
\\[-0.5ex]
\includegraphics[width=0.31 \textwidth]{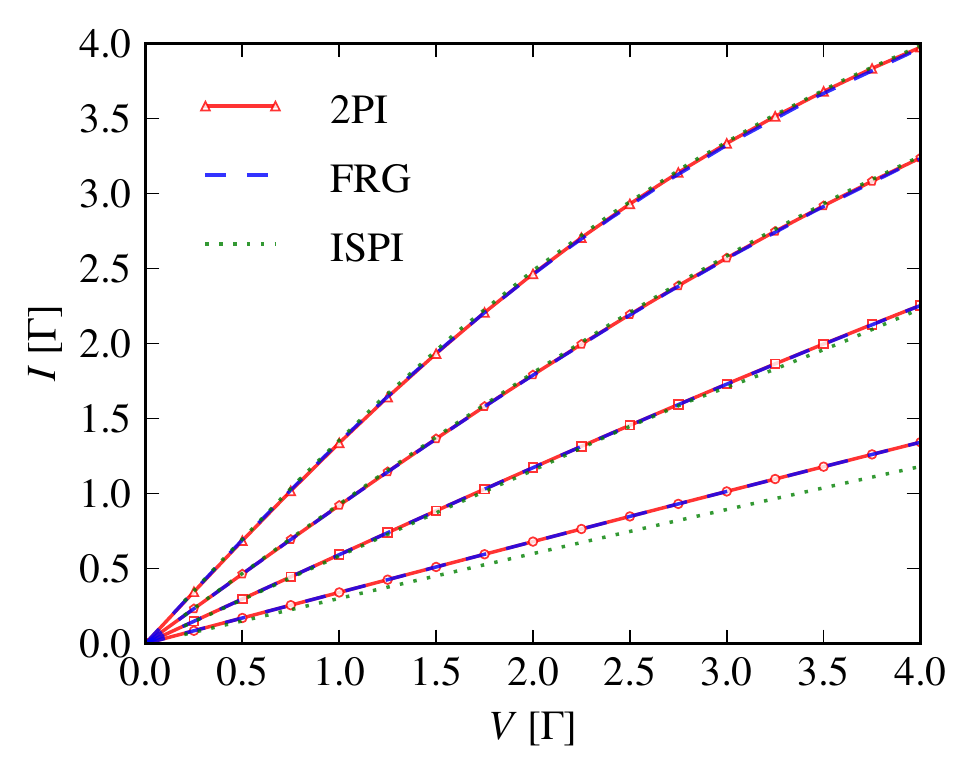}
\includegraphics[width=0.31 \textwidth]{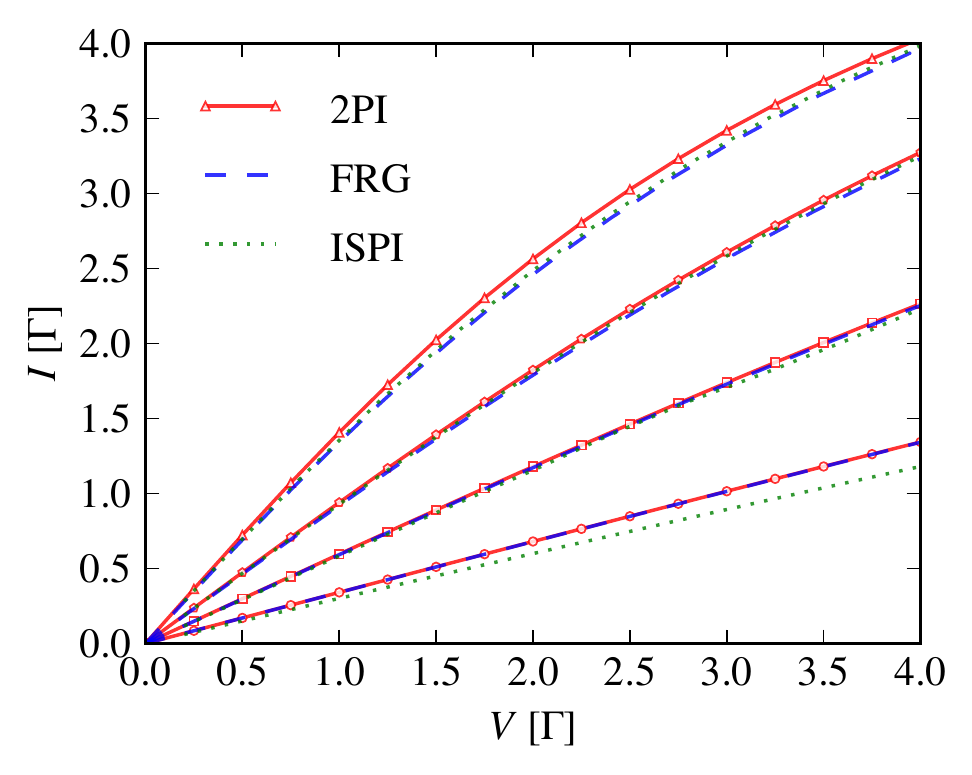}
\includegraphics[width=0.31 \textwidth]{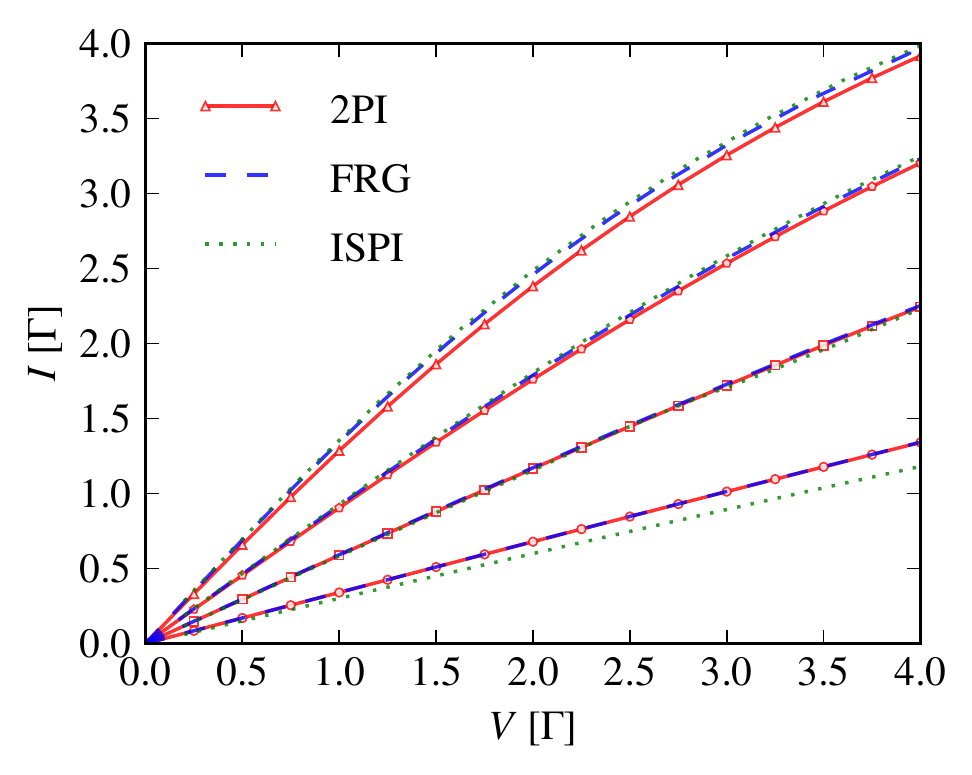}
\caption{(Color online)
Top row: the transient current $I(t)$ through a quantum dot coupled to leads at a temperature $T=0.1\,\Gamma$, as obtained in the $s$- (left), $t$- (middle), and $u$-channels (right), respectively, for different applied bias voltages $V$.
The interaction strength is $U=4\,\Gamma$; the time-step size is $\Delta t = (300\,\Gamma)^{-1}$. 
The particle-hole symmetric case, $E_0=-U/2$, is assumed.
$t^{-1}$, $I$, and $eV$ are given in units of $\Gamma$.
Bottom row: stationary current $I(t_\mathrm{max})$ through the quantum dot coupled to leads at a temperature $T/\Gamma=0.4$, $1.0$, $2.0$, $4.0$ (from top to bottom curve), for $E_0=-U/2$, as functions of the voltage $V$, obtained in the 
$s$- (left), $t$- (middle), and $u$-channels (right), respectively, with $U=2\,\Gamma$.
We compare with the FRG and ISPI results of Ref.~\onlinecite{eckel2010}. 
The system was evolved to the total time $t_\text{max}=6\,\Gamma^{-1}$  with a time-step size $\Delta t = (300\,\Gamma)^{-1}$.
}
\label{fig:iv_t_s-t-u}
\end{center}
\end{figure*}

\begin{figure*}[h!]
\begin{center}
\includegraphics[width=0.24 \textwidth]{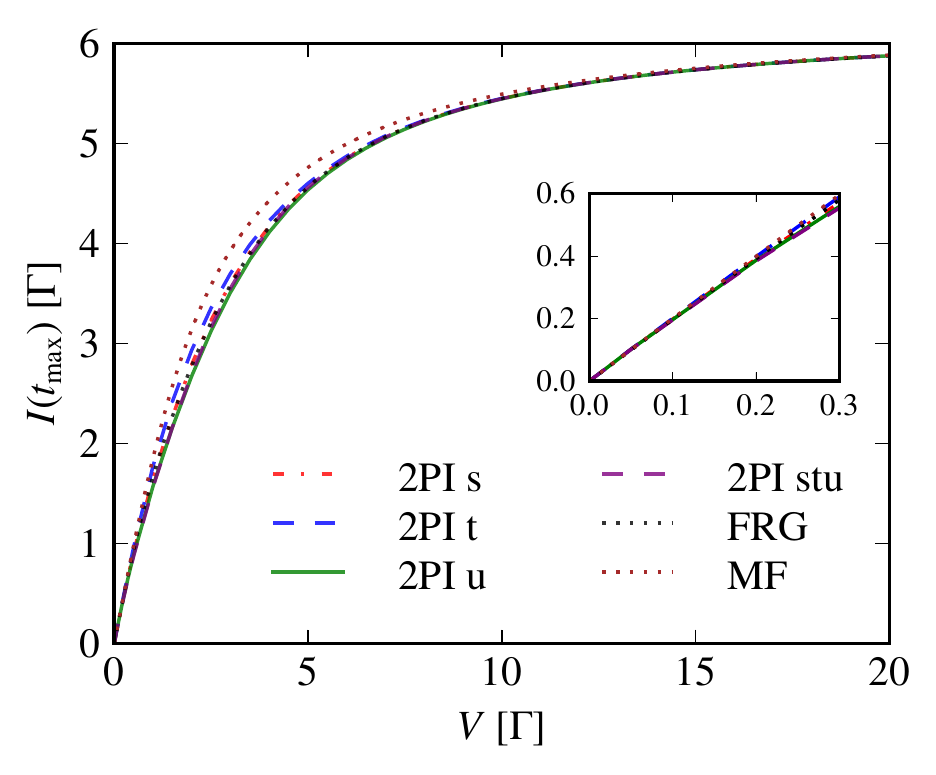}
\includegraphics[width=0.25 \textwidth]{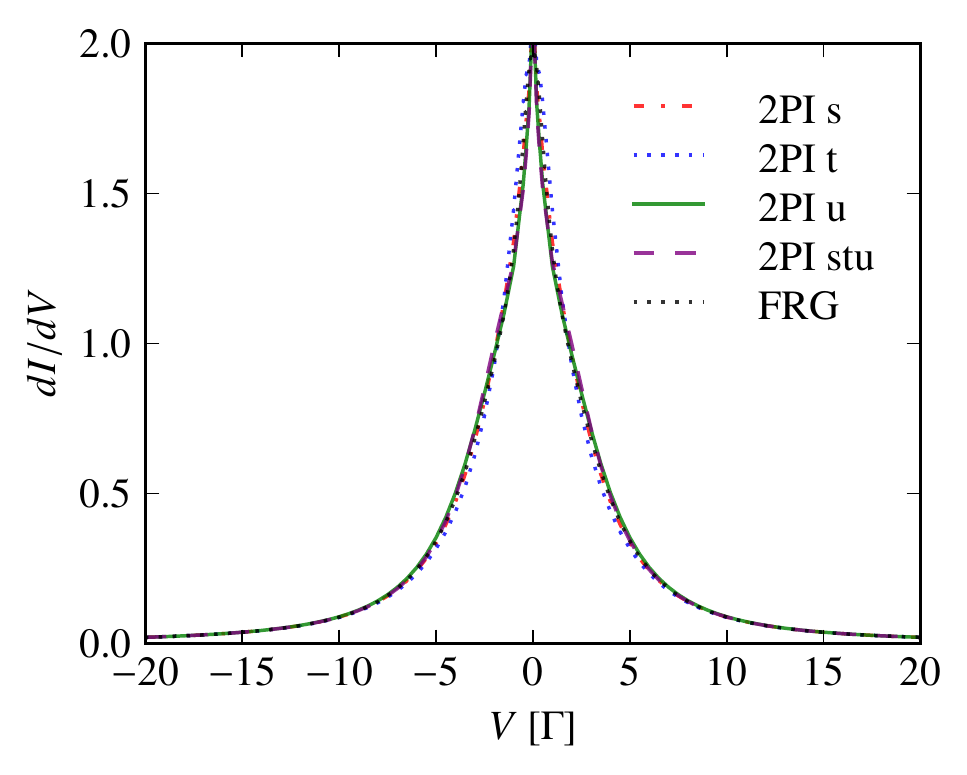}
\includegraphics[width=0.24 \textwidth]{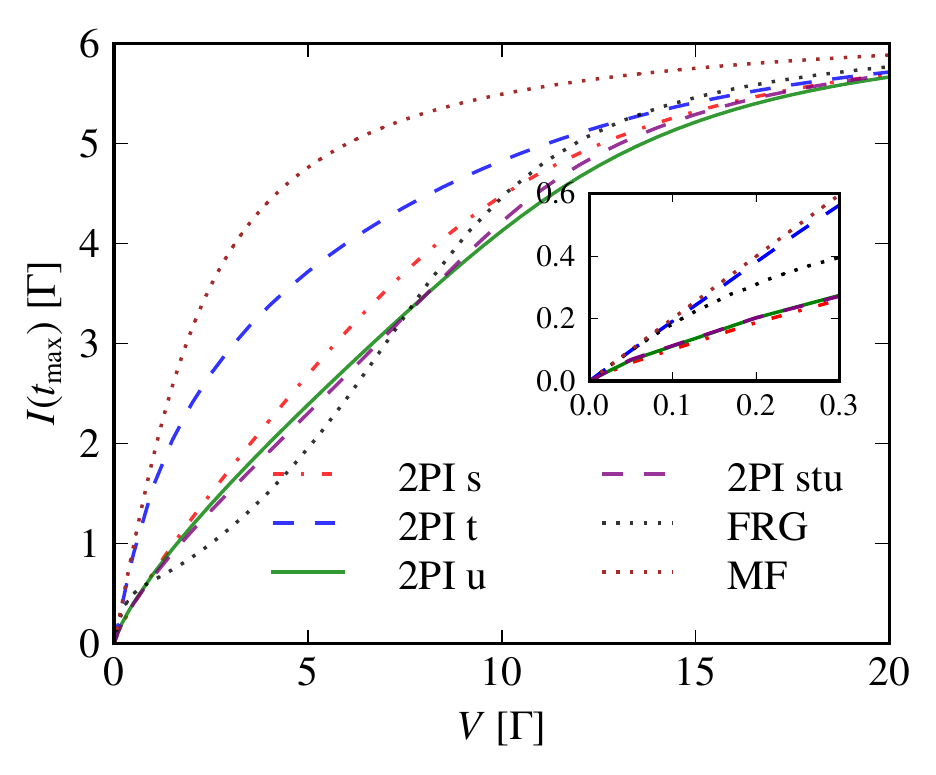}
\includegraphics[width=0.25 \textwidth]{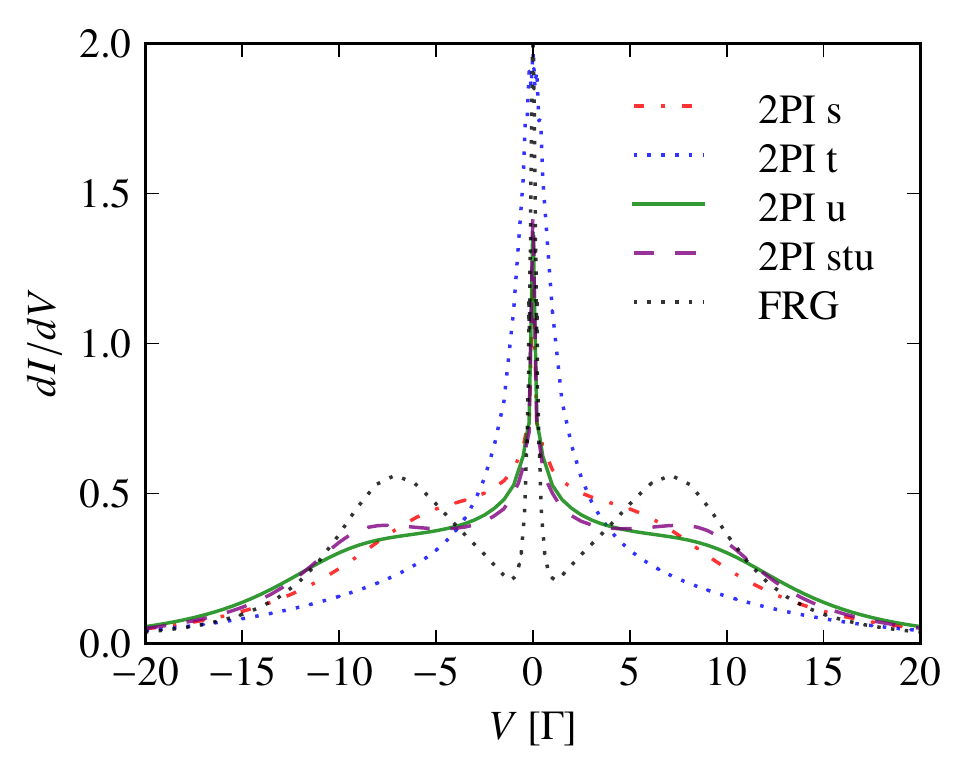}
\caption{
The stationary current $I(t_\mathrm{max})$ through the dot (first and third panels) and the differential conductance (second and fourth panels) as functions of the bias voltage $V$, obtained within the different resummation channels  in the particle-hole symmetric case, $E_0=-U/2$.
We compare our results with those obtained within the functional renormalization-group (FRG) scheme of Ref.~\onlinecite{eckel2010}.
``MF'' denotes the mean-field result, \Eq{ItMF}, for the current.
The interaction strength is $U = 2\,\Gamma$ in the left two panels and  $U = 8\,\Gamma$ in the right two ones. The leads have a temperature $T=0.1\,\Gamma$. 
The system was evolved to the total time $t_\text{max}=40\,\Gamma^{-1}$, for bias voltages $V \leq 0.5\,\Gamma$ and $t_\text{max}=6\,\Gamma^{-1}$, otherwise, with a time-step size of $\Delta t = (300\,\Gamma)^{-1}$.
For smaller $U$, the agreement between the different approximation schemes increases.
}
\label{fig:i_v_u_U2-8}
\end{center}
\end{figure*}

\begin{figure*}[h!]
\begin{center}
\includegraphics[width=0.24 \textwidth]{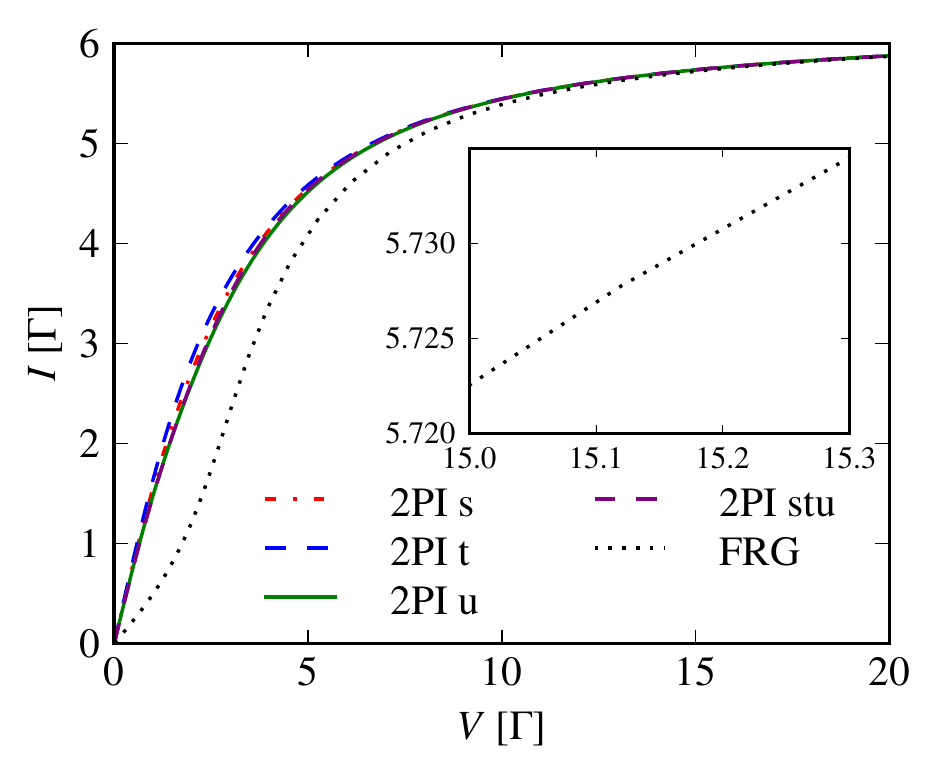}
\includegraphics[width=0.25 \textwidth]{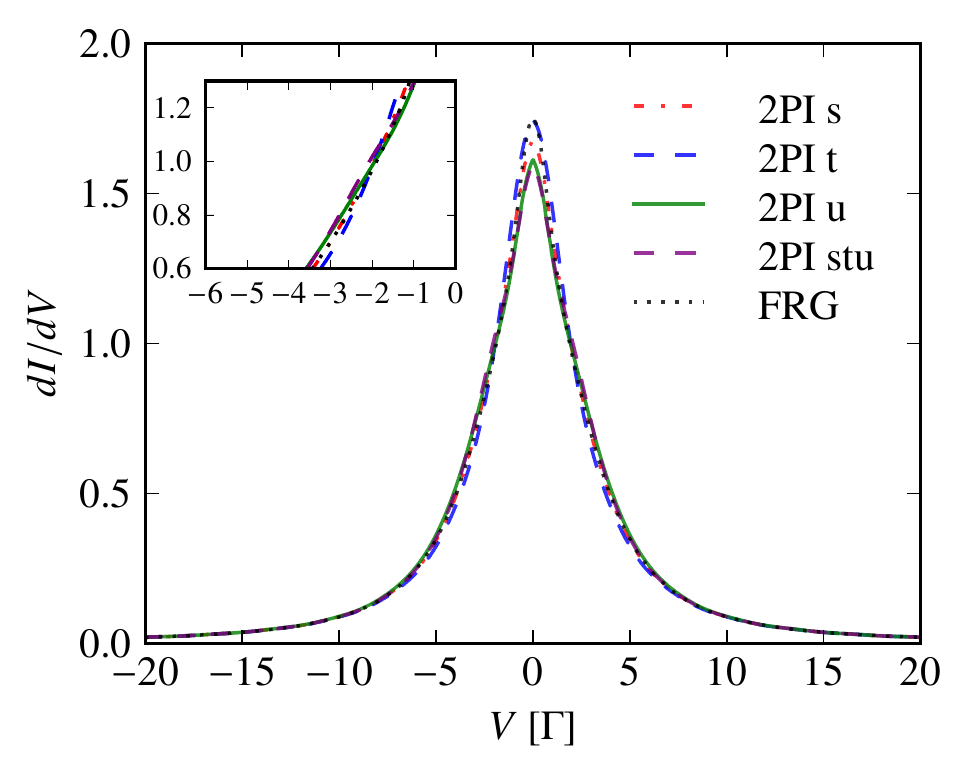}
\includegraphics[width=0.24 \textwidth]{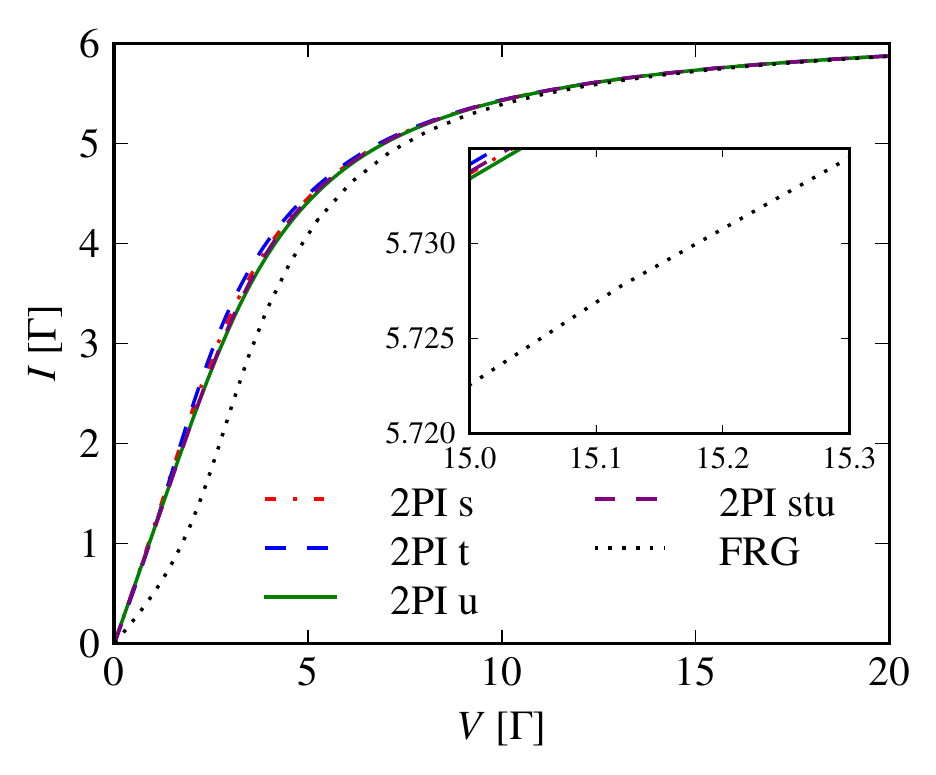}
\includegraphics[width=0.25 \textwidth]{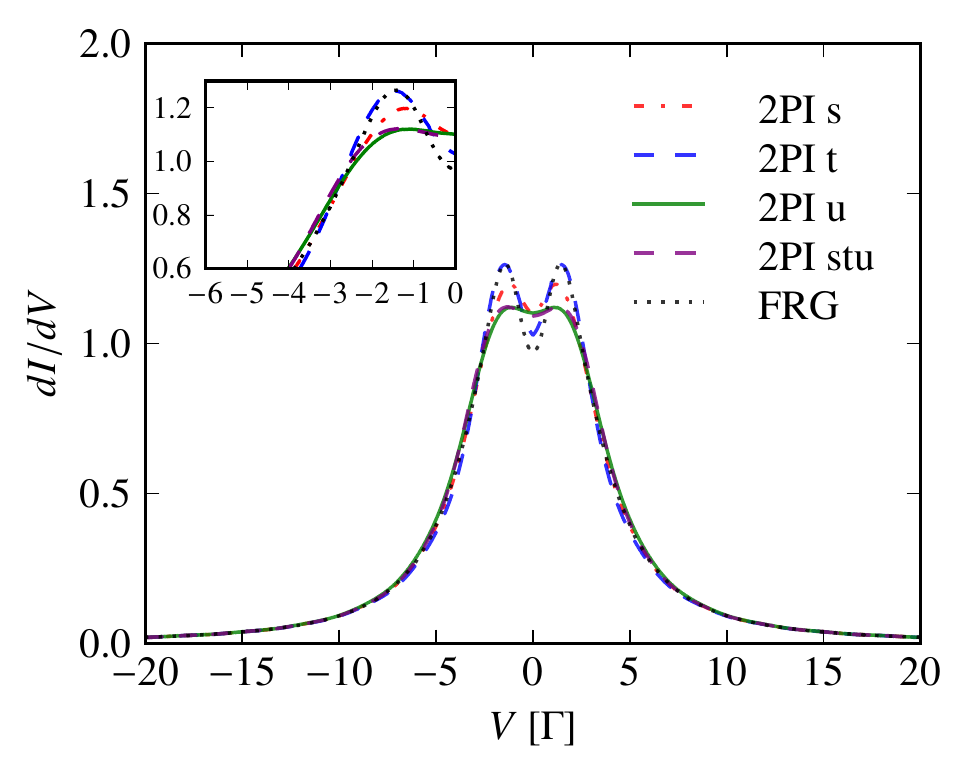}
\caption{
The stationary current $I(t_\mathrm{max})$ through the dot (first and third panels) and the differential conductance (second and fourth panels) as functions of the bias voltage $V$, obtained within the different resummation channels  in the particle-hole symmetric case, $E_0=-U/2$.
``FRG'' denotes the functional renormalization-group results of Ref.~\onlinecite{eckel2010}.
The interaction strength is $U = 2\,\Gamma$.  
An external magnetic fields $B = 0.4\,\Gamma$ is applied in the left two panels and  $B = 1.2\,\Gamma$ in the right two ones. 
The system was evolved to  $t_\text{max}=6\,\Gamma^{-1}$, with a time-step size of $\Delta t = (300\,\Gamma)^{-1}$.
}
\label{fig:i_v_b_U4-12}
\end{center}
\end{figure*}

\begin{figure*}[h!]
\begin{center}
\includegraphics[width=0.245 \textwidth]{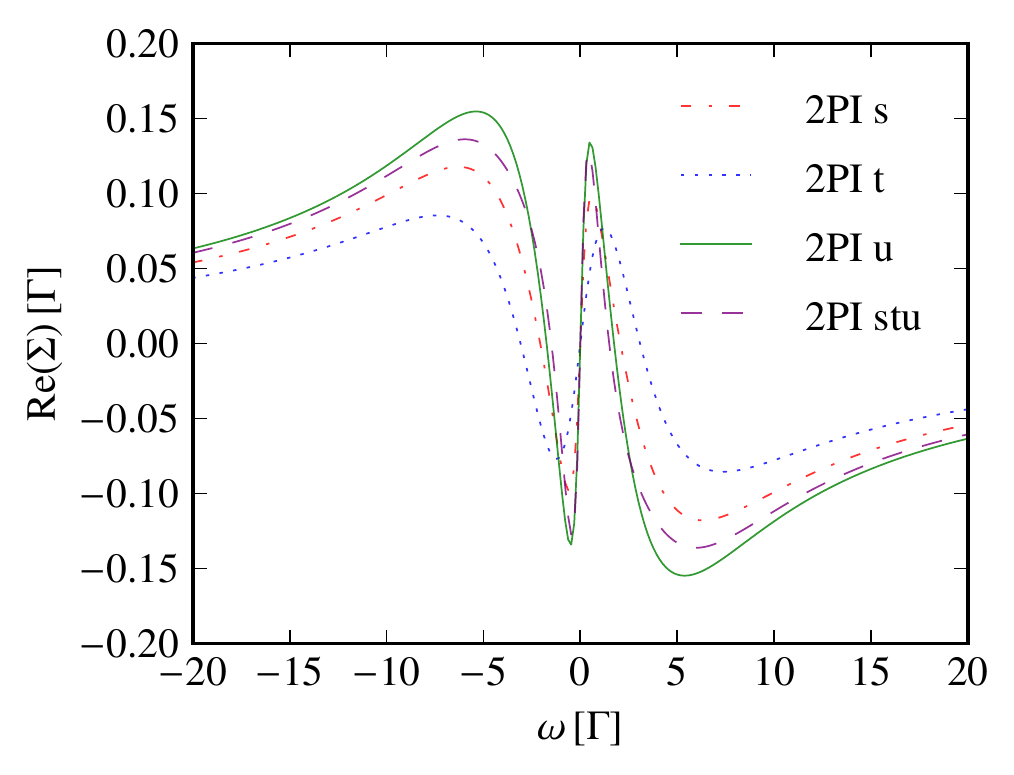}
\includegraphics[width=0.245 \textwidth]{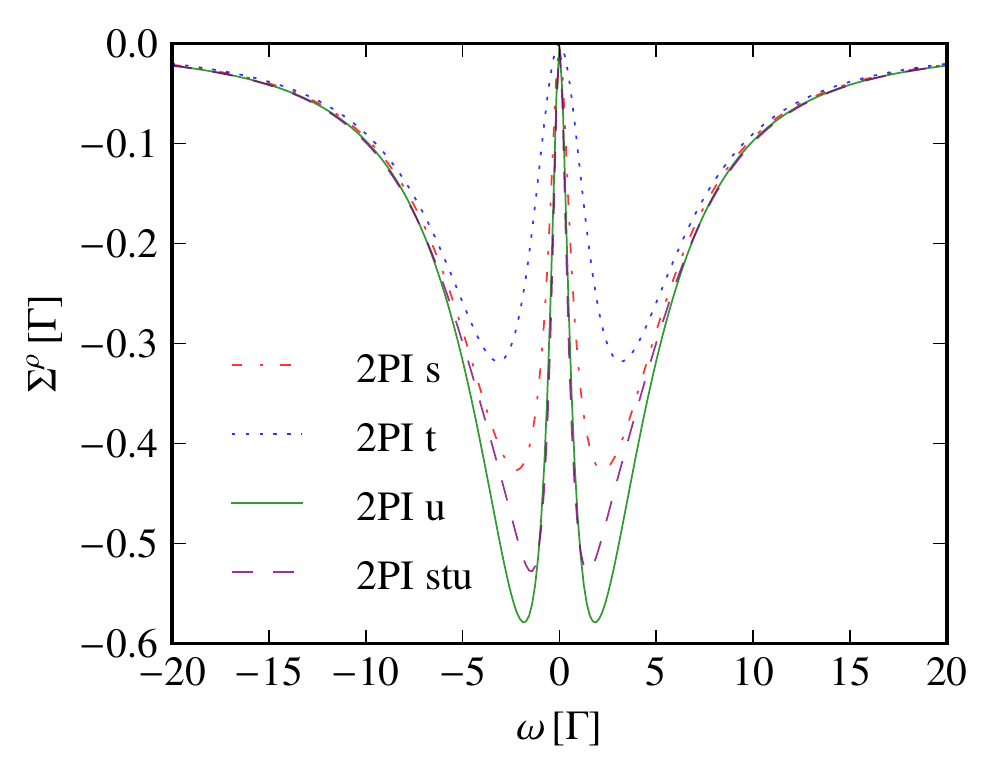}
\includegraphics[width=0.245 \textwidth]{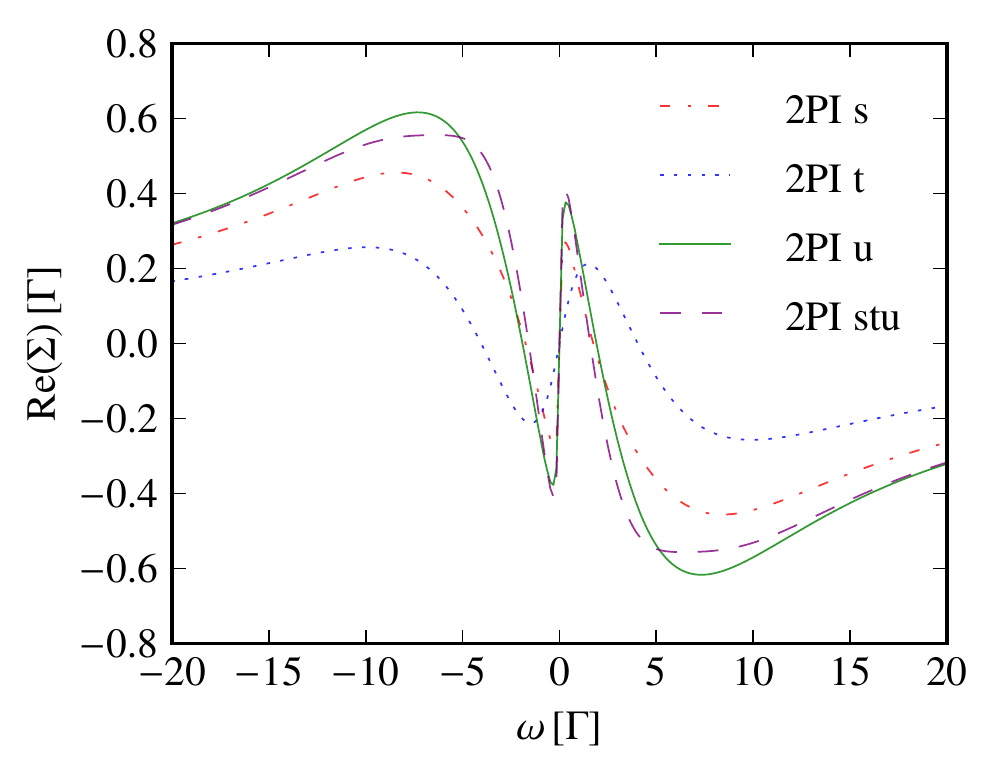}
\includegraphics[width=0.245 \textwidth]{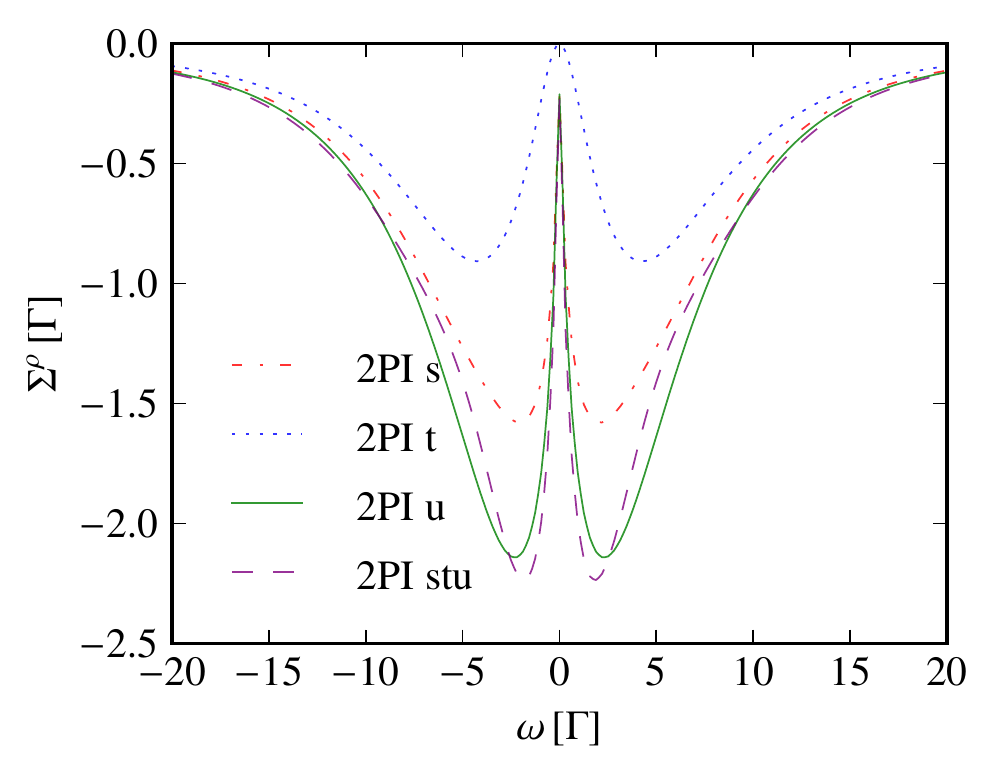}
\caption{
The stationary self-energies $\Sigma^{\rho}$ and $\textrm{Re} \Sigma$ obtained in the different resummation schemes, for $U=2\,\Gamma$ (left two panels) and  $U=4\,\Gamma$ (right two panels).
The system was evolved to the total time $t_\text{max}=20\,\Gamma^{-1}$, with a time-step size of $\Delta t = (800\,\Gamma)^{-1}$.
Neither a magnetic field nor a bias voltage is applied.
\ \vspace*{3cm}
\\
}
\label{fig:SelfEnergies}
\end{center}
\end{figure*}

In this appendix we complement the numerical results presented in the main part with data obtained within the other resummation schemes, for comparison.
This includes, in particular, in \Fig{rho-t_s-t-u}, the dependence of the spectral function $\rho(t,t')$ on the difference of its time arguments, its Fourier transform which shows the narrowing Kondo resonance, as discussed in \Sect{SpecFuncKondo}, 
and the dependence of the Kondo temperature on the interaction strength, cf.~\Sect{KondoTemp}.
\Fig{sumrule-deltat_s-t-u} shows the $s$-, $t$-, and $u$-channel data for the value of the spectral function at zero frequency, $\omega=0$, demonstrating to what extent the Friedel sum rule is obeyed, as discussed in \Sect{Friedel}.
Furthermore, we present in Figs.~\fig{iv_t_s-t-u} and \fig{i_v_u_U2-8} the data for the transient current induced by a nonzero voltage between the leads, for the long-time stationary current induced for different voltages and in the different resummation schemes, and for the differential conductance, as discussed in the main text in \Sect{CurrentVoltage}.
\Fig{i_v_b_U4-12} complements the data on the current-voltage characteristics for a magnetically split dot as discussed in  \Sect{CurrentVoltageBfield}.
Finally, Fig.~\ref{fig:SelfEnergies} shows exemplary graphical representations of $\Sigma^{\rho}$ and $\textrm{Re} \Sigma$ derived in Appendix \ref{sec:KineticTheory}, cf.~Eqs.~(\ref{eq:app:spectralfunckin}) and (\ref{eq:app:realsigma}).

%

%
%

%

%
 
%
\balancecolsandclearpage

\end{appendix}

%
%
\balancecolsandclearpage

\end{document}